\documentclass[aos]{imsart}

\usepackage[utf8]{inputenc} 
\usepackage[T1]{fontenc}    
\usepackage{microtype}
\usepackage{booktabs}       
\usepackage{nicefrac}       
\usepackage{times}
\usepackage{tikz,pgf}
\usetikzlibrary{positioning,babel}
\usepackage{epsfig,ifthen,comment}
\usepackage{dsfont}
\usepackage{xr}
\usepackage{xcolor}
\usepackage[colorlinks,citecolor=blue,urlcolor=blue]{hyperref}
\usepackage{graphicx}
\usepackage{amsthm,amsmath,amsfonts,amssymb}
\usepackage[parfill]{parskip}
\usepackage{caption}
\usepackage{subcaption}
\usepackage[american]{babel}
\usepackage{mathtools}
\usepackage{comment}

\usepackage{cancel}

\usepackage[square,numbers,sort&compress]{natbib}
\allowdisplaybreaks

\renewcommand{\cite}{\citep}
\startlocaldefs

\theoremstyle{plain}

\theoremstyle{remark}

\endlocaldefs 

\def\ci{\perp\!\!\!\perp}

\newenvironment{prf}{\noindent\textit{Proof:}\begin{mdseries}}{\end{mdseries}{\hfill\scriptsize$\Box$}}

\newcommand{\G}{{\cal G}}

\newcommand{\E}{\mathbb{E}}

\newcommand{\I}{\mathbb{I}}

\renewcommand{\o}{^{(1)}}

\newcommand{\aff}{{\text{aff}}}

\newtheorem{thm}{Theorem}

\newtheorem{asmp}{Assumption}

\usepackage{mathtools}

\DeclareMathOperator{\doo}{do}
\DeclareMathOperator{\dis}{dis}

\DeclareMathOperator{\pa}{pa}

\DeclareMathOperator{\ch}{ch}
\DeclareMathOperator{\an}{an}

\DeclareMathOperator{\mb}{mb}

\DeclareMathOperator{\pre}{pre}

\DeclareMathOperator{\pas}{pa^s}

\newenvironment{thma}[1]{\par\noindent{\bf Theorem #1\ }\em}{\em}

\usepackage{xcolor}

\usepackage[normalem]{ulem}
\newcommand{\stkout}[1]{\ifmmode\text{\sout{\ensuremath{#1}}}\else\sout{#1}\fi}

\begin{document}
	
	\begin{frontmatter}
		\title{Graphical Models of Entangled Missingness}
		\runtitle{Entangled Missingness}
		
		\begin{aug}
		
			\author[A]{\fnms{Ranjani}~\snm{Srinivasan}\ead[label=e1]{ranjani@jhu.edu}},
			\author[B]{\fnms{Rohit}~\snm{Bhattacharya}\ead[label=e2]{rb17@williams.edu}},
			\author[C]{\fnms{Razieh}~\snm{Nabi}\ead[label=e3]{razieh.nabi@emory.edu}},
			\author[D]{\fnms{Elizabeth}~\snm{L. Ogburn}\ead[label=e4]{eogburn@jhsph.edu}},
			\and
			\author[E]{\fnms{Ilya}~\snm{Shpitser}\ead[label=e5]{ilyas@cs.jhu.edu}}
			\address[A]{Department of Electrical and Computer Engineering, 
				Johns Hopkins University\printead[presep={,\ }]{e1}}
			
			\address[B]{Department of Computer Science, 
				Williams College\printead[presep={,\ }]{e2}}
			
			\address[C]{Department of Biostatistics and Bioinformatics, 
				Emory University\printead[presep={,\ }]{e3}}
			
			\address[D]{Department of Biostatistics, 
				Johns Hopkins University\printead[presep={,\ }]{e4}}
			
			\address[E]{Department of Computer Science, 
				Johns Hopkins University\printead[presep={,\ }]{e5}}

		\end{aug}

		\begin{abstract}
			Despite the growing interest in causal and statistical inference for settings with data dependence, 
			few methods currently exist to account for missing data in dependent data settings; most classical missing data methods in statistics and causal inference treat data units as independent and identically distributed (i.i.d.). We develop a graphical modeling based framework for causal inference in the presence of \textit{entangled missingness}, defined as missingness with data dependence. We distinguish three different types of \textit{entanglements} that can occur, supported by real-world examples. We give sound and complete identification results for all three settings. We show that existing missing data models may be extended to cover entanglements arising from (1) \textit{target law dependence} and (2) \textit{missingness process dependence},  
			while those arising from (3) \textit{missingness interference} require a novel approach.  
			We demonstrate the use of our entangled missingness framework on synthetic data.
			Finally, we discuss how, subject to a certain reinterpretation of the variables in the model, our model for missingness interference extends missing data methods to novel missing data patterns in i.i.d. settings. 
		\end{abstract}
		
%
		
	\end{frontmatter}

	\section{Introduction}
	\label{sec:intro}
	
	Missing data is a ubiquitous problem: data may be missing due to survey non-response, dropout or loss-to-followup, imperfect data collection, or other complications. Data records missing \emph{systematically} can substantially bias subsequent analyses if not properly addressed. Most of the literature on missing data treats data units as independent and identically distributed (i.i.d.) \citep{rubin1976, little2021missing, little2002statistical, glymour2006using, daniel2012using, martel2013definition, mohan13missing, thoemmes2014cautious, tian2015missing, shpitser2016consistent, bhattacharya19mid, nabi2020full, mohan2021graphical, scharfstein2021markov, nabi2022testability, nabi2022causal}. This assumption is reasonable in many types of problems where interactions between units are negligible and can be ignored. However, in recent years there has been increasing recognition that  data in many settings are subject to dependence across units and, in particular, to \textit{interference}, where variables measured on one unit may have a causal effect on those measured on another unit. The canonical example is infectious disease settings where treating one unit may have a protective effect on others, but more recently attention has turned to the myriad ways in which humans 
	influence one another through their social networks (broadly construed to include familial, work, social, neighborhood, or online relationships) \cite{aronow2013ace, athey2018network, basse2019interference, basse2018network, bowers2013interference, cai2019idcontagion, eck2022randomization, eckles2017design, forastiere2021identification, nabi2022ads, graham2010measuring, halloran1995causal, halloran2012causal, hong2006evaluating, hudgens2008toward, jagadeesan2020designs, toulis2013estimation, leung2020treatment, liu2014large, papadogeorgou2019causal, puelz2019graph, rosenbaum2007interference, rubin1990comment, savje2021causal, savje2021average, sobel2006randomized,tchetgen2012causal, toulis2018propensity, vanderweele2010direct}. 
	
	The current literature lags behind in recognizing that data dependence and missing data might occur simultaneously, with the exception of a few pieces of work: \cite{chang2020multiple} investigated multiple imputation techniques for missing data in health data networks; 
	\cite{smith2017network} empirically characterized bias in the analysis of network data with systematic missing data; \cite{gile2017analysis} used a likelihood-based modeling approach for health studies with partially observed data; \cite{almquist2018dynamic} proposed estimation methods for network logistic regression models in the presence of missing data.
	
	To our knowledge, this paper presents the first nonparametric\footnote{Here, we use the term nonparametric to mean that we place no restrictions on the functional forms of relationships between variables. 
	} method to deal with settings with \textit{entangled missingness}, i.e. with both missingness and data dependence, and also the first approach to such settings based on causal graphical models. We develop a graphical modeling framework to deal with {three different types of} entangled missingness {termed \textit{target law dependence}, \textit{missingness process dependence}, and \textit{missingness interference}. }
	We 
	derive {sound and complete nonparametric identification} results for {all the three scenarios} in which missingness and data dependence may interact. Our results offer full generality for causal estimands in the presence of entangled missingness. 
	
	\textit{Target law dependence} occurs when the full data distribution exhibits either statistical or causal dependence. For example, consider a study of vaccine effectiveness in which one unit's vaccination status may help to protect their friends or family members from contracting an infectious disease. 
	The underlying full data distribution is non-iid and failing to account for this dependence when imputing or otherwise controlling for missing data would generally introduce bias.
	
	\textit{Missingness process dependence} occurs when the missingness status of one unit may depend on (variables of) other units. 
	For example, an individual's decision to participate in a survey may depend on their friends' or family members' decisions. Similarly, consider 
	a mass public health surveillance effort, where demographic and family data are collected on a large population of individuals from which a smaller group is invited to participate in substudy where more detailed data are collected.  An individual's choice to enroll in the substudy may depend both on their own characteristics and on the characteristics of, or choices made by, other individuals in their social network.  
	
	\textit{Missingness interference} occurs 
	when there exist multiple potential outcomes of a variable depending on whether outcomes of other units in the network are measured or not (corresponding to participation of these other units in the study.)
	Consider an evaluation of an event or program, e.g. a course evaluation, using a survey filled out by the attendees. Answers to survey questions in this case would certainly be unobserved for any individual who did not attend or who dropped out before the end of the course.  However, because classes have social and collaborative components, the underlying values of responses of a particular individual, \emph{had they attended,} will potentially differ depending on the attendance of that individual's peers. A similar story could be told about student performance assessments, where there is evidence that classroom composition affects the performance of individual students \cite{sacerdote2011peer}. These settings have not previously been addressed in any literature on missing data or interference. We propose modeling multiple versions of underlying satisfaction or achievement variables in a way that is structurally similar to counterfactual variables or potential outcomes in causal inference.
	

{The paper is organized as follows}. We present a review of relevant background on graphical causal models, interference, and current graphical models for missing data in Sec.~\ref{sec:prelim}. We introduce new graphical models representing entangled missingness 
in Sec.~\ref{sec:graphical-entangled-missing}. Our new identification results for entangled missingness are in Sec.~\ref{sec:generalized-id}. In Sec.~\ref{sec:iid} we show that our model for missingness interference extends missing data methods to novel missing data patterns in i.i.d. settings. Sec.~\ref{sec:experiments}  illustrates the utility of our framework in simulations.  
We conclude the paper in Sec.~\ref{sec:discussion} by outlining remaining open challenges for non-parametric identification of arbitrary estimands in settings with entangled missingness.
	

\section{Graphical Causal Models, Interference, and Missing Data}
\label{sec:prelim}

Causal inference is traditionally concerned with \textit{counterfactual} or \textit{potential} outcomes like $Y(a)$, defined as the outcome that would have been observed if, possibly contrary to fact, the unit had received treatment or exposure value $A=a$.  For simplicity we will assume binary treatment throughout, but our results generalize to multi-level and, with some additional subtleties, continuous $A$. Causal effects are typically defined as contrasts of counterfactual outcomes, e.g. $\beta \coloneqq \E[Y(1) - Y(0)]$ is a comparison of the expected value of counterfactual outcomes in a world in which every unit receives treatment $A=1$ compared with the expected value of counterfactual outcome in a world in which every unit receives treatment $A=0$. Since potential outcomes are not necessarily observed in available data realizations, assumptions are needed to link the counterfactual distributions and the \textit{observed data distribution}, from which available samples are actually drawn.

A standard set of assumptions used to express $\beta$ as a functional of observed data are: (i) \emph{consistency}, stating that the observed outcome realization is equal to the counterfactual outcome realization had treatment been set to the actually observed value, or $Y(a) = Y$ if $A=a$, (ii) \emph{positivity}, stating that all treatment assignments, possibly conditioned on a set of baseline covariates ${\bf C}$, have positive support, or if  $p({\bf C} = {\bf c}) >0$ then $p(a \mid {\bf C} {= {\bf c}}) > 0$ for all $a$ in the support of $A$, and (iii) \emph{conditional ignorability}, stating that potential outcomes are independent of the treatment assignment, possibly conditioned on a set of baseline covariates ${\bf C}$, or $Y(a) \ci A \mid {\bf C}$ for all $a$. Under these assumptions, the parameter $\beta$ of the full data distribution $p(Y(1), Y(0), A, {\bf C})$ is identified from the observed data distribution $p(Y, A, {\bf C})$ via the \emph{adjustment functional}: $\E[\E[Y | A=1,{\bf C}] - \E[Y | A=0 ,{\bf C}]]$. 

{Conditional independence assumptions like conditional ignorability can be encoded by graphical models; this is the approach that we will take in this paper to present identification results in the presence of entangled missingness. We briefly introduce graphical models notation, terminology and concepts that will be used in this work. We have attempted to contain all required concepts within this section, but for a more thorough treatment of graphical models, see \cite{pearl09causality, maathuis2018handbook, thomas13swig}. 

	\subsection{Graphical Causal Models}
	A directed acyclic graph (DAG) $\G$ with vertex set ${\bf V}$ representing variables ${\bf V}$, consists of only directed  edges ($\rightarrow$), with no directed cycles. 
	The statistical model of a DAG ${\cal G}({\bf V})$ is the set of distributions $p({\bf V})$  that factorize as $p({\bf V}) = \prod_{V \in {\bf V}} p(V \mid \pa_{\cal G}(V))$ where $\pa_{\cal G}(V)$ denotes parents of $V$ in $\G({\bf V})$; $U$ is a parent of $V$ if $U \rightarrow V$, with $U,V \in {\bf V}$. All conditional independence restrictions in $p({\bf V})$ encoded in the DAG $\G({\bf V})$ can be read by examining paths in $\G({\bf V})$ via d-separation rules \citep{pearl88probabilistic}. 
	
	Causal models of a DAG combine a generative model of $p({\bf V})$ with the theory of interventions to yield distributions over counterfactual  random variables.  These models are often defined using \emph{structural equations} (which are a mathematical formulation of an invariant causal mechanism), as follows. Each variable $V$ in a causal model is determined from values of its parents $\pa_{\cal G}(V)$ and an exogenous noise variable $\epsilon_V$ via a {structural equation} $f_V(\pa_{\cal G}(V), \epsilon_V)$. Given an intervention on $V \in {\bf V}$ that sets $V$ to $v$, the structural equation $f_V(\pa_{\cal G}(V), \epsilon_V)$ is replaced by one that outputs the constant value $v$.  Structural equations and noise variables together produce the observed data distribution $p({\bf V})$. 
	If any subset ${\bf A} \subseteq {\bf V}$ are set to values ${\bf a}$, $p({\bf Y}({\bf a}))$ denotes the joint distribution over potential outcomes
	${\bf Y} = {\bf V} \setminus {\bf A}$ to this intervention, written as $p({\bf Y} \mid \text{do}({\bf a}))$ in \citep{pearl09causality}.
	For the purposes of this paper, we will assume the structural equations and noise terms are such that the resulting $p({\bf V})$ is a positive distribution.
	
	A popular causal model called the \emph{structural causal model}, or the \emph{non-parametric structural equation model with independent errors (NPSEM-IE)} \citep{pearl2009causality} assumes that the joint distribution of all exogenous terms are marginally independent: $p(\epsilon) = \prod_{V \in {\bf V}} p(\epsilon_V)$.  The NPSEM-IE implies the DAG factorization of $p({\bf V})$ with respect to ${\cal G}({\bf V})$, and a modified DAG factorization known as the \emph{g-formula} that identifies every joint intervention distribution as the following functional of the observed data distribution $p({\bf V})$: 
	\begin{align}
		p({\bf Y}({\bf a})) =
		\prod_{Y \in {\bf Y}} p(Y \mid \pa_{\cal G}(Y)) \Big\vert_{{\bf A} = {\bf a}},
		\label{eqn:g}
	\end{align}%
	for every ${\bf A} \subset {\bf V}$ and ${\bf Y} = {\bf V} \setminus {\bf A}$. 
	
	The conditionally ignorable model described earlier can be represented via the DAG shown  in Fig.~\ref{fig:dags_intro}(a). 
	With ${\bf V} = \{ Y, A, C \}$, ${\bf A} = \{ A \}$ and ${\bf V} \setminus {\bf A} = \{Y, C\}$ in (\ref{eqn:g}), $p(Y(a),C(a)) = p(Y(a),C) = p(Y \mid A = a,C) \times p(C)$. As a consequence, the counterfactual mean $\E[Y(a)]$ is identified via the adjustment formula as $\E[Y(a)] = \sum_C \E[Y|A=a,C]\times p(C)$.\footnote{It is common practice in missing data and some causal inference literature to identify expectations as the target of interest, but we will discuss identification of distributions in this paper to better connect with causal graphical models identification literature.}
	{Equivalently}, the interventional target may be obtained using an \textit{inverse probability weighting} (IPW) functional $\E[ \{\I(A=a)/p(A \mid C)\} \times Y]$, which involves using propensity score $p(A \mid C)$ weights \cite{hernan2020whatif}. 
	The weights approach will be particularly useful when we consider identification in missing data models, where the missingness mechanism will be akin to the propensity scores in this functional (see Sec.~\ref{sec:graphical-missing-iid}). 
	
	Causal models with hidden variables may be associated with a DAG ${\cal G}({\bf V} \cup {\bf H})$, where ${\bf V}$ are observed and ${\bf H}$ are unobserved variables.  In such models not every causal effect is identified under the standard assumptions, and identification theory may be expressed on a latent projection graph ${\cal G}({\bf V})$ \citep{verma90equiv}, containing only vertices corresponding to observed variables.
	Latent projections are a type of mixed graph called an acyclic directed mixed graph (ADMG) containing directed edges, representing direct causation, and bidirected edges ($\leftrightarrow$) representing unobserved confounding.  As an example, the latent projection of a DAG in Fig.~\ref{fig:dags_intro}(a) if $C$ was an unobserved variable, is given in Fig.~\ref{fig:dags_intro}(b).
	Conditional independence statements from a distribution $p({\bf V})$ associated with an ADMG ${\cal G}({\bf V})$ may be read off by a generalization of d-separation called \textit{m-separation} \cite{richardson03markov}. 
	
	Identified causal effects in hidden variable models are given by the ID algorithm, which may be viewed as a generalization of the g-formula in the sense that causal effects identified by the ID algorithm are modified factorizations of ADMGs, just as the g-formula is the modified factorization of DAGs.
	For general identification results for average causal effects in causal 
	models 
	represented by DAGs and ADMGs
	, see \citep{tian02on,shpitser06id,huang06do,richardson17nested, bhattacharya2022semiparametric}.
	
	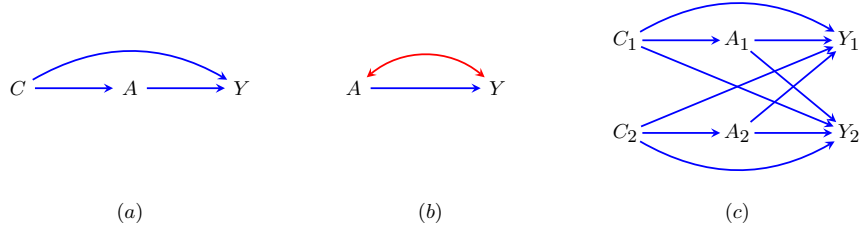
\begin{figure}[!t]
		\begin{center}
			\scalebox{0.85}{
				\begin{tikzpicture}[>=stealth, node distance=1.75cm]
					\tikzstyle{format} = [draw=none, thick, circle, minimum size=5.0mm,	inner sep=0pt]
					
					\begin{scope}[yshift=0cm]
						\path[->,  thick]
						node[format] (c) {$C$}
						node[format, right of=c] (a) {$A$}
						node[format, right of=a] (y) {$Y$}
						
						(c) edge[blue] (a)
						(a) edge[blue] (y)
						(c) edge[blue, bend left =30] (y)
						
						node[below of=a, yshift=-0.2cm, xshift=0cm] (l) {$(a)$}				;
					\end{scope}
					
					\begin{scope}[yshift=0cm, xshift=3.5cm]
						\path[->,  thick]
						node[] (c) {}
						node[format, right of=c] (a) {$A$}
						node[format, right of=a, xshift=0.5cm] (y) {$Y$}
						
						(a) edge[blue] (y)
						
						(a) edge[<->, red, bend left=40] (y)
						
						node[below of=a, yshift=-0.2cm, xshift=1.2cm] (l) {$(b)$}				;
					\end{scope}
					
					\begin{scope}[yshift=0.75cm, xshift=9.5cm]
						\path[->,  thick]
						node[format] (c1) {$C_1$}
						node[format, below of=c1, xshift=0.cm, yshift=0.3cm] (c2) {$C_2$}
						node[format, right of=c1] (a1) {$A_1$}
						node[format, right of=c2] (a2) {$A_2$}
						node[format, right of=a1] (y1) {$Y_1$}
						node[format, right of=a2] (y2) {$Y_2$}
						
						(c1) edge[blue] (a1)
						(c1) edge[blue, bend left=30] (y1)
						(c2) edge[blue] (a2)
						(c2) edge[blue, bend right=30] (y2)
						(a1) edge[blue] (y1)
						(a2) edge[blue] (y2)
						(a1) edge[blue] (y2)
						(a2) edge[blue] (y1)
						(c1) edge[blue] (y2)
						(c2) edge[blue] (y1)				
						
						node[below of=a2, yshift=0.5cm, xshift=0cm] (l) {$(c)$}				;
					\end{scope}
					
				\end{tikzpicture}
			}
		\end{center}
		\caption{(a) Example of a DAG representation in a conditionally ignorable model;	
			(b) A latent projection of the DAG in (a) if $C$ is an unobserved variable;
			(c) Example of a DAG illustration for interference in a dyad.
		}
		\label{fig:dags_intro}
	\end{figure}

	\subsection{Interference}
	\label{subsec:interference}
	
	Historically, causal inference approaches assumed all units to be causally and statistically independent. This is, however, frequently violated and we say that \textit{interference} is present when one unit may causally affect other units \cite{aronow2013ace, athey2018network, basse2019interference, basse2018network, bowers2013interference, cai2019idcontagion, eck2022randomization, eckles2017design, forastiere2021identification, nabi2022ads, graham2010measuring, halloran1995causal, halloran2012causal, hong2006evaluating, hudgens2008toward, jagadeesan2020designs, toulis2013estimation, leung2020treatment, liu2014large, papadogeorgou2019causal, puelz2019graph, rosenbaum2007interference, rubin1990comment, savje2021causal, savje2021average, sobel2006randomized,tchetgen2012causal, toulis2018propensity, vanderweele2010direct}. 
	Sometimes interference is more narrowly defined as the causal effect of one unit's treatment on another's outcome; our use of the term is more general and may apply to covariates, missingness indicators, treatments, or outcomes.

	In this paper, in order to isolate issues of nonparametric identification from issues of statistical dependence, we will assume \textit{partial interference}: interference may occur within collections or \textit{blocks} of units of fixed finite size, but the blocks are mutually independent. We will additionally assume that the blocks are identically distributed. This simplifies our presentation considerably, but in principle all of our identification results can be extended to full interference or to blocks with heterogeneous structure for different realizations. We say that a full data estimand (such as the full data distribution) is identified if it can be expressed as a functional of the observed data distribution; this is agnostic to partial versus full interference. A separate but related issue concerns statistical estimability, which typically requires  identification plus a large number of independent (or weakly dependent) observations, and non-i.i.d. blocks could  pose a challenge for estimability.  Although the purpose of this paper is identification rather than estimation, we note that in the setting of i.i.d. blocks of fixed size,
	estimators may be derived using standard techniques for i.i.d data, simply by treating each \emph{block of units} as a \emph{unit} for the purposes of estimation. Traditional estimation results will hold immediately under an asymptotic regime that lets the number of blocks go to infinity while the block composition remains fixed and i.i.d. This suggests that we may choose to view each block either as a collection of units or as a single multivariate observation; we switch between these two views throughout.
	
	Let there be $b$ blocks, each block consisting of $m = n/b$ units, where $n$ is the total number of units. 
	Let ${\bf A} \equiv \left(A_{1}, \ldots, A_{m}\right)$ be the vector of treatment assignments for units $i=1, \ldots m$ in a block, and the m-dimensional vector $\bf a$ be a realization in the support of ${\bf A}$.  Similarly, let ${\bf Y} \equiv \left(Y_{1}, \ldots, Y_{m}\right)$ and ${\bf C} \equiv \left({\bf C}_{1}, \ldots, {\bf C}_{m}\right)$ be the vector of outcomes and covariates respectively. Since units within the same block may interact, we must index a counterfactual outcome of unit $i$ by interventions not only performed on unit $i$, but also on other units that belong to the same block. In particular, we define $Y_i({\bf a})$ to be the counterfactual outcome of unit $i$, where the treatment vector ${\bf A}$ is intervened on and set to ${\bf a}$. Compare this to the counterfactual $Y(a)$ in (\ref{eqn:g}) for a setting with no interference, where the outcome of a unit is also a response to intervention setting $A$ to $a$, but where $A$ represents only the unit's own treatment.
	
	Counterfactuals $Y_i({\bf a})$ that are indexed by treatments of every unit in the network may be used to define causal effects for individual units, or an average of effects for all units in the network.
	Parameters of interest in interference problems and related estimation strategies are described in detail in \cite{ogburn14interference, hudgens08toward, tchetgen12on, tchetgen17auto}.
	
	\citet*{ogburn14interference} extended causal models of DAGs from the i.i.d. setting to interference problems.  Fig.~\ref{fig:dags_intro}(c) is a typical example of how DAGs are used to represent interference, in this case in a block of $2$ units (a dyad).  In this DAG, the tuple $(C_1,A_1,Y_1)$ corresponds to variables of unit $1$ and the tuple$(C_2,A_2,Y_2)$ corresponds to variables of unit $2$.  The presence of edges between these two tuples encodes the causal influence of one unit's variables on another unit's variables, i.e. the presence of interference. 
	
	The principles of graphical causal models generalize from the i.i.d. setting to settings with interference, the only difference being that the graph now represents an entire block of units rather than a single unit representing an i.i.d. realization. As noted above, it is just a matter of switching our lens between two views: that of a collection of units or of a single unit's multivariate observation. Further, we are able to extend graph-based identification algorithms from i.i.d settings to the partial interference setting.

\subsection{Graphical Representation of Missing Data Models}

\label{sec:graphical-missing-iid}
A missing data model encodes assumptions about how missingness arises and how it relates to underlying variables of interest. The model is a set of distributions defined over a set of random variables $\{{\bf O}, {\bf Z}^{(1)},{\bf R},{\bf Z}\}$, where ${\bf O}$ denotes the set of variables that are always observed, ${\bf Z}^{(1)}$ denotes the set of variables that are potentially missing, ${\bf R}$ denotes the set of missingness indicators of the variables in ${\bf Z}^{(1)}$, and ${\bf Z}$ denotes the set of \textit{observed proxies} of the variables in ${\bf Z}^{(1)}$. Missingness indicator $R_Z \in {\bf R}$, corresponding to $Z^{(1)} \in {\bf Z}^{(1)}$, takes the value 1 when the variable $Z$ is observed, and 0 when it is not. 
The observed proxy $Z$ is deterministically defined as follows: 
\begin{align}
	Z = \begin{cases}Z^{(1)} &\text{ if } R_Z = 1, \\ \text{ ? } &\text{ if } R_Z = 0. \end{cases}
	\label{eqn:consistency-m}
\end{align}%

Using the terminology of causal models, each missingness indicator $R \in {\bf R}$ may be viewed as a treatment variable that can be intervened on, and each $Z \in {\bf Z}$ as an observed outcome. Thus, $Z^{(1)}$ is a counterfactual random variable had we, possibly contrary to fact, intervened and set the corresponding missingness indicator $R_Z$ to $1$. We can {thus} view (\ref{eqn:consistency-m}) as missing data equivalent of the consistency assumption in causal inference.

We call the distribution $p({\bf O}, {\bf Z}\o, {\bf R})$ the \emph{full law}, the distribution $p({\bf O},{\bf Z}\o)$ the \emph{target law}, the distribution $p({\bf O}, {\bf Z}, {\bf R})$ the \emph{observed data law}, and the conditional distribution $p({\bf R} \mid {\bf Z}\o,{\bf O})$ the \emph{missingness process}, or the \emph{missing data propensity score}. 

Typically we are interested in a functional of the target law, often of the form  $h({\bf O}, {\bf Z}\o)$, 
and the goal of missing data methods is to nonparametrically identify such functionals 
in terms of the observed data law.
This requires assumptions, and it is common to assume the missingness consistency assumption (\ref{eqn:consistency-m}) and a missing data version of the positivity assumption:
\begin{align}
	p({\bf R}=1 \mid {\bf Z}\o,{\bf O}) > 0. 
\end{align} 

As is the case with causal inference, consistency and positivity do not suffice for nonparametric identification of target law parameters, but the addition of certain independence assumptions may suffice for nonparametric identification of the full law and therefore of any functionals thereof. 
The current literature considers missing data models where such assumptions can be encoded on a DAG $\G({\bf V})$, where vertices ${\bf V}$ correspond to random variables in ${\bf O} \cup {\bf Z}^{(1)} \cup {\bf R} \cup {\bf Z}$, and certain additional restrictions are placed on ${\cal G}({\bf V})$: (i) each $Z \in {\bf Z}$ has only two parents,  $R_Z$ and $Z^{(1)}$, and (ii) variables in ${\bf R}$ cannot point to variables in ${\bf O} \cup {\bf Z}^{(1)}$. Restriction (i) is imposed by definition of $Z$ via (\ref{eqn:consistency-m}). 
In order to distinguish the deterministic relations implied by (\ref{eqn:consistency-m}) from probabilistic relations, we draw edges pointing into $Z \in {\bf Z}$ in gray. 
Restriction (ii) is imposed to ensure that, while changes in $R_Z$ cause changes in the observed proxies $Z \in \bf Z$, they do not result in changes to the underlying full data $Z^{(1)} \in {\bf Z}^{(1)}$.  
Our proposed framework, outlined in the following sections, relaxes the edge restrictions in (ii); see Sec.~\ref{sec:iid} for more details. 

\begin{figure}[!t]
	\begin{center}
		\scalebox{0.85}{
			\begin{tikzpicture}[>=stealth, node distance=1.5cm]
				\tikzstyle{format} = [draw=none, thick, circle, minimum size=5mm,	inner sep=0pt]
				
				\begin{scope}[xshift=0cm]
					\path[->,  thick]
					
					node[format] (c2) {$Y^{(1)}$}
					node[format, right of=c2, xshift=-0.4cm] (a2) {$R_Y$}
					node[format, right of=a2, xshift=-0.4cm] (y2) {$Y$}
					node[format, below of=c2, yshift=0.25cm] (c1) {$O$}
					
					(c2) edge[gray, bend left=30] (y2)
					(a2) edge[gray] (y2)
					(c1) edge[blue] (c2)
					
					node[below of=c1, yshift=1cm, xshift=1.2cm] (l) {$(a)$}				;
				\end{scope}
				
				\begin{scope}[xshift=3.8cm]
					\path[->,  thick]
					
					node[format] (c2) {$Y^{(1)}$}
					node[format, right of=c2, xshift=-0.4cm] (a2) {$R_Y$}
					node[format, right of=a2, xshift=-0.4cm] (y2) {$Y$}
					node[format, below of=c2,  yshift=0.25cm] (c1) {$O$}
					
					(c2) edge[gray, bend left=30] (y2)
					(a2) edge[gray] (y2)
					(c1) edge[blue] (a2)
					(c1) edge[blue] (c2)
					
					node[below of=c1, yshift=1cm, xshift=1.1cm] (l) {$(b)$}				;
				\end{scope}
				
				\begin{scope}[xshift=7.6cm]
					\path[->,  thick]
					
					node[format] (c2) {$Y^{(1)}$}
					node[format, right of=c2, xshift=-0.4cm] (a2) {$R_Y$}
					node[format, right of=a2, xshift=-0.4cm] (y2) {$Y$}
					node[format, below of=c2,  yshift=0.25cm] (c1) {$O$}
					
					(c2) edge[gray, bend left=30] (y2)
					(a2) edge[gray] (y2)
					(c1) edge[blue] (a2)
					(c2) edge[blue] (a2)
					(c1) edge[blue] (c2)
					
					node[below of=c1, yshift=1cm, xshift=1.1cm] (l) {$(c)$}				;
				\end{scope}
				
				\begin{scope} [xshift=9cm, yshift=-0cm] 
					\path[->, thick]
					
					node[format, xshift=2cm] (c) {${\bf C}$}
					node[format, right of=c, yshift=0.7cm, xshift=-0.2cm] (a1) {\small{$A^{(1)}_{\phantom{A}}$}}
					node[format, right of=c, yshift=-0.7cm, xshift=-0.2cm] (y1) {\small{$\:\:Y^{(1)_{\phantom{Y}}}$}}
					node[format, right of=a1, xshift=-0.4cm] (ra) {$R_{A}^{\phantom{(1)}}$}
					node[format, right of=y1, xshift=-0.4cm] (ry) {$R_{Y}^{\phantom{(1)}}$}
					
					node[format, right of=ra, xshift=-0.4cm] (a) {$A$}
					node[format,right of=ry, xshift=-0.4cm] (y) {$Y$}
					(c) edge[blue] (y1)
					(c) edge[blue] (a1)
					(c) edge[blue, opacity= 1] (ra)
					(c) edge[blue, opacity= 1] (ry)
					(ra) edge[gray, opacity=0.7] (a)
					(ry) edge[gray, opacity=0.7] (y)
					(y1) edge[gray, opacity=0.7, bend right = 25] (y)
					(a1) edge[gray, opacity=0.7, bend left = 25] (a)
					(y1) edge[<-,blue] (a1)

					node[below of=y1, yshift=0.45cm, xshift=0.7cm] (la) {$(d)$}
					;
				\end{scope}
				
			\end{tikzpicture}
		}
	\end{center}
	\caption{
		A missing data DAG example corresponding to (a) a MCAR model, 
		(b) a MAR model, and 
		(c) a MNAR model. 
		(d) A missing data MAR model with $Z \equiv \{A,Y\}$.
	}
	\label{fig:m_dags_intro}
\end{figure}
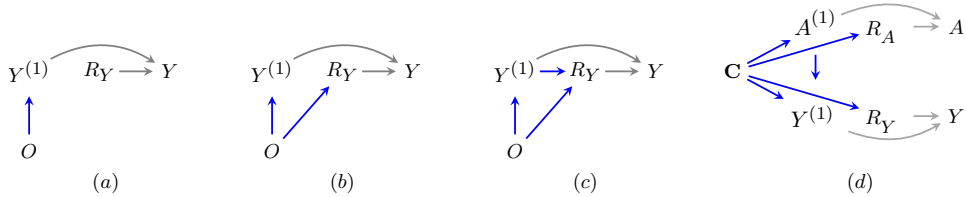

A missing data DAG model is a set of distributions defined over variables in ${\bf O} \cup {\bf Z}^{(1)} \cup {\bf R} \cup {\bf Z}$ that factorize with respect to a DAG with the above restrictions, as follows:  
\begin{align}
	p({\bf O}, {\bf Z}^{(1)}, {\bf R}, {\bf Z}) = \prod_{V \in {\bf O} \cup {\bf Z}^{(1)} \cup {\bf R}} p(V \mid \pa_{\G}(V)) \times \prod_{Z \in {\bf Z}} p(Z \mid R_Z, Z^{(1)}).
	\label{eqn:m-dag-fact}
\end{align}
Examples of missing data DAGs are shown in Fig.~\ref{fig:m_dags_intro}: Fig.~\ref{fig:m_dags_intro}(a) corresponds to a \textit{missing completely at random} (MCAR) model where $R_Y \ci O, Y^{(1)}$, Fig.~\ref{fig:m_dags_intro}(b)  corresponds to a \textit{missing at random} (MAR) model where $R_Y \ci Y^{(1)} \mid O$, and Fig.~\ref{fig:m_dags_intro}(c)  corresponds to a \textit{missing not at random} (MNAR) model where neither independence holds. 

If a missing data model contains hidden variables $\bf H$, it may be represented by the latent projection ADMG  ${\cal G}({\bf O},{\bf Z},{\bf Z}\o,{\bf R})$ of the DAG ${\cal G}({\bf H},{\bf O},{\bf Z},{\bf Z}\o,{\bf R})$, just like in the case of hidden variable causal models.  In such models, the target of inference is some function of the full data law $p({\bf O},{\bf Z}\o,{\bf R})$ where $\bf H$ is marginalized out. 

There is a substantial literature on non-graphical approaches to dealing with missing data.
A simple approach is to use only fully observed rows in complete-case analysis, justifiable only if the underlying mechanism is MCAR \cite{rubin1976}. For MAR scenarios, many approaches including expectation maximization \cite{dempster77maximum, horton1999maximum, little2002statistical}, multiple imputation \cite{rubin1988overview, schafer1999multiple}, inverse probability weighting \cite{robins94estimation, li2013weighting} have been developed. For MNAR problems, often parametric
or semiparametric restrictions have been imposed on the underlying data distribution and missingness selection model, such that they yield identification \cite{little2002statistical, tchetgen2018discrete,wu1988est,wang2014instrumental,miao2016identifiability,miao2016varieties,sun2018semiparametric}. In this work we use graphical models to derive conditions for nonparametric identification.  

The full law in a missing data model is identified \textit{if and only if} the missingness mechanism is identified by some functional of the observed data law $g(p({\bf R},{\bf O},{\bf Z}))$, because
\begin{align}
	p({\bf O}, {\bf Z}\o, {\bf R}) =  p({\bf O},{\bf Z}\o) \times \underbrace{p({\bf R} \mid {\bf O}, {\bf Z}\o)}_{g(p({\bf R},{\bf O},{\bf Z}))}
	= \frac{p({\bf O}, {\bf Z}\o, {\bf R}=1)}{\underbrace{p({\bf R}=1 \mid {\bf O}, {\bf Z}\o)}_{g(p({\bf R},{\bf O},{\bf Z}))\vert_{{\bf R}=1}}} \times \underbrace{p({\bf R} \mid {\bf O}, {\bf Z}\o)}_{g(p({\bf R},{\bf O},{\bf Z}))}.
	\label{eq:propensity-missing-id}
\end{align}
Note that this functional mirrors the IPW-based estimator we outlined for causal models earlier, with the missingness mechanism serving as the inverse weight. 

As a quick example, the full law in the MAR missing data DAG model of Fig. \ref{fig:m_dags_intro}(d) is identified because the missingness mechanism $p(R_A, R_Y | {\bf C}, A\o, Y\o) = p(R_A, R_Y | {\bf C})$ is a function of observed data. 

The hierarchy of MCAR, MAR and MNAR mechanisms  holds relevance in the context of identification simply due to the fact that if missingness is not at random, the full law may not be identified. Hence, when we provide illustrative examples and graphs in Sec.~\ref{sec:graphical-entangled-missing}, we will often point out what kind of mechanism is at play.

\section{Graphical Representation of Entangled Missingness}
\label{sec:graphical-entangled-missing}
	
Having discussed existing models for missing data and interference, we now propose three types of \emph{entanglements} by which the two phenomena can occur together. In naming and describing these phenomena, we hope to avert the common practice 
of simply ignoring missing data in settings with interference.

The following definitions pertain to the law $p({\bf R,O,Z\o,Z})$ with all hidden variables marginalized out. 
\begin{enumerate}
	\item Target Law Dependence ($\mathcal{E}_1$): Counterfactuals ${\bf Z}_i\o$ and ${\bf Z}_j\o$ of units $i, \:j$ in a block depend on each other. This means the
	target law $p({\bf O},{\bf Z}\o)$ does not factorize into unit-specific marginal distributions.
	Such a situation arises in problems where the underlying distribution, had there been no missing data, exhibits data dependence or interference. 
	
	\item Missingness Process Dependence ($\mathcal{E}_2$):
	Missingness indicators ${\bf R}_i$ for unit $i$ depend on variables corresponding to unit $j$, and thus the missingness process $p({\bf R} \mid   {\bf O, Z\o})$ does not factorize into unit-specific factors. 
	Similar situations have been considered in causal inference applications \cite{jackson2020adjusting}. 
	
	\item Missingness Interference ($\mathcal{E}_3$): Observed proxies ${\bf Z}_i$ for unit $i$ depend on missingness indicators ${\bf R}_j$ for unit $j.$\footnote{This can be interpreted as an extension of classical interference, where one unit's treatment affects another unit's outcome, except that the treatment has been replaced by the missingness indicator. Hence the name missingness interference.}
	Under $\mathcal{E}_3$, the counterfactual variable corresponding to  variable $Z_i$ 
	must be indexed not only by $R_{Z_i}$ being set to $1$, but also by the missingness status of (variables of) other units. For instance the counterfactual $Z^{(R_{Z_i} = 1, R_{Z_j}=0)}_i$ would correspond to variable $Z$ of unit $i$, had it been observed ($R_{Z_i}=1$) and had variable $Z$ of unit $j$ been missing. 
\end{enumerate}


These three types of entanglement can occur in any combination, and the seven possible situations where at least one entanglement occurs collectively describe all scenarios where data dependence and missingness occur together. 
In describing these scenarios in detail, we use the notation $\mathcal{E}_i$ to indicate that the $i$-th kind of entanglement is present and use $\bar {\mathcal{E}_i}$ to indicate its absence. For example, we will use ${\mathcal{E}_1} \bar {\mathcal{E}_2}\bar {\mathcal{E}_3}$ to denote a scenario when there is target law dependence but no missingness process dependence or missingness interference.

We first present entanglements \textit{without} missingness interference in Sec.~\ref{subsec:scalar-ctfl}, as it turns out that these forms of entanglements share a graphical representation with existing i.i.d. missing data models. In Sec.~\ref{subsec:vector-ctfl}, we describe scenarios \emph{with} missingness interference; these require the development of new notation and graphical representations. 

\subsection{Entanglements Without Missingness Interference}
\label{subsec:scalar-ctfl}

Consider studying the effect of drug $A$ on disease status $Y$ from a community hospital database. 
Assume hospital records are incomplete and missingness indicators $R_{A_i},\: R_{Y_i}$ denote whether the treatment and disease status values were recorded for patient $i$, with counterfactuals $A\o_{i} \: Y\o_{i},$  denoting the true (but possibly unrecorded) values of these variables. Baseline covariates, denoted by ${\bf C}_i$ for unit $i$, are always observed and include age and alcohol consumption. In this example, ${\bf Z} \equiv \{A,Y\}$ and ${\bf O} \equiv \{{\bf C}\}$. For simplicity of presentation we will work with dyads. 

In order to provide a contrast between settings with i.i.d. missing data models and those with entanglement, we first start with a scenario which involves no kind of entanglement ($\bar {\mathcal{E}_1} \bar {\mathcal{E}_2}\bar {\mathcal{E}_3}$) within this setup and build the examples further to illustrate how entanglements might arise. 

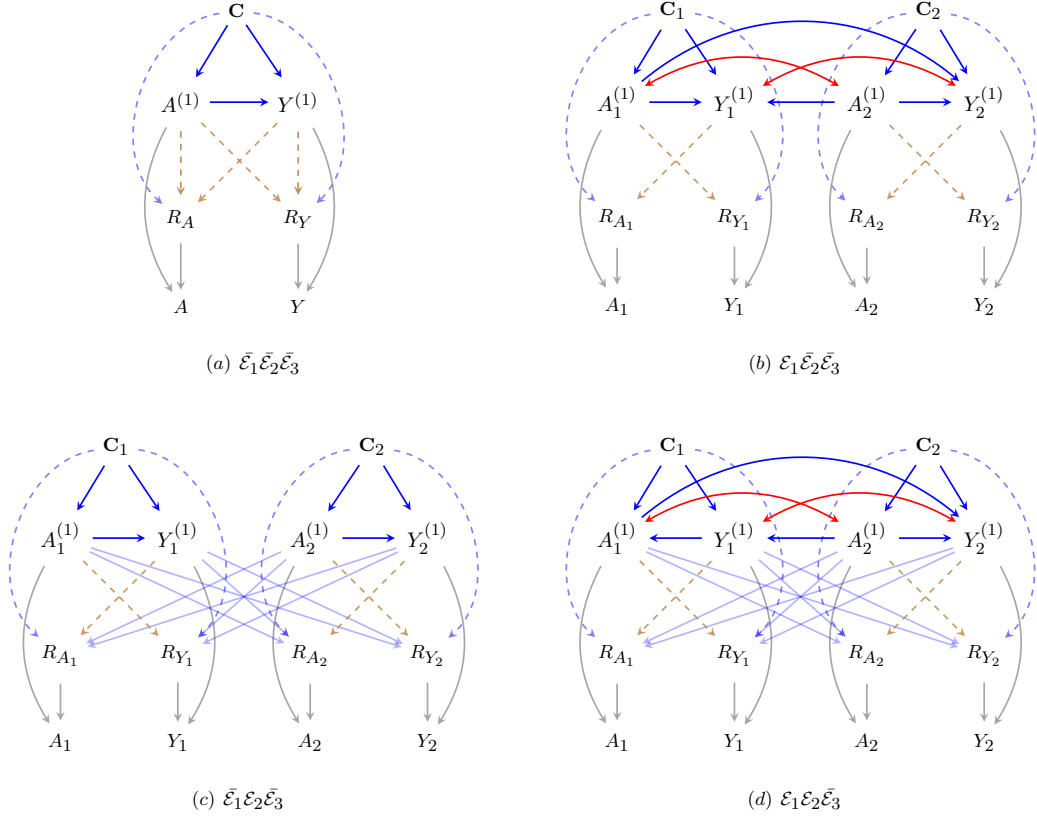
\begin{figure}[!t]
	\begin{center}
		\scalebox{0.8}{
			\begin{tikzpicture}[>=stealth, node distance=1.45cm]
				\tikzstyle{format} = [draw=none, thick, circle, minimum size=4mm, inner sep=2pt]
				
				\begin{scope} [xshift=0cm, yshift=0cm] 
					\path[->, thick]
					
					node[format, xshift=2cm] (o1) {${\bf C}$}
					node[format, below right of=o1, yshift=-0.5cm] (y111) {\small{$Y^{(1)}$}}
					node[format, left of=y111, xshift=-0.5cm] (y121) {\small{$A^{(1)}$}}
					node[format, below of=y121,  yshift=-0.5cm] (r12) {$R_{A}$}
					node[format, below of=y111, yshift=-0.5cm] (r11) {$R_{Y}$}
					node[format, below of=r12] (y12) {$A$}
					node[format, below of=r11] (y11) {$Y$}
					
					(o1) edge[blue] (y111)
					(o1) edge[blue] (y121)
					
					(o1) edge[blue, dashed, opacity= 0.5, bend left=65] (r11)
					(o1) edge[blue, dashed, opacity= 0.5, bend right=65] (r12)
					
					(r11) edge[gray, opacity=0.7] (y11)
					(r12) edge[gray, opacity=0.7] (y12)
					
					(y111) edge[gray, opacity=0.7, bend left = 30] (y11)
					(y121) edge[gray, opacity=0.7, bend right = 30] (y12)
					
					(y111) edge[<-,blue] (y121)
					
					(y111) edge[brown, dashed, opacity=0.8] (r12)	
					(y121) edge[brown, dashed, opacity=0.8] (r11)
					
					(y111) edge[brown, dashed, opacity=0.9] (r11)
					(y121) edge[brown, dashed, opacity=0.9] (r12)
					
					node[below of=y11, yshift=0.5cm, xshift=-0.75cm] (la) {$(a) \:\: \bar {\mathcal{E}_1} \bar {\mathcal{E}_2}\bar {\mathcal{E}_3}$}
					;
				\end{scope}

				\begin{scope} [xshift=9.25cm, yshift=0cm] 
					\path[->, thick]
					
					node[format] (o1) {${\bf C}_1$}
					node[format, below right of=o1, yshift=-0.5cm] (y111) {\small{$Y_{1}^{(1)}$}}
					
					node[format, below of=y111, yshift=-0.5cm] (r11) {$R_{Y_1}$}
					node[format, below of=r11] (y11) {$Y_{1}$}
					
					node[format, left of=y111, xshift=-0.5cm] (y121) {\small{$A_{1}^{(1)}$}}
					node[format, below of=y121,  yshift=-0.5cm] (r12) {$R_{A_1}$}
					node[format, below of=r12] (y12) {$A_{1}$}
					
					(o1) edge[blue] (y111)
					(o1) edge[blue] (y121)
					
					(o1) edge[blue, dashed, opacity= 0.5, bend left=65] (r11)
					(o1) edge[blue, dashed, opacity= 0.5, bend right=65] (r12)
					
					(r11) edge[gray, opacity=0.7] (y11)
					(r12) edge[gray, opacity=0.7] (y12)
					
					(y111) edge[gray, opacity=0.7, bend left = 30] (y11)
					(y121) edge[gray, opacity=0.7, bend right = 30] (y12)
					
					(y111) edge[<-,blue] (y121)

					
					node[format, right of=y111,	xshift=0.75cm] (y211) {\small{$A_{2}^{(1)}$}}
					node[format, above right of=y211, yshift=0.5cm] (o2) {${\bf C}_2$}
					
					node[format, below of=y211, yshift=-0.5cm] (r21) {$R_{A_2}$}
					node[format, below of=r21] (y21) {$A_{2}$}

					node[format, right of=y211, xshift=0.5cm] (y221) {\small{$Y_{2}^{(1)}$}}
					node[format, below of=y221,  yshift=-0.5cm] (r22) {$R_{Y_2}$}
					node[format, below of=r22] (y22) {$Y_{2}$}
					
					(o2) edge[blue] (y211)
					(o2) edge[blue] (y221)
					
					(o2) edge[blue, dashed, opacity= 0.5, bend right=65] (r21)
					(o2) edge[blue, dashed, opacity= 0.5, bend left=65] (r22)
					
					(r21) edge[gray, opacity=0.7] (y21)
					(r22) edge[gray, opacity=0.7] (y22)
					
					(y211) edge[gray, opacity=0.7, bend right = 30] (y21)
					(y221) edge[gray, opacity=0.7, bend left = 30] (y22)
					(y121) edge[->,blue, bend left=40] (y221)
					
					(y211) edge[->,blue] (y221)
					
					(y211) edge[<->,red, bend right=30] (y121)
					
					(y111) edge[<-,blue] (y211)
					(y111) edge[<->,red, bend left=30] (y221)
					
					(y111) edge[brown, dashed, opacity=0.8] (r12)	
					(y121) edge[brown, dashed, opacity=0.8] (r11)
					(y211) edge[brown, dashed, opacity=0.8] (r22)	
					(y221) edge[brown, dashed, opacity=0.8] (r21)

					node[below of=y11, yshift=0.5cm, xshift=1.cm] (la) {$(b) \:\:  {\mathcal{E}_1} \bar {\mathcal{E}_2}\bar {\mathcal{E}_3}$}
					;
				\end{scope}

				\begin{scope} [xshift=0cm, yshift=-7.25cm] 
					\path[->, thick]
					
					node[format] (o1) {${\bf C}_1$}
					node[format, below right of=o1, yshift=-0.5cm] (y111) {\small{$Y_{1}^{(1)}$}}
					
					node[format, right of=y111,	xshift=0.75cm] (y211) {\small{$A_{2}^{(1)}$}}
					node[format, above right of=y211, yshift=0.5cm] (o2) {${\bf C}_2$}
					node[format, below of=y111, yshift=-0.5cm] (r11) {$R_{Y_1}$}
					node[format, below of=y211, yshift=-0.5cm] (r21) {$R_{A_2}$}
					node[format, below of=r11] (y11) {$Y_{1}$}
					node[format, below of=r21] (y21) {$A_{2}$}
					
					node[format, left of=y111, xshift=-0.5cm] (y121) {\small{$A_{1}^{(1)}$}}
					node[format, below of=y121,  yshift=-0.5cm] (r12) {$R_{A_1}$}
					node[format, below of=r12] (y12) {$A_{1}$}
					
					node[format, right of=y211, xshift=0.5cm] (y221) {\small{$Y_{2}^{(1)}$}}
					node[format, below of=y221,  yshift=-0.5cm] (r22) {$R_{Y_2}$}
					node[format, below of=r22] (y22) {$Y_{2}$}
					
					(o1) edge[blue] (y111)
					(o2) edge[blue] (y211)
					(o1) edge[blue] (y121)
					(o2) edge[blue] (y221)
					
					(o1) edge[blue, dashed, opacity= 0.5, bend left=65] (r11)
					(o1) edge[blue, dashed, opacity= 0.5, bend right=65] (r12)
					(o2) edge[blue, dashed, opacity= 0.5, bend right=65] (r21)
					(o2) edge[blue, dashed, opacity= 0.5, bend left=65] (r22)
					
					(r11) edge[gray, opacity=0.7] (y11)
					(r12) edge[gray, opacity=0.7] (y12)
					(r21) edge[gray, opacity=0.7] (y21)
					(r22) edge[gray, opacity=0.7] (y22)
					
					(y111) edge[gray, opacity=0.7, bend left = 30] (y11)
					(y121) edge[gray, opacity=0.7, bend right = 30] (y12)
					(y211) edge[gray, opacity=0.7, bend right = 30] (y21)
					(y221) edge[gray, opacity=0.7, bend left = 30] (y22)
					
					(y111) edge[<-,blue] (y121)
					(y211) edge[->,blue] (y221)
					
					(y211) edge[blue,opacity=0.3] (r11)
					(y211) edge[blue, opacity=0.3] (r12)	
					(y221) edge[blue, opacity=0.3] (r11)
					(y221) edge[blue, opacity=0.3] (r12)	
					(y111) edge[blue,  opacity=0.3] (r21)
					(y111) edge[blue,  opacity=0.3] (r22)	
					(y121) edge[blue, opacity=0.3] (r21)
					(y121) edge[blue, opacity=0.3] (r22)
					
					(y111) edge[brown, dashed, opacity=0.8] (r12)	
					(y121) edge[brown, dashed, opacity=0.8] (r11)
					(y211) edge[brown, dashed, opacity=0.8] (r22)	
					(y221) edge[brown, dashed, opacity=0.8] (r21)

					node[below of=y11, yshift=0.5cm, xshift=1.cm] (la) {$(c) \:\: \bar {\mathcal{E}_1}  {\mathcal{E}_2}\bar {\mathcal{E}_3}$}
					;
				\end{scope}

				\begin{scope} [xshift=9.25cm,yshift=-7.25cm]
					\path[->, thick]
					
					node[format] (o1) {${\bf C}_1$}
					node[format, below right of=o1, yshift=-0.5cm] (y111) {\small{$Y_{1}^{(1)}$}}
					node[format, right of=y111,	xshift=0.75cm] (y211) {\small{$A_{2}^{(1)}$}}
					node[format, above right of=y211, yshift=0.5cm] (o2) {${\bf C}_2$}
					node[format, below of=y111, yshift=-0.5cm] (r11) {$R_{Y_1}$}
					node[format, below of=y211, yshift=-0.5cm] (r21) {$R_{A_2}$}
					node[format, below of=r11] (y11) {$Y_{1}$}
					node[format, below of=r21] (y21) {$A_{2}$}
					
					node[format, left of=y111, xshift=-0.5cm] (y121) {\small{$A_{1}^{(1)}$}}
					node[format, below of=y121,  yshift=-0.5cm] (r12) {$R_{A_1}$}
					node[format, below of=r12] (y12) {$A_{1}$}
					
					node[format, right of=y211, xshift=0.5cm] (y221) {\small{$Y_{2}^{(1)}$}}
					node[format, below of=y221,  yshift=-0.5cm] (r22) {$R_{Y_2}$}
					node[format, below of=r22] (y22) {$Y_{2}$}
					
					(o1) edge[blue] (y111)
					(o2) edge[blue] (y211)
					(o1) edge[blue] (y121)
					(o2) edge[blue] (y221)
					
					(o1) edge[blue, dashed, opacity= 0.5, bend left=65] (r11)
					(o1) edge[blue, dashed, opacity= 0.5, bend right=65] (r12)
					(o2) edge[blue, dashed, opacity= 0.5, bend right=65] (r21)
					(o2) edge[blue, dashed, opacity= 0.5, bend left=65] (r22)
					
					(r11) edge[gray, opacity=0.7] (y11)
					(r12) edge[gray, opacity=0.7] (y12)
					(r21) edge[gray, opacity=0.7] (y21)
					(r22) edge[gray, opacity=0.7] (y22)
					
					(y111) edge[gray, opacity=0.7, bend left = 30] (y11)
					(y121) edge[gray, opacity=0.7, bend right = 30] (y12)
					(y211) edge[gray, opacity=0.7, bend right = 30] (y21)
					(y221) edge[gray, opacity=0.7, bend left = 30] (y22)
					
					(y111) edge[->,blue] (y121)
					(y211) edge[->,blue] (y221)
					(y111) edge[<->,red, bend left=30] (y221)
					(y211) edge[<->,red, bend right=30] (y121)
					
					(y111) edge[<-,blue] (y211)
					(y121) edge[->,blue, bend left=40] (y221)
					
					(y211) edge[blue, opacity=0.3] (r11)
					(y211) edge[blue, opacity=0.3] (r12)	
					(y221) edge[blue, opacity=0.3] (r11)
					(y221) edge[blue, opacity=0.3] (r12)	
					(y111) edge[blue, opacity=0.3] (r21)
					(y111) edge[blue, opacity=0.3] (r22)	
					(y121) edge[blue, opacity=0.3] (r21)
					(y121) edge[blue, opacity=0.3] (r22)
					
					(y111) edge[brown, dashed, opacity=0.8] (r12)	
					(y121) edge[brown, dashed, opacity=0.8] (r11)
					(y211) edge[brown, dashed, opacity=0.8] (r22)	
					(y221) edge[brown, dashed, opacity=0.8] (r21)

					node[below of=y11, yshift=0.5cm, xshift=1.cm] (la) {$(d)\:\: {\mathcal{E}_1}  {\mathcal{E}_2}\bar  {\mathcal{E}_3}$}
					;
				\end{scope}
				
			\end{tikzpicture}
		}
	\end{center}
	\caption{Four scenarios representing all possible ways target law dependence and missingness process dependence may arise in a dyadic partial interference setting without  missingness interference arising.}
	\label{fig:scalar-graphs}
\end{figure}

\textbf{Scenario 1.1 ($\bar {\mathcal{E}_1} \bar {\mathcal{E}_2}\bar {\mathcal{E}_3}$).}
When the disease being investigated is not contagious, and the missingness process, pertaining to non-response, deficiencies in data collection, and so on, does not exhibit dependence across patients, we might be able to assume that a patient's record does not influence the record of any other patient, i.e.,  data is i.i.d. 

The graph in Fig.~\ref{fig:scalar-graphs}(a) depicts this situation. Including different set of edges in Fig.~\ref{fig:scalar-graphs}(a) yields either MCAR where $R_{A}, R_{Y} \ci  A\o, Y\o, {\bf C}$ (if only solid edges are included), MAR where $R_{A}, R_{Y} \ci A\o, Y\o \mid {\bf C}$ (if in addition to the solid lines, blue dashed edges are also included) \footnote{Note that the MAR version is identical to that in Fig.~\ref{fig:m_dags_intro}(d) as all of Scenario 1.1 pertains to i.i.d. settings.}, or MNAR (if at least one brown dashed edge is included). Edges $A \rightarrow R_A$ and $Y \rightarrow R_Y$ are called \textit{self-censoring} or \textit{self-masking} edges \cite{brown1990protecting, mohan2018estimation,nabi2020full}.%

\textbf{Scenario 1.2 ($\mathcal{E}_1 \bar {\mathcal{E}_2} \bar {\mathcal{E}_3}$).} Suppose the outcome $Y$ is an infectious disease, like COVID-19. Since patients can infect one another, data on patients residing in the same household or geographic area are likely to exhibit dependence in outcomes. In addition, surging infections may lead to treatment shortages, which will lead to dependence among treatments due to \emph{allocational interference} \cite{ogburn14interference}; geographically localized treatment shortages occurred at several points early in the COVID-19 pandemic. Finally, successful prevention or treatment of an infectious outcome for unit $i$ may influence outcomes for other units $j$ by preventing potential transmission from $i$ to $j$. Despite these complications, missingness mechanisms remain independent across units.

Fig.~\ref{fig:scalar-graphs}(b) depicts this setting. Bidirected arrows encode dependence (due to shared hidden causes) between $A_1^{(1)}$ and $A_2^{(1)}$ and between $Y_1^{(1)}$ and $Y_2^{(1)}$, respectively. For simplicity of visualization, we omit connections among {${\bf C}_i$ and $\{A_j, Y_j\}$. } 
The edge subgraph of Fig.~\ref{fig:scalar-graphs}(b) consisting only of solid edges corresponds to a MCAR model, since ${\bf R} \ci {\bf Y}\o, {\bf A}\o,  {\bf C}$ by the m-separation criterion  \cite{richardson03markov}. Similarly, the edge subgraph of Fig.~\ref{fig:scalar-graphs}(b) consisting of all edges other than dashed brown corresponds to a MAR model, since ${\bf R} \ci {\bf Y}\o, {\bf A}\o \mid {\bf C}$. Finally, including dashed brown edges in the MAR model yields an MNAR model.  

\textbf{Scenario 1.3 ($\bar {\mathcal{E}_1} {\mathcal{E}_2} \bar {\mathcal{E}_3}$).} Assume that we are again dealing with a non-contagious disease, and that treatment allocation for one unit is not influenced by any features of other units. However, the missingness process for one unit does depend on other units. For example, data includes members of a family who may jointly decide to change healthcare providers, resulting in missing treatment and outcome data for all members of the family. 

Fig.~\ref{fig:scalar-graphs}(c) depicts this setting, and the light blue edges between the counterfactuals $A_{1}\o, Y_{1}\o$ and missingness indicators $R_{A_2}, R_{Y_2}$, as well as $Y_{2}\o, Y_{2}\o$ and $R_{A_1}, R_{Y_1}$ represent missingness process dependence. Edges from ${\bf C}_2$ to variables of unit $1$, and from ${\bf C}_1$ to variables of unit $2$ are omitted for simplicity. 
An MCAR version of this scenario simply reduces to scenario $\bar {\mathcal{E}_1} \bar {\mathcal{E}_2} \bar {\mathcal{E}_3}$, representing i.i.d. data.  
An example of a MAR mechanism in this scenario corresponds to the absence of light solid blue edges in Fig.~\ref{fig:scalar-graphs}(c) and instead having edges ${\bf C}_1 \rightarrow R_{A_2}$, ${\bf C}_1 \rightarrow R_{Y_2}$, ${\bf C}_2 \rightarrow R_{A_1}$ and ${\bf C}_2 \rightarrow R_{Y_1}$ (not shown). 
{Missingness process dependence can also be represented as a bidirected edge between variables of unit $i$ and missingness indicators of unit $j$, coming from a hidden common cause of the corresponding variable and the missingness indicator.}


\textbf{Scenario 1.4 (${\mathcal{E}_1} {\mathcal{E}_2} \bar {\mathcal{E}_3}$).} 
This scenario simply combines complications in scenarios 1.2 and 1.3, such that both target law dependence and missingness process dependence are present. The edges can be interpreted exactly as before, with scenarios 1.2 and 1.3. Specifically, the light blue solid edges between counterfactuals and missingness indicators make the missingness process dependence MNAR. Instead, if the dependence was on observed covariates, the mechanism would be MAR. Dashed edges, both brown and blue operate within a unit and simply induce dependence that is MNAR or MAR, respectively, without involving other units.  

Finally, we note that the graphs in Fig.~\ref{fig:scalar-graphs} can be extended in a straightforward fashion to a network where blocks are not just dyads, but are of arbitary size $m$, to illustrate the different types of entanglement discussed here. Identification results corresponding to the settings described until now are discussed in Sec.~\ref{subsec:id-wo-missingness-interference}.

\subsection{Entanglements With Missingness Interference}
\label{subsec:vector-ctfl}

In some situations where missingness and data dependence occur together, observed realizations of underlying variables are influenced by missingness indicators of multiple units, leading to a missing data analogue of
\emph{interference problems} in causal inference.

Consider the following example, where investigators are interested in evaluating outcomes of a health and wellness program that involves a diet and workout regimen of participants, based on their baseline health indicators. Participation in the study is through voluntary registration. Bob and Anne are neighbors, and see the advertisement for enrollment. Imagine two different scenarios: (1) Both Bob and Anne feel strongly about the program and enroll together. Since they participate together, they are able to easily plan their meals and physical activity, and motivate each other. As a result, Anne records a successful outcome, improved metabolism and better BMI. Alternatively, (2) Bob enrolls and his baseline variables are collected but he does not actually participate. Anne goes through the program but finds it hard to remain motivated
and follow effectively, and her results are not as good. Bob's final outcome (BMI) is not measured as he does not participate. In this example, Anne has two possible hypothetical program results under the two different scenarios: one where Bob participates, and the other where he does not. To put it another way, the observed results for Anne depend on the missingness status of Bob's results in the study. Similar arguments might be made for Bob.

Formally, for Anne ({the unit} indexed by $1$), the outcome BMI has two possible values: $Y_1^{(r_{Y_1}=1,r_{Y_2}=0)}$ for when Bob does not participate and his BMI is, hence, not recorded ($r_{Y_2}=0$) and $Y_1^{(r_{Y_1}=1,r_{Y_2}=1)}$ for when Bob does participate and his BMI is recorded ($r_{Y_2}=1$). We make two simplifications here. One, we assume that the only missing variable for any unit is the outcome, and hence denote $r_{Y_1}$ by $r_1$ and $r_{Y_2}$ by $r_2$. This assumption is only being made to simplify this example, and will be relaxed in the more general case (see Sec.~\ref{subsec:beyond-dyads}). Second, we drop $r_1$ from the superscript and denote the counterfactuals as $Y_1^{(1, r_2=0)}$ and $Y_1^{(1, r_2=1)}$. This latter simplification can be done without ambiguity because Anne's outcomes are defined only when $r_1=1$. 
{Anne's} missingness status $R_1$ and observable BMI $Y_1$ are recorded in the study data. This observed outcome is not just a function of her counterfactuals $Y_1^{(1,r_2=0)}$, $Y_1^{(1,r_2=1)}$ and her missingess status $R_1$, but also the missingness status of Bob's outcome, $R_2$. One can think of $R_2$ as a switch, selecting which counterfactual,  $Y_1^{(1,r_2=0)}$ or $Y_1^{(1,r_2=1)}$, is realized in the observation. By symmetry in this particular example, we assume the same is true for Bob. 
Thus, we have the following relationships between observed values $Y_1,Y_2$ of neighbors Bob and Anne and the corresponding counterfactuals:
\begin{align*}
	Y_1 \leftarrow
	\left\{ \begin{array}{cc}
		(1-r_2)Y_1^{(1,r_2=0)} + r_2Y_1^{(1,r_2=1)} &\text{ if }r_1 = 1\\
		?& \text{ if }r_1 = 0
	\end{array} \right. \\
	Y_2 \leftarrow 
	\left\{ \begin{array}{cc}
		(1-r_1)Y_2^{(1,r_1=0)} + r_1Y_2^{(1,r_1=1)} &\text{ if }r_2 = 1\\
		?& \text{ if }r_2 = 0
	\end{array} \right.
\end{align*}

{Since} this setting {involves multiple counterfactual versions of $Y_1$, it}
cannot be captured by missing data models described so far, nor described by the graphs in Fig.~\ref{fig:scalar-graphs}, which correspond to the full data law with a single counterfactual version of every variable.

\begin{figure}[!t]
	\begin{center}
		\scalebox{0.75}{
			\begin{tikzpicture}[>=stealth, node distance=1.35cm]
				\tikzstyle{format} = [draw=none, thick, circle, minimum size=3.5mm,
				inner sep=0pt]
				
				\begin{scope} [xshift=-5.5cm]
					\path[->, thick]
					
					node[format] (o1) {$O_1$}
					node[format, below right of=o1, yshift=-1cm] (y111) {\small{$Y_{1}^{(1,r_{2}=1)}$}}
					node[format, right of=y111, xshift=0.75cm] (y211) {\small{$Y_{2}^{(1,r_{1}=1)}$}}
					node[format, above right of=y211, yshift=1cm,xshift=0cm] (o2) {$O_2$}
					node[format, below of=y111, xshift=-1.2cm,yshift=-0.5cm] (r11) {$R_{1}$}
					node[format, below of=r11] (y11) {$Y_{1}$}
					
					node[format, left of=y111, xshift=-1cm] (y121) {\small{$Y_{1}^{(1,r_{2}=0)}$}}
					node[format, right of=y211, xshift=1cm] (y221) {\small{$Y_{2}^{(1,r_{1}=0)}$}}
					node[format, below of=y221,  xshift=-1.1cm, yshift=-0.5cm] (r21) {$R_{2}$}
					node[format, below of=r21] (y21) {$Y_{2}$}
					
					(o1) edge[blue] (y111)
					(o2) edge[blue] (y211)
					(o1) edge[blue] (y121)
					(o2) edge[blue] (y221)
					
					(r21) edge[black, opacity=0.3] (y11)
					(r11) edge[black, opacity=0.3] (y21)
					
					(o1) edge[blue, dotted, opacity= 0.4] (r11)
					(o2) edge[blue, dotted, opacity= 0.4] (r21)
					
					(r11) edge[black, opacity=0.3] (y11)
					(r21) edge[black, opacity=0.3] (y21)
					
					(y111) edge[black, opacity=0.3, bend left = 30] (y11)
					(y121) edge[black, opacity=0.3, bend right = 30] (y11)
					(y211) edge[black, opacity=0.3, bend right = 30] (y21)
					(y221) edge[black, opacity=0.3, bend left = 30] (y21)
					
					(y111) edge[<->,red] (y121)
					(y211) edge[<->,red] (y221)			
					(y111) edge[brown, dashed, opacity=0.8] (r11)	
					(y121) edge[brown, dashed, opacity=0.8] (r11)
					(y211) edge[brown, dashed, opacity=0.8] (r21)	
					(y221) edge[brown, dashed, opacity=0.8] (r21)
					
					node[below of=y11, yshift=0.4cm, xshift=2.5cm] (la) {$(a) \:\:\bar{\mathcal{E}_1} \bar{\mathcal{E}_2} {\mathcal{E}_3}$}
					;
				\end{scope}

				\begin{scope} [xshift=4.cm]
					\path[->, thick]
					
					node[format] (o1) {$O_1$}
					node[format, below right of=o1, yshift=-1cm] (y111) {\small{$Y_{1}^{(1,r_{2}=1)}$}}
					node[format, right of=y111, xshift=1.25cm] (y211) {\small{$Y_{2}^{(1,r_{1}=1)}$}}
					node[format, above right of=y211, yshift=1cm,xshift=-0cm] (o2) {$O_2$}
					node[format, below of=y111, xshift=-1.2cm,yshift=-0.5cm] (r11) {$R_{1}$}
					node[format, below of=r11] (y11) {$Y_{1}$}
					
					node[format, left of=y111, xshift=-1cm] (y121) {\small{$Y_{1}^{(1,r_{2}=0)}$}}
					node[format, right of=y211, xshift=1cm] (y221) {\small{$Y_{2}^{(1,r_{1}=0)}$}}
					node[format, below of=y221,  xshift=-1.1cm, yshift=-0.5cm] (r21) {$R_{2}$}
					node[format, below of=r21] (y21) {$Y_{2}$}
					
					(o1) edge[blue] (y111)
					(o2) edge[blue] (y211)
					(o1) edge[blue] (y121)
					(o2) edge[blue] (y221)
					
					(r21) edge[black, opacity=0.3] (y11)
					(r11) edge[black, opacity=0.3] (y21)
					
					(o1) edge[blue, dotted, opacity= 0.4] (r11)
					(o2) edge[blue, dotted, opacity= 0.4] (r21)
					
					(r11) edge[black, opacity=0.3] (y11)
					(r21) edge[black, opacity=0.3] (y21)
					
					(y111) edge[black, opacity=0.3, bend left = 30] (y11)
					(y121) edge[black, opacity=0.3, bend right = 30] (y11)
					(y211) edge[black, opacity=0.3, bend right = 30] (y21)
					(y221) edge[black, opacity=0.3, bend left = 30] (y21)
					
					(y111) edge[<->,red] (y121)
					(y211) edge[<->,red] (y221)
					
					(y111) edge[<->,red] (y211)
					(y111) edge[<->,red, bend left=20] (y221)
					(y211) edge[<->,red, bend right=20] (y121)
					(y121) edge[<->,red, bend left=40] (y221)
					
					(y111) edge[brown, dashed, opacity=0.8] (r11)	
					(y121) edge[brown, dashed, opacity=0.8] (r11)
					(y211) edge[brown, dashed, opacity=0.8] (r21)	
					(y221) edge[brown, dashed, opacity=0.8] (r21)
					
					node[below of=y11, yshift=0.4cm, xshift=2.5cm] (la) {$(b) \:\: \mathcal{E}_1 \bar{\mathcal{E}_2} {\mathcal{E}_3}$}
					;
				\end{scope}
				
				\begin{scope} [xshift=-5.5cm, yshift=-7.25cm]
					\path[->, thick]
					
					node[format] (o1) {$O_1$}
					node[format, below right of=o1, yshift=-1cm] (y111) {\small{$Y_{1}^{(1,r_{2}=1)}$}}
					node[format, right of=y111, xshift=0.75cm] (y211) {\small{$Y_{2}^{(1,r_{1}=1)}$}}
					node[format, above right of=y211, yshift=1cm,xshift=-0cm] (o2) {$O_2$}
					node[format, below of=y111, xshift=-1.2cm,yshift=-0.5cm] (r11) {$R_{1}$}
					node[format, below of=r11] (y11) {$Y_{1}$}
					
					node[format, left of=y111, xshift=-1cm] (y121) {\small{$Y_{1}^{(1,r_{2}=0)}$}}
					node[format, right of=y211, xshift=1cm] (y221) {\small{$Y_{2}^{(1,r_{1}=0)}$}}
					node[format, below of=y221,  xshift=-1.1cm, yshift=-0.5cm] (r21) {$R_{2}$}
					node[format, below of=r21] (y21) {$Y_{2}$}
					
					(o1) edge[blue] (y111)
					(o2) edge[blue] (y211)
					(o1) edge[blue] (y121)
					(o2) edge[blue] (y221)
					
					(r21) edge[black, opacity=0.3] (y11)
					(r11) edge[black, opacity=0.3] (y21)
					
					(o1) edge[blue, dotted, opacity= 0.4] (r11)
					(o2) edge[blue, dotted, opacity= 0.4] (r21)
					
					(r11) edge[black, opacity=0.3] (y11)
					(r21) edge[black, opacity=0.3] (y21)
					
					(y111) edge[black, opacity=0.3, bend left = 30] (y11)
					(y121) edge[black, opacity=0.3, bend right = 30] (y11)
					(y211) edge[black, opacity=0.3, bend right = 30] (y21)
					(y221) edge[black, opacity=0.3, bend left = 30] (y21)
					
					(y111) edge[<->,red] (y121)
					(y211) edge[<->,red] (y221)
					
					(y211) edge[blue, opacity=1] (r11)
					(y221) edge[blue, opacity=1] (r11)
					(y111) edge[blue, opacity=1] (r21)
					(y121) edge[blue, opacity=1] (r21)
					
					(y111) edge[brown, dashed, opacity=0.8] (r11)	
					(y121) edge[brown, dashed, opacity=0.8] (r11)
					(y211) edge[brown, dashed, opacity=0.8] (r21)	
					(y221) edge[brown, dashed, opacity=0.8] (r21)
					
					node[below of=y11, yshift=0.4cm, xshift=2.5cm] (la) {$(c)\:\: \bar{\mathcal{E}_1} {\mathcal{E}_2} {\mathcal{E}_3}$}
					;
				\end{scope}
				
				\begin{scope} [xshift=4.cm,yshift=-7.25cm]
					\path[->, thick]
					
					node[format] (o1) {$O_1$}
					node[format, below right of=o1, yshift=-1cm] (y111) {\small{$Y_{1}^{(1,r_{2}=1)}$}}
					node[format, right of=y111, xshift=1.25cm] (y211) {\small{$Y_{2}^{(1,r_{1}=1)}$}}
					node[format, above right of=y211, yshift=1cm,xshift=-0cm] (o2) {$O_2$}
					node[format, below of=y111, xshift=-1.2cm,yshift=-0.5cm] (r11) {$R_{1}$}
					node[format, below of=r11] (y11) {$Y_{1}$}
					
					node[format, left of=y111, xshift=-1cm] (y121) {\small{$Y_{1}^{(1,r_{2}=0)}$}}
					node[format, right of=y211, xshift=1cm] (y221) {\small{$Y_{2}^{(1,r_{1}=0)}$}}
					node[format, below of=y221,  xshift=-1.1cm, yshift=-0.5cm] (r21) {$R_{2}$}
					node[format, below of=r21] (y21) {$Y_{2}$}
					
					(o1) edge[blue] (y111)
					(o2) edge[blue] (y211)
					(o1) edge[blue] (y121)
					(o2) edge[blue] (y221)
					
					(r21) edge[black, opacity=0.3] (y11)
					(r11) edge[black, opacity=0.3] (y21)
					
					(o1) edge[blue, dotted, opacity= 0.4] (r11)
					(o2) edge[blue, dotted, opacity= 0.4] (r21)
					
					(r11) edge[black, opacity=0.3] (y11)
					(r21) edge[black, opacity=0.3] (y21)
					
					(y111) edge[black, opacity=0.3, bend left = 30] (y11)
					(y121) edge[black, opacity=0.3, bend right = 30] (y11)
					(y211) edge[black, opacity=0.3, bend right = 30] (y21)
					(y221) edge[black, opacity=0.3, bend left = 30] (y21)
					
					(y111) edge[<->,red] (y121)
					(y211) edge[<->,red] (y221)
					
					(y111) edge[<->,red] (y211)
					(y111) edge[<->,red, bend left=20] (y221)
					(y211) edge[<->,red, bend right=20] (y121)
					(y121) edge[<->,red, bend left=40] (y221)
					
					(y211) edge[blue, opacity=1] (r11)
					(y221) edge[blue, opacity=1] (r11)
					(y111) edge[blue, opacity=1] (r21)
					(y121) edge[blue, opacity=1] (r21)
					
					(y111) edge[brown, dashed, opacity=0.8] (r11)	
					(y121) edge[brown, dashed, opacity=0.8] (r11)
					(y211) edge[brown, dashed, opacity=0.8] (r21)	
					(y221) edge[brown, dashed, opacity=0.8] (r21)
					
					node[below of=y11, yshift=0.4cm, xshift=2.5cm] (la) {$(d)\:\: {\mathcal{E}_1} {\mathcal{E}_2} {\mathcal{E}_3}$}
					;
				\end{scope}
				
			\end{tikzpicture}
		}
	\end{center}
	\caption{
		Four scenarios representing all possible ways target law dependence and missingness process dependence may arise in a dyadic partial interference setting when missingness interference is present. 
		We assume that only outcomes $Y_1, Y_2$ are missing, and use $R_i$ as a shorthand for $R_{Y_i}$. 
	}
	\label{fig:vector-graphs}
\end{figure}
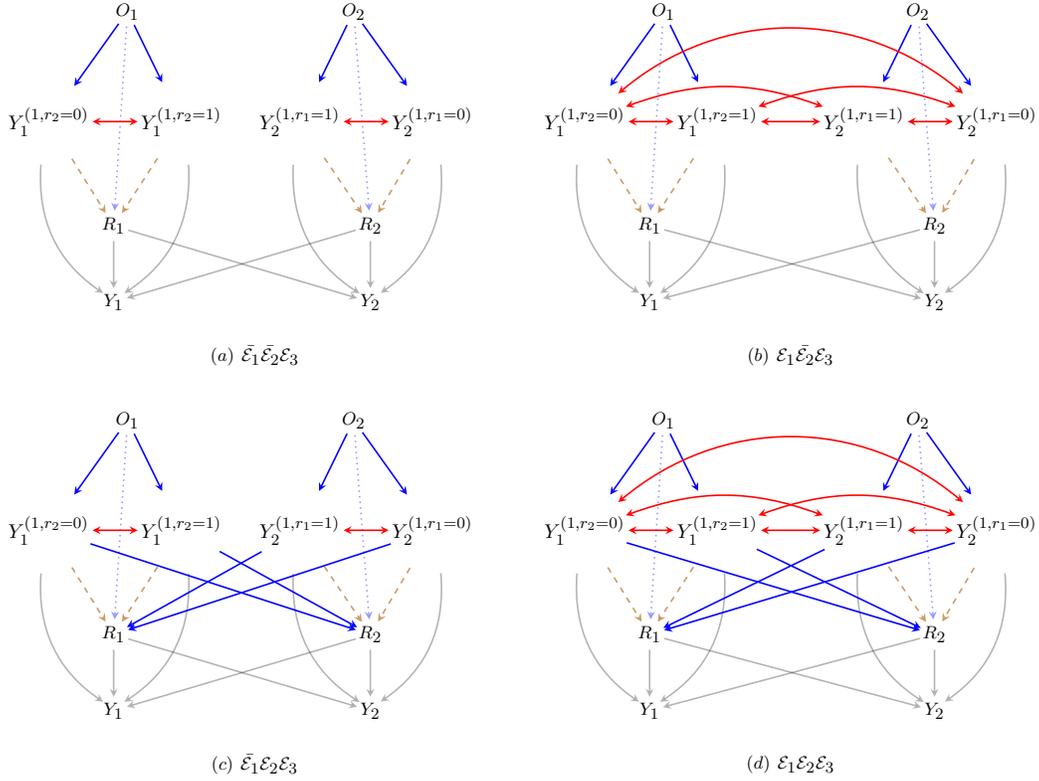

{We call, for every unit $i$, the set of units whose missingness indicators index the counterfactuals of unit $i$, as \emph{affectors} of $i${, or $\text{aff}(i)$,} in the network, inspired by the use of the word in neuroscience \cite{ebeling2011physics}. As one can imagine, this relationship need not be symmetric, and one unit can be the affector of another, without the reverse being true. 
With these considerations in mind, we now describe the four scenarios that arise when missingness interference may occur. }

\textbf{Scenario 2.1 ($\bar{\mathcal{E}_1} \bar{\mathcal{E}_2} {\mathcal{E}_3}$).} First, consider a setting where there is neither target law dependence, nor missingness process dependence, but only missingness interference.
Let us assume that diets and exercise regimens are recommended to participants, based solely on their measured baseline covariates. As an approximation, we assume that a person's BMI is affected only by their actions, and hence the target laws are not directly influenced by each other.
As described, Anne is more likely to workout if her friend Bob does ($R_j \rightarrow Y_i$), affecting her BMI. We assume that $Y_i^{(1,r_j=0)}$ and $Y_i^{(1,r_j=1)}$ are associated for any unit $i$ via a hidden common cause $H_i$, resulting in a bidirected $(\leftrightarrow)$ edge between the counterfactuals in the latent projection graphs. See Fig.~\ref{fig:vector-graphs}(a). Brown dashed edges {depict} self-censoring mechanisms.

\textbf{Scenario 2.2 (${\mathcal{E}_1} \bar{\mathcal{E}_2} {\mathcal{E}_3}$).} Now, assume that Bob and Anne are siblings. One might imagine that the effect of diet and exercise on an individual are mediated by their genetics, which siblings share, giving rise to target law dependence. Additionally, assume that those conducting the study randomize who gets to participate, from the list of participants who register (and hence, whether or not someone is included in the study is independent of their sibling). This can give rise to a structure as shown in Fig.~\ref{fig:vector-graphs}(b). {Adherence to prescribed regimens} is still dependent on a person's sibling being included in the study, giving rise to two underlying counterfactual BMIs for each unit. Here too, the dependence between counterfactuals is assumed to be due to a shared common cause $H$, resulting in target law dependence being represented by bidirected edges. 

\textbf{Scenario 2.3 ($\bar{\mathcal{E}_1} {\mathcal{E}_2} {\mathcal{E}_3}$).} 
Assume Bob and Anne are not genetically related. Further, assume that knowledge of Bob's potential BMI if he enrolled in the study (or not) makes Anne more likely to enroll herself, and the investigators include everyone who registers, without randomly selecting a subset of the entries, giving rise to missingness process dependence. This would result in a graph like in Fig.~\ref{fig:vector-graphs}(c). 
The blue edge $Y_{2}^{(1,r_1=1)} \rightarrow R_{1}$ introduces a more general type of self-censoring, which we call 
\textit{affector-censoring}. This corresponds to the situation where a counterfactual variable of a unit (which is indexed by its affector's missingness status) censors that very same missingness status. 

\textbf{Scenario 2.4 (${\mathcal{E}_1} {\mathcal{E}_2} {\mathcal{E}_3}$).} Finally assume a situation where Bob and Anne share genetics and are also likely to enroll based on their knowledge of how fit the other person is likely to become as part of the study, combining all three possible types of entanglements in one setup. This scenario is illustrated via the graph in Fig.~\ref{fig:vector-graphs}(d). 

\subsubsection{Beyond Dyads} \label{subsec:beyond-dyads}
Interactions in many realistic settings are not restricted to {dyads} and hence  
we describe counterfactuals (and graphs) for networks with missingness interference. 
{For simplicity of presentation, we assume the following: given units $i$ and $j$, if $R_{Z_j} \rightarrow Z_i$ exists, then $R_{Z'_j} \rightarrow Z_i$ also exists for all $Z'_j \in {\bf Z}_j$. We write this assumption as $j \in \aff(i)$. This enables us to index counterfactuals corresponding to unit $i$ by ${\bf R}_j$, which determines whether ${\bf Z}_j$ is missing, instead of individual variables $Z_j \in {\bf Z}_j$. We further assume within a unit, the edge $R_{Z_i} \rightarrow Z'_i$ is absent, for all distinct variables $Z, Z'$. } 
This simply allows us to replace $R_{Z_i}=1$ by $1$, in the index of counterfactuals of unit $i$. To understand the implications of removing these restrictions, see the note at the end of this section.


We assume a network of $n$ individuals. 
{The full law is defined over variables in $\bigcup_{i=1}^n {\bf O}_i \cup \tilde{\bf Z}_i \cup {\bf R}_{{\bf Z}_i} \cup {\bf H}_i$
where $\tilde{\bf Z}_i \equiv \{ {\bf Z}_i^{(1,{\bf r}_{\aff(i)})} : {\bf r}_{\aff(i)} \text{ a value of }{\bf R}_{\aff(i)} \}$ represents counterfactual versions of {variables} ${\bf Z}_i$ for unit $i$, the set ${\aff(i)}$ refers to the units that are affectors of unit $i$, and ${\bf R}_{\aff(i)} \equiv \{{\bf R}_{{\bf Z}_j} : j \in \aff(i) \}$ where ${\bf R}_{{\bf Z}_j}$ refers to the set of all missingness indicators $R_{Z_j}$  for all variables $Z_j\in {\bf Z}_j$ of unit $j$. In other words, variable{s} ${\bf Z}_i^{(1,{\bf r}_{\aff(i)})} \in \tilde{\bf Z}_i$, for every value ${\bf r}_{\aff(i)}$, reads ``the value assumed by variable{s} ${\bf Z}_i$, had {they} been observed, and had observability status of the affectors of $i$ in the network been set to ${\bf r}_{\aff(i)}$.'' 

As a generalization of (\ref{eqn:consistency-m}), every observed proxy $Z_i \in {\bf Z}_i$ is defined as:
\begin{align}
	\label{eqn:consistency-m-2}
	Z_i \leftarrow
	\left\{ \begin{array}{cc}
		Z_i^{(1,{\bf R}_{\aff(i)})}
		&\text{ if }R_{Z_i} = 1\\
		?& \text{ if }R_{Z_i} = 0
	\end{array} \right.,
\end{align}
{where ${\bf R}_{\aff(i)}$ in the superscript is interpreted to mean interventions that set ${\bf R}_{\aff(i)}$ to their naturally occurring values.}
Just as in causal interference, all counterfactuals corresponding to one variable $Z_i$ are 
dependent on each other in general. Next, we discuss graphs for these networks.

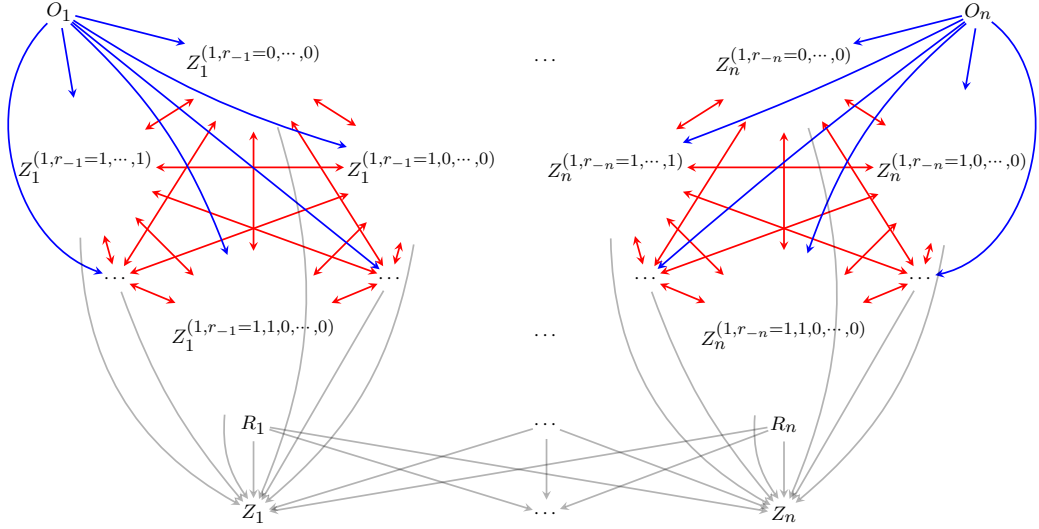
\begin{figure}[!t]
	\begin{center}
		\scalebox{0.82}{
			\begin{tikzpicture}[>=stealth, node distance=1.45cm]
				\tikzstyle{format} = [draw=none, thick, circle, minimum size=3.5mm,
				inner sep=0pt]
				
				\begin{scope} [xshift=-6cm]
					\path[->, thick]
					
					node[format] (x1) {$Z_1^{(1, r_{-1}= 0, \cdots ,0)}$}
					node[format, left of =x1, yshift=0.8cm,xshift=-1.7cm] (o1) {$O_1$}
					node[format, below right of =x1, xshift=1.7cm, yshift=-0.7cm] (x2) {$Z_1^{(1, r_{-1}= 1,0,\cdots ,0)}$}
					node[format, below right of =x1, yshift=-2.5cm, xshift=1.2cm] (x3) {$\cdots$}
					node[format, below of =x1, yshift=-3cm, xshift=0cm] (x4) {$Z_1^{(1, r_{-1}= 1,1,0, \cdots ,0)}$}
					node[format, below left of=x1, yshift=-2.5cm, xshift=-1.2cm] (x5) {$\cdots$}
					node[format, below left of =x1, yshift=-0.7cm, xshift=-1.7cm] (x6) {$Z_1^{(1, r_{-1}= 1,\cdots ,1)}$}
					node[format, below of =x4] (r1) {$R_1$}
					node[format, below of =r1] (z1) {$Z_1$}
					
					node[format, right of =x1, xshift = 3.3cm] (d2) {$\cdots$}
					node[format, right of =x4, xshift = 3.3cm] (d3) {$\cdots$}
					node[format, right of =r1, xshift = 3.3cm] (d4) {$\cdots$}
					node[format, right of =z1, xshift = 3.3cm] (d5) {$\cdots$}

					(x1) edge[<->,red] (x2)
					(x1) edge[<->,red] (x4)
					(x1) edge[<->,red] (x6)
					(x2) edge[<->,red] (x4)
					(x2) edge[<->,red] (x6)
					(x4) edge[<->,red] (x6)
					
					(x1) edge[<->,red] (x3)
					(x1) edge[<->,red] (x5)
					(x2) edge[<->,red] (x3)
					(x2) edge[<->,red] (x5)
					(x4) edge[<->,red] (x3)
					(x4) edge[<->,red] (x5)
					(x6) edge[<->,red] (x3)
					(x6) edge[<->,red] (x5)
					
					(o1) edge[blue] (x1)
					(o1) edge[blue, bend right=8] (x2)
					(o1) edge[blue] (x3)
					(o1) edge[blue, bend left=13] (x4)
					(o1) edge[blue,bend right=58] (x5)
					(o1) edge[blue] (x6)

					(r1) edge[black, opacity=0.3] (z1)
					(x1) edge[black, opacity=0.3, bend left=20] (z1)
					(x2) edge[black, opacity=0.3, bend left=20] (z1)
					(x3) edge[black, opacity=0.3, bend left=0] (z1)
					(x4) edge[black, opacity=0.3, bend right=20] (z1)
					(x5) edge[black, opacity=0.3, , bend right=10] (z1)
					(x6) edge[black, opacity=0.3, bend right=30] (z1)
					
					;
				\end{scope}

				\begin{scope} [xshift=2.6cm]
					\path[->, thick]
					node[format] (xx1) {$Z_n^{(1, r_{-n}= 0, \cdots ,0)}$}
					node[format, right of =xx1,yshift=0.8cm,xshift=1.7cm] (ox1) {$O_n$}
					node[format, below right of =xx1, xshift=1.7cm, yshift=-0.7cm] (xx2) {$Z_n^{(1, r_{-n}= 1,0,\cdots ,0)}$}
					node[format, below right of =xx1, yshift=-2.5cm, xshift=1.2cm] (xx3) {$\cdots$}
					node[format, below of =xx1, yshift=-3cm, xshift=0cm] (xx4) {$Z_n^{(1, r_{-n}= 1,1,0, \cdots ,0)}$}
					node[format, below left of=xx1, yshift=-2.5cm, xshift=-1.2cm] (xx5) {$\cdots$}
					node[format, below left of =xx1, yshift=-0.7cm, xshift=-1.7cm] (xx6) {$Z_n^{(1, r_{-n}= 1,\cdots ,1)}$}
					node[format, below of =xx4] (rx1) {$R_n$}
					node[format, below of =rx1] (zx1) {$Z_n$}
					
					(xx1) edge[<->,red] (xx2)
					(xx1) edge[<->,red] (xx4)
					(xx1) edge[<->,red] (xx6)
					(xx2) edge[<->,red] (xx4)
					(xx2) edge[<->,red] (xx6)
					(xx4) edge[<->,red] (xx6)
					
					(xx1) edge[<->,red] (xx3)
					(xx1) edge[<->,red] (xx5)
					(xx2) edge[<->,red] (xx3)
					(xx2) edge[<->,red] (xx5)
					(xx4) edge[<->,red] (xx3)
					(xx4) edge[<->,red] (xx5)
					(xx6) edge[<->,red] (xx3)
					(xx6) edge[<->,red] (xx5)
					
					(ox1) edge[blue] (xx1)
					(ox1) edge[blue] (xx2)
					(ox1) edge[blue, bend left=65] (xx3)
					(ox1) edge[blue, bend right=15] (xx4)
					(ox1) edge[blue, bend right=2] (xx5)
					(ox1) edge[blue, bend left=3] (xx6)

					(rx1) edge[black, opacity=0.3] (zx1)
					(xx1) edge[black, opacity=0.3, bend left=20] (zx1)
					(xx2) edge[black, opacity=0.3, bend left=20] (zx1)
					(xx3) edge[black, opacity=0.3, bend left=0] (zx1)
					(xx4) edge[black, opacity=0.3, bend right=20] (zx1)
					(xx5) edge[black, opacity=0.3, , bend right=10] (zx1)
					(xx6) edge[black, opacity=0.3, bend right=30] (zx1)

					(r1) edge[black, opacity=0.3] (zx1)
					(rx1) edge[black, opacity=0.3] (z1)
					(r1) edge[black, opacity=0.3] (d5)
					(rx1) edge[black, opacity=0.3] (d5)
					
					(d4) edge[black, opacity=0.3] (d5)
					(d4) edge[black, opacity=0.3] (z1)
					(d4) edge[black, opacity=0.3] (zx1)
					
					;
				\end{scope}

			\end{tikzpicture}
		}
	\end{center}
\caption{
	A simple example of a general network with $n$ units undergoing missingness interference. This graph represents an MCAR model, ${\bf R} \ci {\bf O}, {\bf Z}_i^{(1, {\bf r}_{\aff(i)})}$. 
}
\label{fig:vector-graphs-components}
\end{figure}

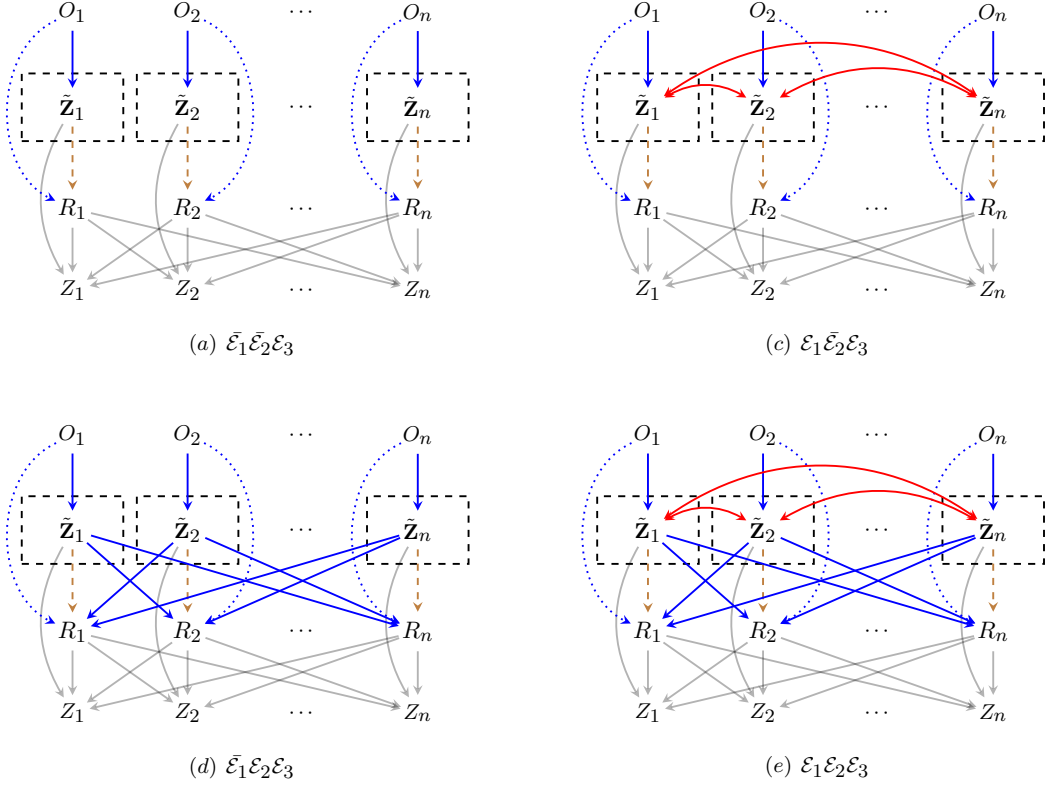
\begin{figure}[!t]
	\begin{center}
		\scalebox{0.9}{
			\begin{tikzpicture}[>=stealth, node distance=1.2cm]
				\tikzstyle{format} = [draw=none, thick, circle, minimum size=3.5mm,
				inner sep=0pt]
				
				\begin{scope} [xshift=-6cm]
					\path[->, thick]
					
					node[format, shape=rectangle, draw, color=black, opacity=1, dashed, minimum width=15mm, minimum height=10mm, xshift=0cm, yshift=-1.4cm] (u1) {$ $}
					
					node[format, shape=rectangle, draw, color=black, opacity=1, dashed, minimum width=15mm, minimum height=10mm, xshift=1.7cm, yshift=-1.4cm] (u2) {$ $}
					
					node[format, shape=rectangle, draw, color=black, opacity=1, dashed, minimum width=15mm, minimum height=10mm, xshift=5.1cm, yshift=-1.4cm] (ud) {$ $}
					
					node[format] (o1) {$O_1$}
					node[format, right of=o1, xshift=0.5cm] (o2) {$O_2$}
					node[format, right of=o2, xshift=0.5cm] (o3) {$ \cdots$}
					node[format, right of=o3, xshift=0.5cm] (od) {$O_n$}
					
					node[format, below of=o1, yshift=-0.2cm] (y1) {$\small{\tilde{\bf Z}_1}$}
					node[format, right of=y1,xshift=0.5cm] (y2) {$\small{\tilde{\bf Z}_2}$} 
					node[format, right of=y2,xshift=0.5cm] (y3) {$\cdots$}
					node[format, right of=y3,xshift=0.5cm] (yd) {$\small{\tilde{\bf Z}_n}$}
					
					node[format, below of=y1, yshift=-0.3cm] (r1) {\small{$R_1$}}
					node[format, below of=y2, yshift=-0.3cm] (r2) {\small{$R_2$}}
					node[format, below of=y3, yshift=-0.3cm] (r3) {$\cdots$}
					node[format, below of=yd, yshift=-0.3cm] (rd) {\small{$R_n$}}
					
					node[format, below of=r1] (yp1) {$Z_1$}
					node[format, below of=r2] (yp2) {$Z_2$}
					node[format, below of=r3] (yp3) {$\cdots$}
					node[format, below of=rd] (ypd) {$Z_n$}

					(o1) edge[blue] (y1)
					(o2) edge[blue] (y2)
					(od) edge[blue] (yd)
					
					(r1) edge[black, opacity=0.3] (yp1)
					(r2) edge[black, opacity=0.3] (yp2)
					(rd) edge[black, opacity=0.3] (ypd)
					
					(r1) edge[black, opacity=0.3] (yp2)
					(r1) edge[black, opacity=0.3] (ypd)
					
					(r2) edge[black, opacity=0.3] (yp1)
					(r2) edge[black, opacity=0.3] (ypd)
					
					(rd) edge[black, opacity=0.3] (yp1)
					(rd) edge[black, opacity=0.3] (yp2)
					
					(y1) edge[black, opacity=0.3, bend right=30] (yp1)
					(y2) edge[black, opacity=0.3, bend right=30] (yp2)
					(yd) edge[black, opacity=0.3, bend right=30] (ypd)
					
					(o1) edge[blue, dotted, bend right=65] (r1)
					(o2) edge[blue, dotted, bend left=65] (r2)
					(od) edge[blue, dotted, bend right=65] (rd)
					
					(y1) edge[brown, dashed, bend right=0] (r1)
					(y2) edge[brown, dashed, bend left=0] (r2)
					(yd) edge[brown, dashed, bend right=0] (rd)

					node[below of=y1, yshift=-2.3cm, xshift=2.5cm] (la) {$(a) \:\: \bar{\mathcal{E}_1} \bar{\mathcal{E}_2} \mathcal{E}_3$}
					;
				\end{scope}

				\begin{scope} [xshift=2.5cm]
					\path[->, thick]
					
					node[format, shape=rectangle, draw, color=black, opacity=1, dashed, minimum width=15mm, minimum height=10mm, xshift=0cm, yshift=-1.4cm] (u1) {$ $}
					
					node[format, shape=rectangle, draw, color=black, opacity=1, dashed, minimum width=15mm, minimum height=10mm, xshift=1.7cm, yshift=-1.4cm] (u2) {$ $}
					
					node[format, shape=rectangle, draw, color=black, opacity=1, dashed, minimum width=15mm, minimum height=10mm, xshift=5.1cm, yshift=-1.4cm] (ud) {$ $}
					
					node[format] (o1) {$O_1$}
					node[format, right of=o1, xshift=0.5cm] (o2) {$O_2$}
					node[format, right of=o2, xshift=0.5cm] (o3) {$ \cdots$}
					node[format, right of=o3, xshift=0.5cm] (od) {$O_n$}
					
					node[format, below of=o1, yshift=-0.2cm] (y1) {$\small{\tilde{\bf Z}_1}$}
					node[format, right of=y1,xshift=0.5cm] (y2) {$\small{\tilde{\bf Z}_2}$} 
					node[format, right of=y2,xshift=0.5cm] (y3) {$\cdots$}
					node[format, right of=y3,xshift=0.5cm] (yd) {$\small{\tilde{\bf Z}_n}$}
					
					node[format, below of=y1, yshift=-0.3cm] (r1) {\small{$R_1$}}
					node[format, below of=y2, yshift=-0.3cm] (r2) {\small{$R_2$}}
					node[format, below of=y3, yshift=-0.3cm] (r3) {$\cdots$}
					node[format, below of=yd, yshift=-0.3cm] (rd) {\small{$R_n$}}
					
					node[format, below of=r1] (yp1) {$Z_1$}
					node[format, below of=r2] (yp2) {$Z_2$}
					node[format, below of=r3] (yp3) {$\cdots$}
					node[format, below of=rd] (ypd) {$Z_n$}

					(o1) edge[blue] (y1)
					(o2) edge[blue] (y2)
					(od) edge[blue] (yd)
					
					(r1) edge[black, opacity=0.3] (yp1)
					(r2) edge[black, opacity=0.3] (yp2)
					(rd) edge[black, opacity=0.3] (ypd)
					
					(r1) edge[black, opacity=0.3] (yp2)
					(r1) edge[black, opacity=0.3] (ypd)
					
					(r2) edge[black, opacity=0.3] (yp1)
					(r2) edge[black, opacity=0.3] (ypd)
					
					(rd) edge[black, opacity=0.3] (yp1)
					(rd) edge[black, opacity=0.3] (yp2)
					
					(y1) edge[black, opacity=0.3, bend right=30] (yp1)
					(y2) edge[black, opacity=0.3, bend right=30] (yp2)
					(yd) edge[black, opacity=0.3, bend right=30] (ypd)
					
					(o1) edge[blue, dotted, bend right=65] (r1)
					(o2) edge[blue, dotted, bend left=65] (r2)
					(od) edge[blue, dotted, bend right=65] (rd)
					
					(y1) edge[brown, dashed, bend right=0] (r1)
					(y2) edge[brown, dashed, bend left=0] (r2)
					(yd) edge[brown, dashed, bend right=0] (rd)
					
					(y1) edge[red, <->, bend left=30] (y2)
					(y1) edge[red, <->, bend left=35] (yd)
					(y2) edge[red, <->, bend left=30] (yd)

					node[below of=y1, yshift=-2.3cm, xshift=2.5cm] (lc) {$(c) \:\: \mathcal{E}_1 \bar{\mathcal{E}_2} \mathcal{E}_3$}
					;
				\end{scope}

				\begin{scope} [xshift=-6cm, yshift=-6.25cm]
					\path[->, thick]
					
					node[format, shape=rectangle, draw, color=black, opacity=1, dashed, minimum width=15mm, minimum height=10mm, xshift=0cm, yshift=-1.4cm] (u1) {$ $}
					
					node[format, shape=rectangle, draw, color=black, opacity=1, dashed, minimum width=15mm, minimum height=10mm, xshift=1.7cm, yshift=-1.4cm] (u2) {$ $}
					
					node[format, shape=rectangle, draw, color=black, opacity=1, dashed, minimum width=15mm, minimum height=10mm, xshift=5.1cm, yshift=-1.4cm] (ud) {$ $}
					
					node[format] (o1) {$O_1$}
					node[format, right of=o1, xshift=0.5cm] (o2) {$O_2$}
					node[format, right of=o2, xshift=0.5cm] (o3) {$ \cdots$}
					node[format, right of=o3, xshift=0.5cm] (od) {$O_n$}
					
					node[format, below of=o1, yshift=-0.2cm] (y1) {$\small{\tilde{\bf Z}_1}$}
					node[format, right of=y1,xshift=0.5cm] (y2) {$\small{\tilde{\bf Z}_2}$} 
					node[format, right of=y2,xshift=0.5cm] (y3) {$\cdots$}
					node[format, right of=y3,xshift=0.5cm] (yd) {$\small{\tilde{\bf Z}_n}$}
					
					node[format, below of=y1, yshift=-0.3cm] (r1) {\small{$R_1$}}
					node[format, below of=y2, yshift=-0.3cm] (r2) {\small{$R_2$}}
					node[format, below of=y3, yshift=-0.3cm] (r3) {$\cdots$}
					node[format, below of=yd, yshift=-0.3cm] (rd) {\small{$R_n$}}
					
					node[format, below of=r1] (yp1) {$Z_1$}
					node[format, below of=r2] (yp2) {$Z_2$}
					node[format, below of=r3] (yp3) {$\cdots$}
					node[format, below of=rd] (ypd) {$Z_n$}

					(o1) edge[blue] (y1)
					(o2) edge[blue] (y2)
					(od) edge[blue] (yd)
					
					(r1) edge[black, opacity=0.3] (yp1)
					(r2) edge[black, opacity=0.3] (yp2)
					(rd) edge[black, opacity=0.3] (ypd)
					
					(r1) edge[black, opacity=0.3] (yp2)
					(r1) edge[black, opacity=0.3] (ypd)
					
					(r2) edge[black, opacity=0.3] (yp1)
					(r2) edge[black, opacity=0.3] (ypd)
					
					(rd) edge[black, opacity=0.3] (yp1)
					(rd) edge[black, opacity=0.3] (yp2)
					
					(y1) edge[black, opacity=0.3, bend right=30] (yp1)
					(y2) edge[black, opacity=0.3, bend right=30] (yp2)
					(yd) edge[black, opacity=0.3, bend right=30] (ypd)
					
					(o1) edge[blue, dotted, bend right=65] (r1)
					(o2) edge[blue, dotted, bend left=65] (r2)
					(od) edge[blue, dotted, bend right=65] (rd)
					
					(y1) edge[brown, dashed, bend right=0] (r1)
					(y2) edge[brown, dashed, bend left=0] (r2)
					(yd) edge[brown, dashed, bend right=0] (rd)
					
					
					(y2) edge[blue, bend right=0] (r1)
					(yd) edge[blue, bend left=0] (r1)
					(y1) edge[blue, bend right=0] (r2)
					(yd) edge[blue, bend right=0] (r2)	
					(y1) edge[blue, bend right=0] (rd)
					(y2) edge[blue, bend right=0] (rd)

					node[below of=y1, yshift=-2.3cm, xshift=2.5cm] (ld) {$(d) \:\: \bar{\mathcal{E}_1} \mathcal{E}_2 \mathcal{E}_3$}
					;
				\end{scope}
				
				\begin{scope} [xshift=2.5cm, yshift=-6.25cm]
					\path[->, thick]
					
					node[format, shape=rectangle, draw, color=black, opacity=1, dashed, minimum width=15mm, minimum height=10mm, xshift=0cm, yshift=-1.4cm] (u1) {$ $}
					
					node[format, shape=rectangle, draw, color=black, opacity=1, dashed, minimum width=15mm, minimum height=10mm, xshift=1.7cm, yshift=-1.4cm] (u2) {$ $}
					
					node[format, shape=rectangle, draw, color=black, opacity=1, dashed, minimum width=15mm, minimum height=10mm, xshift=5.1cm, yshift=-1.4cm] (ud) {$ $}
					
					node[format] (o1) {$O_1$}
					node[format, right of=o1, xshift=0.5cm] (o2) {$O_2$}
					node[format, right of=o2, xshift=0.5cm] (o3) {$ \cdots$}
					node[format, right of=o3, xshift=0.5cm] (od) {$O_n$}
					
					node[format, below of=o1, yshift=-0.2cm] (y1) {$\small{\tilde{\bf Z}_1}$}
					node[format, right of=y1,xshift=0.5cm] (y2) {$\small{\tilde{\bf Z}_2}$} 
					node[format, right of=y2,xshift=0.5cm] (y3) {$\cdots$}
					node[format, right of=y3,xshift=0.5cm] (yd) {$\small{\tilde{\bf Z}_n}$}
					
					node[format, below of=y1, yshift=-0.3cm] (r1) {\small{$R_1$}}
					node[format, below of=y2, yshift=-0.3cm] (r2) {\small{$R_2$}}
					node[format, below of=y3, yshift=-0.3cm] (r3) {$\cdots$}
					node[format, below of=yd, yshift=-0.3cm] (rd) {\small{$R_n$}}
					
					node[format, below of=r1] (yp1) {$Z_1$}
					node[format, below of=r2] (yp2) {$Z_2$}
					node[format, below of=r3] (yp3) {$\cdots$}
					node[format, below of=rd] (ypd) {$Z_n$}

					(o1) edge[blue] (y1)
					(o2) edge[blue] (y2)
					(od) edge[blue] (yd)
					
					(r1) edge[black, opacity=0.3] (yp1)
					(r2) edge[black, opacity=0.3] (yp2)
					(rd) edge[black, opacity=0.3] (ypd)
					
					(r1) edge[black, opacity=0.3] (yp2)
					(r1) edge[black, opacity=0.3] (ypd)
					
					(r2) edge[black, opacity=0.3] (yp1)
					(r2) edge[black, opacity=0.3] (ypd)
					
					(rd) edge[black, opacity=0.3] (yp1)
					(rd) edge[black, opacity=0.3] (yp2)
					
					(y1) edge[black, opacity=0.3, bend right=30] (yp1)
					(y2) edge[black, opacity=0.3, bend right=30] (yp2)
					(yd) edge[black, opacity=0.3, bend right=30] (ypd)
					
					(o1) edge[blue, dotted, bend right=65] (r1)
					(o2) edge[blue, dotted, bend left=65] (r2)
					(od) edge[blue, dotted, bend right=65] (rd)
					
					(y1) edge[brown, dashed, bend right=0] (r1)
					(y2) edge[brown, dashed, bend left=0] (r2)
					(yd) edge[brown, dashed, bend right=0] (rd)
					
					(y1) edge[red, <->, bend left=30] (y2)
					(y1) edge[red, <->, bend left=35] (yd)
					(y2) edge[red, <->, bend left=30] (yd)
					
					(y2) edge[blue, bend right=0] (r1)
					(yd) edge[blue, bend left=0] (r1)
					(y1) edge[blue, bend right=0] (r2)
					(yd) edge[blue, bend right=0] (r2)	
					(y1) edge[blue, bend right=0] (rd)
					(y2) edge[blue, bend right=0] (rd)

					node[below of=y1, yshift=-2.3cm, xshift=2.5cm] (le) {$(e) \:\: \mathcal{E}_1 \mathcal{E}_2 \mathcal{E}_3$}
					;
				\end{scope}
				
			\end{tikzpicture}
		}
	\end{center}
	\caption{
		{
			A general network with $n$ units featuring missingness interference. (a) - (d) enumerate all possible ways target law dependence and missingness process dependence may arise, as a generalization of the dyad in Fig. \ref{fig:vector-graphs}. 
		}
	}
	\label{fig:vector-graphs-generalized}
\end{figure}
{
	The full data distribution can be represented by a DAG, which contains a vertex for every ${ Z}_i^{(1,{\bf r}_{\aff(i)})}$, as well as every element in $\bigcup_{i=1}^n {\bf O}_i \cup {\bf Z}_i \cup {\bf R}_{{\bf Z}_i} \cup {\bf H}_i$. We also assume that the set of all counterfactuals of one variable $Z_i$ of unit $i$ share a common parent $H_{Z_i} \in {\bf H}_i$, since they are all associated. 
	This DAG obeys two restrictions, which generalize restrictions discussed in Sec.~\ref{sec:graphical-missing-iid}: (i) variables in ${\bf R}$ cannot point to variables in
	$\bigcup_{i=1}^n {\bf O}_i \cup \{ { \bf Z}_i^{(1,{\bf r}_{\aff(i)})} \}$, and (ii) every element $Z_i$ in ${\bf Z}$ has all its corresponding counterfactuals, {affectors of $i$}, 
	and ${R}_{{Z}_i}$ as parents. Restriction (ii) is a consequence of the consistency property in (\ref{eqn:consistency-m-2}). 
	
	A latent projection ADMG, obtained by projecting out fully hidden variables $\bigcup_{i=1}^n {\bf H}_i$ from this DAG, will be our preferred choice of graphical representation, just as in Figures \ref{fig:scalar-graphs} and \ref{fig:vector-graphs}. A simple example of such an ADMG for general networks is shown in Fig.~\ref{fig:vector-graphs-components}, where each unit $i$ has only one missing variable $Z_i$. Dependence among counterfactuals ${\bf Z}_i^{(1,{\bf r}_{\aff(i)})}$ is represented by bidirected edges, obtained by latent projecting $H_i$ out. In this example, every unit is an affector of every other unit. For compactness in representation, we have chosen to show the counterfactuals only for two of the units.
	
	When the incoming and outgoing edges of vertices $Z_i^{(1,{\bf r}_{\aff(i)})}$ and $Z_i^{(1,{\bf r'}_{\aff(i)})}$ are identical for all ${\bf r},  {\bf r'}$ in the graph, we can make the graphical representation even more compact by introducing the vertex $\tilde{\bf Z}_i$ corresponding to the set $\tilde{\bf Z}_i$ in place of all its elements $Z_i^{(1,{\bf r}_{\aff(i)})}$, for all {possible indexing corresponding to ${\bf r}_{\aff(i)}$.} Such graphs, which we will denote by $\tilde{\G}$, are shown in Fig.~\ref{fig:vector-graphs-generalized}. 
	In particular, the graph composed of solid edges in Fig.~\ref{fig:vector-graphs-generalized}(a) is a condensed representation of the graph in Fig.~\ref{fig:vector-graphs-components}, and any vertex $\tilde{\bf Z}_i$ (shown in a dashed black box) compactly represents the full set of vertices $\{Z_i^{(1,{\bf r}_{\aff(i)} = 0, \cdots ,0)}, Z_i^{(1,{\bf r}_{\aff(i)} = 1, \cdots ,0)}, \cdots, Z_i^{(1,{\bf r}_{\aff(i)} = 1, \cdots ,1)} \}$, all connected by bidirected edges. In $\tilde{\G}$, we {interpret} the edge $V \rightarrow \tilde{\bf Z}_i$, for any node $V$, {to mean that $V$ influences} all the counterfactuals
	${\bf Z}_i^{(1,{\bf r}_{\aff(i)})}$
	{in $\tilde{\bf Z}_i$}. Similarly $V \leftarrow \tilde{\bf Z}_i$ {is interpreted to mean that} all counterfactuals
	${\bf Z}_i^{(1,{\bf r}_{\aff(i)})}$ {in $\tilde{\bf Z}_i$} influence $V$. 
}

Further, it is helpful to view $\tilde{\G}$ in Fig.~\ref{fig:vector-graphs-generalized}(a) as the generalized version of Fig.~\ref{fig:vector-graphs}(a), where there is no target law dependence or missingness process dependence but only missingness interference. The graph can encode MCAR (only solid edges), MAR (with blue dotted edges) and MNAR (with brown dashed edges) processes, just as before. Graphs in Fig.~\ref{fig:vector-graphs-generalized}(b)-(d) are generalized versions of graphs in Fig.~\ref{fig:vector-graphs}(b)-(d). Graphs $\tilde{\G}$ are useful for illustrative purposes as they are compact, but as we show in Sec. \ref{sec:generalized-id}, identification is more straightforward in models where counterfactuals do \textit{not} share identical edges. Hence, we will use graphs $\G$, like those in Fig.~\ref{fig:vector-graphs-components} (and not $\tilde{\G}$ in Fig.~\ref{fig:vector-graphs-generalized}), in the remainder of the paper.

We have looked at examples where 
	{
		all potentially missing variables of a unit are either all observed or all missing, meaning only a single missingness indicator per unit is needed.
	}
	In general, this may not be true. This gives rise to two different complications: (i) there could be edges from 
	$R_{Z_j}$ to $Z'_i$ {for some units $i,j$ and variables $Z, Z' \in {\bf Z}$},
	and (ii) {edges from a missingness indicator of a unit to another variable in that same unit (e.g. $R_{Z_i} \rightarrow Z'_i$)} might be present. If (i) occurs, we cannot index counterfactuals simply by the missingness indicators of other units as we had done before, but of specific variables of those units. To be consistent with our prior definition, we call a unit the affector of another if \textit{any} missingness indicator of the first unit indexes the counterfactual of the latter unit. If (ii) occurs, the missingness interference is within the unit, and not across units, making the data i.i.d. in the absence of other interactions.
	{See Sec.~\ref{sec:iid} for additional discussion of this issue.} 
	It is worth pointing out that in both of these cases, the graphical structure and the mathematical framework remains the same as what we have discussed in this section so far. What changes is the interpretation of the relationships between units and their recorded variables.

\section{Identification in Entangled Missingness Models}
\label{sec:generalized-id}

Having set up the notation and graphical framework for entanglement, we next discuss identification in these models. 

\subsection{Identification Without Missingness Interference}
\label{subsec:id-wo-missingness-interference}

We recognized in Sec.~\ref{subsec:scalar-ctfl} that the (graphical) models in the absence of missingness interference are essentially identical to existing i.i.d. models for missing data, with the only difference arising from how we view the smallest unit of investigation - whether it is one individual or a set of individuals interacting within a block. 

It is common practice in missing data literature to treat expectations of the form $\beta = \E[h({\bf Z}\o,{\bf O})]$, for some known function $h$ of ${\bf Z}\o, {\bf O}$, as parameters of interest.
In this paper, however, we will discuss identification of distributions in keeping with causal graphical modeling identification literature.
Specifically, we focus on non-parametric identification of the full law $p({\bf Z}\o, {\bf O}, {\bf R})$, because if the full law is nonparametrically identified from the observed data so is any functional of it. 


A \textit{sound} and \textit{complete} algorithm for full law identification in i.i.d. missing data DAG models with fully observed variables has been proposed in \cite{nabi2020full}.\footnote{A sound and complete algorithm corresponds to necessary and sufficient identification assumptions.} The authors also provide full law identification in missing data models with hidden variables in the same work.
{For both types of graphs, DAGs and ADMGs, their identification criterion relies on the notion of the Markov blanket \citep{pearl88probabilistic, richardson17nested}: in an ADMG $\G$, the Markov blanket of a vertex $V$, denoted by $\mb_{\G}(V)$, consists of all variables sharing an edge with $V$ or with a collider path to $V$; a collider path is path where all vertices on the path are colliders of the form $ \rightarrow o \leftarrow, \leftrightarrow o \leftarrow, \leftrightarrow o \leftrightarrow $. In a DAG, which is an ADMG with no bidirected edges, the Markov blanket reduces to the set of variables with an edge in common with $V$, and variables that share a child with $V$.} We provide a brief description of their results in Theorem \ref{thm:m-dag-sound-complete} and Theorem \ref{thm:m-admg-sound-complete} below, as they apply directly to entangled missingness settings without missingness interference.

\begin{thm}{[Full law identification in DAGs \citep{nabi2020full}]}\label{thm:m-dag-sound-complete}		
	
	In a missing data model, represented by DAG $\G$, the full law $p({\bf R},{\bf Z}\o,{\bf O})$ is identified if and only if $Z^{(1)} \not\in \mb_{\G}(R_{Z}), \forall Z^{(1)} \in {\bf Z}^{(1)}$. 
	Thus, for the full law to be identified,	no edge of the form	$Z\o \rightarrow R_Z$ can be present (no \textit{self-censoring}) and no structure of the form $Z\o \rightarrow R_{Z'} \leftarrow Z$ can be present (no \textit{colluders}). The identifying functional is given by Eq. \ref{eq:propensity-missing-id}, where the missingness mechanism $p({\bf R} \mid {\bf Z}\o, {\bf O}) $ is given by an odds ratio parameterization \citep{chen07semiparametric}: 
	\begin{align}
		\frac{1}{\sigma({\bf Z}\o, {\bf O})} \times \prod_{k=1}^{K} p(R_k\mid {\bf R}_{-k}=1, {\bf Z}\o, {\bf O}) 
		\times  \prod_{k=2}^{K}  \text{OR}(R_{k}, R_{\prec k} \mid R_{\succ k}=1, {\bf Z}\o, {\bf O}), 
		\label{eqn:OR-K-vars}
	\end{align}
	where ${\bf R}_{-k} = {\bf R} \setminus R_k, R_{\prec k} = \{R_1, \cdots, R_{k-1}\}, R_{\succ k} = \{R_{k+1}, \cdots, R_{K}\}$, 
	\begin{align*}
		&\text{OR}(R_k, R_{\prec k}\mid R_{\succ k} = 1, {\bf Z\o, O}) \\
		&\hspace{1.5cm} = \frac{p(R_k\mid R_{\succ k} =1, R_{\prec k}, {\bf Z\o, O})}{p(R_k=1\mid R_{\succ k} =1, R_{\prec k}, {\bf Z\o, O})} \times \frac{p(R_k=1\mid {\bf R}_{-k} =1, {\bf Z\o, O})}{p(R_k\mid {\bf R}_{-k} =1, {\bf Z\o, O})}, 
	\end{align*}
	and $\sigma({\bf Z}\o,{\bf O}) = \sum_{r}\{\prod_{k=1}^{K} p(r_k | {\bf R}_{-k} = 1, {\bf Z}\o, {\bf O}) \times \prod_{k=2}^{K} \text{OR}(r_k, r_{\prec k} | R_{\succ k} =1, {\bf Z}\o, {\bf O}) \}$ is the normalizing function.
\end{thm}

Consider the i.i.d. graph in Fig.~\ref{fig:m_dags_intro} (d). The odds ratio parameterization of $p(R_A, R_Y \mid  {\bf C}, A\o, Y\o)$ is identical to its regular DAG parameterization (shown in Sec.~\ref{sec:graphical-missing-iid}) since $p(R_A \mid R_Y = 1, {\bf C}, A\o, Y\o) = p(R_A | {\bf C})$, $p(R_Y | R_A= 1, {\bf C}, A\o, Y\o) = p(R_Y | {\bf C})$, and  $\text{OR}(R_A, R_Y \mid Y^{(1)}, A^{(1)}, {\bf C}) =  \sum_{R_Y, R_A} p(R_A|{\bf C}) \times p(R_Y|{\bf C}) = 1,$ and the normalizing term is one. Self-censoring edges $A^{(1)} \rightarrow R_A$ and $Y^{(1)} \rightarrow R_Y$ prevent identification of the full law in Fig.~\ref{fig:scalar-graphs} (a).

The following theorem summarizes the full law identification results in \citep{nabi2020full} for ADMGs obtained as latent projections of hidden variable missing data DAGs.

\begin{thm}{[Full law identification in ADMGs \citep{nabi2020full}]}\label{thm:m-admg-sound-complete}
	
	In a missing data model represented by a hidden variable DAG $\G({\bf R},{\bf Z}\o,{\bf O},{\bf H})$ and its latent projection ADMG $\G({\bf R},{\bf Z}\o,{\bf O})$ , the full law $p({\bf R},{\bf Z}\o,{\bf O})$ is identified if and only if $Z^{(1)} \not\in \mb_{\G}(R_Z), \forall Z^{(1)} \in {\bf Z}^{(1)}$.  Thus, for the full law to be identified,  no pair $(Z\o,R_Z)$ should be connected directly (a.k.a. no \textit{self-censoring}) or 
	through a collider path (a.k.a. no \textit{colluding paths}). Moreover, the identification of the missingness mechanism is given by the odds ratio parameterization, as stated in Theorem~\ref{thm:m-dag-sound-complete}. 
\end{thm}
None of the graphs in Fig.~\ref{fig:scalar-graphs} have a colluding path. However, as examples, if the path $A_1^{(1)} \rightarrow R_{Y_1} \leftarrow R_{A_1}$ or the path $A_1^{(1)} \leftrightarrow A_2^{(1)} \leftrightarrow R_{A_1}$ were to exist in any of the graphs, then the full law would not be identified.

Settings with entanglements but no missingness interference are amenable to these (existing) identification results, but valid identification requires careful consideration of the dependence engendered by the entanglements, and whether they introduce self-censoring or colluding paths in the corresponding graph.

\subsection{Identification With Missingness Interference}
\label{subsec:id-missingness-interference}

{In settings with missingness interference, the joint distribution over all counterfactuals, often called the target law,} 
	involves counterfactual variables from \textit{multiple worlds}. For example, $Y_1^{(1,r_2=0)}$ and $Y_1^{(1,r_2=1)}$ in the case of Anne in Sec.~\ref{subsec:vector-ctfl}. This implies that the full law
	is not identified without very strong assumptions.\footnote{See \cite{hernan2020whatif, nabi2022causal} for discussions on how rank preservation assumption relates counterfactuals across multiple worlds.} 
	Instead, we will consider identification of 
	{\emph{single world} objects. Before giving a rigorous definition, let us motivate the idea with simple examples. In the case of Bob and Anne, the queries that one might be interested in would be, ``what would Anne's BMI be if Bob drops out?", 
		or ``what would their individual or joint outcomes be if they both followed through?", and not ``what would the \textit{joint} outcome of Bob and Anne be, if the other was not in the study?" as the corresponding events happen in different worlds. 
	
	{For a given network and an assignment of missingness indicators ${\bf R} = {\bf r}$, a single-world query is defined as a marginal distribution over counterfactual variables $\{\bigcup_{i = 1}^n Z_i^{(1,{\bf r}_{\aff(i)})}  \mid r_i = 1 \text{ if unit } i \text{ is inlcuded in this set}\}.$ For example, consider a 3-unit network where ${\bf r}_{\aff(1)} = (r_2,r_3)$, ${\bf r}_{\aff(2)} = (r_1,r_3),$ and ${\bf r}_{\aff(3)} = (r_1,r_2)$. Given the choice of ${\bf r} = (r_1,r_2,r_3) = (1,0,1)$, we consider the query $p(Z_1^{(1,r_2=0,r_3=1)}, Z_3^{(1,r_1=1,r_2=0)})$ to be a valid single-world query, while any marginal distribution that would include $Z_2^{(1,{\bf r}_{\aff(2)})}$ would be invalid. This is because the assignment $r_2=0$ in counterfactuals $Z_1^{(1,r_2=0,r_3=1)}$ and $Z_3^{(1,r_1=1,r_2=0)}$ is inconsistent with the assignment $r_2=1$ in the counterfactual $Z_2^{(1,{\bf r}_{\aff(2)})}$.  
	We denote the single world queries in missing data networks by $h(\tilde{{\bf Z}}; {\bf r})$, where $\tilde{{\bf Z}} \equiv \{\tilde{{\bf Z}}_i: i \in \{1, \ldots, n\}\}$ is the set of all counterfactuals.
	}

	
{Our first identification result shows that in the presence of missingness interference, analogues of MCAR and MAR models yield non-parametric identification for any single-world query $h(\tilde{{\bf Z}}; {\bf r})$.}

{The first identification result we outline shows that, analogues of MCAR and MAR models when missingness interference is present, yield non-parametric identification for any single-world object $h(\tilde{{\bf Z}}; {\bf r})$.} 

\begin{thm}\label{thm-vector-mcar-mar-margins}
	In a missing data ADMG $\G$ with missingness interference, valid single-world objects $h(\tilde{\bf Z};{\bf r})$ consisting of a set of counterfactuals $\bf Z' \equiv \bigcup_i \{Z_i^{(1,{\bf r}_{\aff(i)})}\} $, $i \in \{1, \cdots , n\}$ are identified 
	when either of these two conditions is satisfied: (1) ${\bf R}' \ci {\bf O}, {\tilde{{\bf Z}}}$ (MCAR), or (2) ${\bf R}' \ci  {\tilde{{\bf Z}}} | {\bf O}$ (MAR), where ${\bf R}'$ refers to the set 
	of all missingness indicators $R$ that index counterfactuals in $h(\tilde{\bf Z};{\bf r})$. The object $h(\tilde{\bf Z};{\bf r})$ is a function of $p({\bf Z}', {\bf R}, {\bf O})$, and the identifying functional is given by:
	\begin{align}
		p({\bf Z}', {\bf R}, {\bf O}) =  p({\bf Z}',{\bf O}) \times { p({\bf R}|{\bf O},{\bf Z}') }
		= \frac{p( {\bf Z}', {\bf R}={\bf r}, {\bf O}) }{p({\bf R}={\bf r} \mid {\bf O})} \times {p({\bf R} \mid {\bf O})}
		\label{eq:mcar-id}
	\end{align}
	where propensity scores are obtained by simple m-separation or (d-separation) rules on ADMG (or DAG) factorization. 
	
\end{thm}
\begin{prf}
	See Appendix for proof.
\end{prf}


{In the presence of an MNAR missingness mechanism,} single-world {queries may} not always be identified.  
	{Consider instead} a special single-world query called the \textit{full-observability} law, $p(\tilde{\bf Z}^{({\bf r}={\bf 1})}, {\bf R})$, where $\tilde{\bf Z}^{({\bf r}={\bf 1})} \equiv {\bigcup_{i = 1}^n  Z_i^{(1,{\bf r}_{\aff(i)}={\bf 1})}}$
	is the set of all counterfactuals where missingness status is set to full observability (i.e, missingness pattern is ${\bf R} = {\bf 1}$). 
	The full observability law might be thought of as a close analogue to the full law in i.i.d settings. It corresponds to the distribution
	$p(Y_1^{(1,r_2=1)}, Y_2^{(1,r_1=1)}, R_1,R_2)$ in the example with Anne and Bob.
	The full observability law is in fact identified under certain assumptions.  
	We first outline our main assumption and a few definitions before presenting our result for identifying the full observability law in MNAR models with entangled missingness. 
	
	{We first assume } the only type of counterfactual that can be a parent of any $R_i \in {\bf R}$ has the form $Z_j^{(1,{\bf r}_{\aff(j)} = {\bf 1})}$. {We formalize this assumption as follows.} 
	\begin{asmp} \label{assump:10-not-parent]}
		In a missing data ADMG $\G$ with missingness interference, $\ch_{\G}(\tilde{\bf Z}^{({\bf r}\neq{\bf 1})} ) \cap {\bf R} = \emptyset$. Here, $\tilde{\bf Z}^{({\bf r}\neq{\bf 1})}$ denotes the set of all counterfactuals {corresponding to patterns} with at least one zero, i.e.,  $\tilde{\bf Z}^{({\bf r}\neq{\bf 1})} = \tilde{{\bf Z}} \setminus \tilde{\bf Z}^{({\bf r}={\bf 1})}$. 
		Further, the definition $\ch_{\G}(V)$ applies disjunctively to a set ${\bf V}$, i.e., $\ch_{\G}({\bf V}) = \bigcup_{V \in {\bf V}} \ch_{\G}(V)$. 
	\end{asmp} 
	Next, we define the following new {graphical structures} to be used in our results. 
	\begin{enumerate}
		
		\item \textit{e-self-censoring} \footnote{The `e' in these structures stands for `entangled'.} or \textit{{affector-censoring}}: {For a given $R_k \in {\bf R}_{\aff(i)}$, there exists an affector-censoring mechanism if $Z_i^{(1,{\bf r}_{\aff(i)}={\bf 1})} \rightarrow R_k$.} 
			
		\item \textit{e-colluder}: {For a given $R_k \in {\bf R}_{\aff(i)}$ and $R_j \notin {\bf R}_{\aff(k)}$, there exists an (entangled- or) e-colluder at $R_j$ if} $Z_i^{(1,{\bf r}_{\aff(i)}={\bf 1})} \rightarrow R_j \leftarrow R_k$. 
		
		\item \textit{e-colluding path}: {For a given $R_k \in {\bf R}_{\aff(i)}$}, an (entangled- or) e-colluding path is defined to exist between the pair $(Z_i^{(1,{\bf r}_{\aff(i)}={\bf 1})}, R_k)$ if $Z_i^{(1,{\bf r}_{\aff(i)}={\bf 1})}$ and $R_k$ are connected through at least one colliding path\footnote{A path between vertices $V_i$ and $V_j$ is a colliding path if every vertex $V_k$ on the path is a collider, i.e., bears ones of these forms: $\rightarrow V_k \leftarrow$, $\leftrightarrow V_k \leftrightarrow$, $\rightarrow V_k \leftrightarrow$, or $\leftrightarrow V_k \leftarrow$.} that does not go through an observed proxy. 

	\end{enumerate}

	{
	\begin{thm}\label{thm-vector-mnar-11s-sound-complete}
		In a missing data ADMG $\G$ with missingness interference, under Assumption \ref{assump:10-not-parent]}, the full-observability law $P(\tilde{\bf Z}^{({\bf r}={\bf 1})}, {\bf R})$ is identified 
		{if and only if}
		there is no e-colluding path for any $R \in {\bf R}$ in the ADMG. Further, if $\G$ is a missing data DAG, the full-observability law is identified 
		{if and only if} there is no e-colluder and no e-self-censoring. The identifying functional is given by
		\begin{align}
			p(\tilde{\bf Z}^{({\bf r}={\bf 1})}, {\bf R}) =  p(\tilde{\bf Z}^{({\bf r}={\bf 1})}) \times \underbrace{p({\bf R} \mid \tilde{\bf Z}^{({\bf r}={\bf 1})})}_{g(p({\bf R},{\bf Z}))}
			= \frac{p( \tilde{\bf Z}^{({\bf r}={\bf 1})}, {\bf R}=1)}{\underbrace{p({\bf R}=1 \mid \tilde{\bf Z}^{({\bf r}={\bf 1})})}_{g(p({\bf R},{\bf Z}))\vert_{{\bf R}=1}}} \times \underbrace{p({\bf R} \mid \tilde{\bf Z}^{({\bf r}={\bf 1})})}_{g(p({\bf R},{\bf Z}))}.
			\label{eq:full-observability-id}
		\end{align}
		and missingness mechanism $p({\bf R} \mid \tilde{\bf Z}^{({\bf r}={\bf 1})})$ is identified using the OR parameterization given below:
		\begin{align*}
			p({\bf R} | \tilde{{\bf Z}}^{({\bf r} = {\bf 1})} ) = \frac{1}{\sigma} \times \prod_{k=1}^{K} p(R_k | {\bf R}_{-k} = 1, \tilde{{\bf Z}}^{({\bf r} = {\bf 1})} ) \times \prod_{k=2}^{K}  \text{OR}(R_k, R_{\prec k} | R_{\succ k} = 1, \tilde{{\bf Z}}^{({\bf r} = {\bf 1})} )  
		\end{align*}
		where notation and OR is consistent with Sec.~\ref{subsec:id-wo-missingness-interference}.
	\end{thm}
}

\begin{prf}
	{
		The proof extends the proof for sound and complete identification of the missing data full law in ADMGs in the absence of colluding paths in \cite{nabi2020full}. Proof is in the Appendix.
	}
\end{prf}

	We illustrate Theorem \ref{thm-vector-mnar-11s-sound-complete} using two different examples, one where the full observability law is identified, and one where the graphical criterion is violated, and hence the law is not identified.

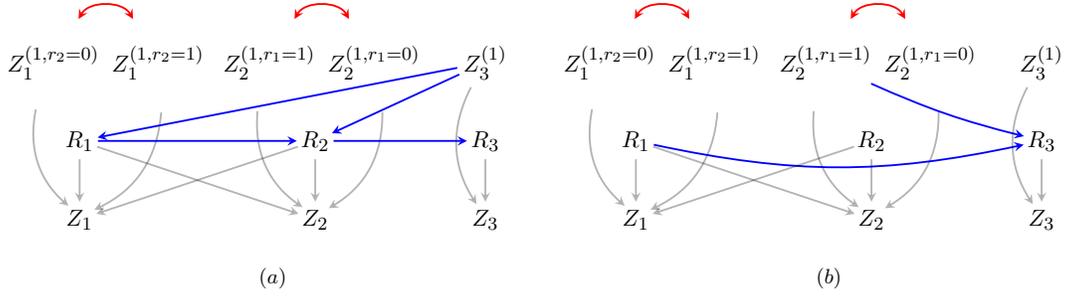
\begin{figure}[!t]
	\begin{center}
		\scalebox{0.87}{
			\begin{tikzpicture}[>=stealth, node distance=1.2cm]
				\tikzstyle{format} = [draw=none, thick, circle, minimum size=4mm,
				inner sep=0pt]

				\begin{scope} [xshift=-1cm]
					\path[->, thick]
					
					

					node[format] (y111) {\small{$Z_{1}^{(1,r_{2}=1)}$}}
					node[format, left of=y111, xshift=-0.4cm] (y110) {\small{$Z_{1}^{(1,r_{2}=0)}$}}
					node[format, below of=y111, xshift=-1.2cm] (r1) {\small{$R_1$}}
					node[format, below of=r1, xshift=0cm] (y1) {\small{$Z_1$}}
					
					node[format, right of= y111, xshift=0.5cm] (y211) {\small{$Z_{2}^{(1,r_{1}=1)}$}}
					node[format, right of=y211, xshift=0.4cm] (y210) {\small{$Z_{2}^{(1,r_{1}=0)}$}}
					node[format, below of=y211, xshift=+0.7cm] (r2) {\small{$R_2$}}
					node[format, below of=r2, xshift=0cm] (y2) {\small{$Z_2$}}
					
					node[format, right of= y210, xshift=0.5cm] (y31) {\small{$Z_{3}^{(1)}$}}
					node[format, below of=y31, xshift=0cm] (r3) {\small{$R_3$}}
					node[format, below of=r3, xshift=0cm] (y3) {\small{$Z_3$}}
					

					(y111) edge[black, opacity=0.3, bend left = 30] (y1)
					(y110) edge[black, opacity=0.3, bend right = 30] (y1)
					(y211) edge[black, opacity=0.3, bend right = 30] (y2)
					(y210) edge[black, opacity=0.3, bend left = 30] (y2)
					(y31) edge[black, opacity=0.3, bend right = 30] (y3)

					(y31) edge[blue] (r1)
					(y31) edge[blue] (r2)
					
					(r1) edge[blue] (r2)
					(r2) edge[blue] (r3)

					(r1) edge[black, opacity=0.3] (y1)
					(r2) edge[black, opacity=0.3] (y2)
					(r3) edge[black, opacity=0.3] (y3)

					(r2) edge[black, opacity=0.3] (y1)
					(r1) edge[black, opacity=0.3] (y2)
					
					(y111) edge[<->,red, bend right=60] (y110)
					(y211) edge[<->,red,bend left=60] (y210)
					
					node[below of=y111, yshift=-2.1cm, xshift=1.75cm] (la) {$(a)$}
					;
				\end{scope}
				
				\begin{scope} [xshift=7.5cm, yshift=-0cm]
					\path[->, thick]
					
					node[format] (y111) {\small{$Z_{1}^{(1,r_{2}=1)}$}}
					node[format, left of=y111, xshift=-0.4cm] (y110) {\small{$Z_{1}^{(1,r_{2}=0)}$}}
					node[format, below of=y111, xshift=-1.2cm] (r1) {\small{$R_1$}}
					node[format, below of=r1, xshift=0cm] (y1) {\small{$Z_1$}}
					
					node[format, right of= y111, xshift=0.5cm] (y211) {\small{$Z_{2}^{(1,r_{1}=1)}$}}
					node[format, right of=y211, xshift=0.4cm] (y210) {\small{$Z_{2}^{(1,r_{1}=0)}$}}
					node[format, below of=y211, xshift=+0.7cm] (r2) {\small{$R_2$}}
					node[format, below of=r2, xshift=0cm] (y2) {\small{$Z_2$}}
					
					node[format, right of= y210, xshift=0.5cm] (y31) {\small{$Z_{3}^{(1)}$}}
					node[format, below of=y31, xshift=0cm] (r3) {\small{$R_3$}}
					node[format, below of=r3, xshift=0cm] (y3) {\small{$Z_3$}}
					

					(y111) edge[black, opacity=0.3, bend left = 30] (y1)
					(y110) edge[black, opacity=0.3, bend right = 30] (y1)
					(y211) edge[black, opacity=0.3, bend right = 30] (y2)
					(y210) edge[black, opacity=0.3, bend left = 30] (y2)
					(y31) edge[black, opacity=0.3, bend right = 30] (y3)

					(y211) edge[blue, bend right=5] (r3)
					(r1) edge[blue, bend right=12] (r3)

					(r1) edge[black, opacity=0.3] (y1)
					(r2) edge[black, opacity=0.3] (y2)
					(r3) edge[black, opacity=0.3] (y3)

					(r2) edge[black, opacity=0.3] (y1)
					(r1) edge[black, opacity=0.3] (y2)
					
					(y111) edge[<->,red, bend right=60] (y110)
					(y211) edge[<->,red,bend left=60] (y210)
					
					node[below of=y111, yshift=-2.1cm, xshift=1.75cm] (la) {$(b)$}
					;
				\end{scope}
					
			\end{tikzpicture}
		}
	\end{center}
	\caption{Two examples used for illustrations of Theorem \ref{thm-vector-mnar-11s-sound-complete}. ADMG in (a) does not have an e-colluding path. ADMG in (b) has the e-colluding path $Z_{2}^{(1,r_{1}=1)} \rightarrow R_3 \leftarrow R_1$. }
	\label{fig:thm-examples}
\end{figure}

In the ADMG shown in Fig.~\ref{fig:thm-examples}(a), there are no e-colluding paths. Thus, by Theorem \ref{thm-vector-mnar-11s-sound-complete}, the full observability law $P(Z_{1}^{(1,r_{2}=2)}, Z_{2}^{(1,r_{1}=1)}, Z_3^{(1)}, R_1, R_2, R_3)$ is identified. We show how in the following paragraph.

The propensity score for $R_3$ is identified from observed data in a straightforward manner: $p(R_3 | \pa(R_3)) = p(R_3|R_2)$. But this is not true for propensity scores of $R_2, R_1$. For $R_1$, the propensity score is $p(R_1|Z_3^{(1)})$. We see that $R_1 \ci R_3 | Z_3^{(1)}, R_2$, but that the independence would not hold without $R_2$. That is, we can identify $p(R_1|Z_3^{(1)},R_2,R_3=1)$. For $R_2$, we have the propensity score $p(R_2|Z_3^{(1)},R_1)$. Here, we cannot insert $R_3$ behind the conditioning bar since the conditional independence does not hold. But, we could \textit{fix} $R_3=1$, an operation that removes the dependence of $R_3$ on $R_2$ and yields a world where $R_2 \ci R_3| Z_3^{(1)}, R_1$ (see Appendix for more details). But we can use the odds ratio parameterization to yield identification of the joint score $p(R_1,R_2|Z_3^{(1)})$. 
\begin{align*}
p(R_1, R_2|Z_3\o) & = \frac{1}{Z} \times p(R_2|R_1=1,Z_3\o) p(R_1|R_2=1,Z_3\o) \times \text{OR}(R_2,R_1|Z_3\o)
\end{align*}
Here, $p(R_2|R_1=1,Z_3\o)$ and $p(R_1|R_2=1,Z_3\o)$ can be identified as described in the previous paragraph. And, we can expand the odds ratio as follows: 
\begin{align*}
 \text{OR}(R_2,R_1|Z_3\o) & = \frac{p(R_1=r_1|R_2=r_2,Z_3\o)}{p(R_1=1|R_2=r_2,Z_3\o )} \times \frac{p(R_1=1|R_2=1,Z_3\o)}{p(R_1=r_1|R_2=1,Z_3\o)} \\
 & = \frac{p(R_1=r_1|R_2=r_2,Z_3\o,R_3=1)}{p(R_1=1|R_2=r_2,Z_3\o ,R_3=1)} \times \frac{p(R_1=1|R_2=1,Z_3\o,R_3=1)}{p(R_1=r_1|R_2=1,Z_3\o,R_3=1)} 
\end{align*}
which is also identified. The normalizing term $Z$ is also identified since the conditional pieces and the OR piece are all individually identified. This means that $p(R_1, R_2|Z_3\o)$ is identified. By Eqn. \ref{eq:full-observability-id}, the full observability law is identified as all the pieces are individually a function of the observed data.

%

	The full observability law is however not identified in Fig.~\ref{fig:thm-examples}(b) because of the colluding path $Z_{2}^{(1,r_{1}=1)} \rightarrow R_3 \leftarrow R_1$. 
	{Intuitively,} the missingness mechanism $p({\bf R} \mid \pa({\bf R}))$ should be identified for all levels of $\bf R$. Writing the missingness mechanism as $p(R_1 ) \times p(R_2) \times p(R_3  \mid R_1, Z_2^{(1,r_1=1)})$, we realize that we cannot identify the last term unless we are able to set \textit{both} $R_1$ and $R_2$ to the value 1, and by consistency, replace the counterfactual $Z_2^{(1,r_1=1)}$ by observed proxy $Z_2$. Doing so in the joint would mean that we cannot identify this quantity for levels of $\bf R$ that set $R_1$ or $R_2$ to 0. The rigorous argument for non-identification involves {providing counterexamples where two missing data ADMGs disagree on the full observability law but agree on the  observed data law}. {Such an argument can be achieved by, for instance,} counting the parameters required to characterize the full observability law against the observed law in a binary model, and showing that the observed law has fewer parameters and hence it is not possible to uniquely map back to the full observability law. A detailed account of parameter counting in examples with e-colluding paths has been deferred to the Appendix, under the completeness section of the proof for Theorem~\ref{thm-vector-mnar-11s-sound-complete} to follow the procedure for parameter counting, knowledge of the M\"{o}bius parameterization of the nested Markov model for binary variable ADMGs is required, which we also discuss in the Appendix.}


\section{Implications of the Missingness Interference Model for I.I.D. Settings}
\label{sec:iid}

In what we have presented so far, graphical models of entangled missingness are viewed as representing a block of units subject to dependence or interference. The same graphical structures, however, can also be viewed as representing multivariate observations for a single unit in i.i.d. settings, e.g. settings where a single unit has multivariate treatment and/or outcome measurements. 

As we have shown, target law dependence and missingness process dependence can be handled by existing i.i.d. missing data methods. However, to the best of our knowledge, the i.i.d. analogue of \textit{missingness interference} entanglement has never been discussed in the prior literature on missing data in i.i.d. settings. In fact, the absence of  missingness interference is typically required by an edge restriction ubiquitous in classical graphical approaches to missing data -- namely that no missingness indicator is a cause/parent of any counterfactual or fully observed variable (restriction (ii) in Sec.~\ref{sec:graphical-missing-iid}). Relaxing this assumption implies that changes in a missingness indicator may cause changes in the underlying full data law.  Our proposed framework on missingness interference entanglement can be readily adapted to relax the imposed edge restriction in i.i.d. settings. 

\begin{figure}[!t]
	\begin{center}
		\scalebox{0.65}{
			\begin{tikzpicture}[>=stealth, node distance=1.5cm]
				\tikzstyle{format} = [draw=none, thick, circle, minimum size=4mm, inner sep=2pt]
				
				\begin{scope} [xshift=0cm, yshift=0cm] 
					\path[->, thick]
		
					node[format, yshift=-0.cm] (L11) {\small{$L^{(1)}_1$}}
					node[format, right of=L11, xshift=0.75cm] (L21-1) {\small{$L^{(1, r_1=1)}_2$}}
					node[format, right of=L11, xshift=0.75cm, yshift=-0.25cm] (L21-1-t) {}
					node[format, below right of=L21-1, xshift=-0.7cm] (L21) {}
					node[format, right of=L21-1, xshift=0.95cm] (L21-0) {\small{$L^{(1, r_1=0)}_2$}}
					node[format, right of=L21-1, xshift=0.5cm, yshift=-0.25cm] (L21-0-t) {}
						
					node[format, below of=L11, yshift=-0.75cm] (R1) {\small{$R_1$}}
					node[format, below of=L21-1, xshift=1.2cm, yshift=-0.75cm] (R2) {\small{$R_2$}}

					node[format, left of=R1] (L1) {\small{$L_1$}}
					node[format, right of=R2] (L2) {\small{$L_2$}}
					
					(L21-1-t) edge[gray, bend left=10] (L2)
					(L21-0-t) edge[gray, bend left] (L2)
					(R2) edge[gray] (L2)
					(R1) edge[gray, bend left=25] (L2)
					
					(R1) edge[gray] (L1)
					(L11) edge[gray, bend right=0] (L1)

					(L11) edge[blue, bend left] (L21-0)
					(L11) edge[blue] (L21-1) 
					
					(L21-1) edge[red, <->] (L21-0)
					
					node[below right of=R1, yshift=0.35cm, xshift=0.9cm] (la) {$(a)$}
					;
				\end{scope}
			
				\begin{scope} [xshift=8cm, yshift=0cm] 
					\path[->, thick]
					
					node[format, yshift=-0.cm] (L11) {\small{$L^{(1)}_1$}}
					node[format, right of=L11, xshift=0.75cm] (L21-1) {\small{$L^{(1, r_1=1)}_2$}}
					node[format, right of=L11, xshift=0.5cm, yshift=-0.25cm] (L21-1-t) {}
					node[format, below right of=L21-1, xshift=0.15cm] (L21) {\small{$L^{(1)}_2$}}
					node[format, above right of=L21, xshift=0.15cm] (L21-0) {\small{$L^{(1, r_1=0)}_2$}}
					node[format, above right of=L21, xshift=0.15cm, yshift=-0.25cm] (L21-0-t) {}
					
					node[format, below of=L11, yshift=-0.75cm] (R1) {\small{$R_1$}}
					node[format, below of=L21, xshift=0cm, yshift=0.25cm] (R2) {\small{$R_2$}}
					
					node[format, left of=R1] (L1) {\small{$L_1$}}
					node[format, right of=R2] (L2) {\small{$L_2$}}
					
					(L11) edge[gray, bend right=0] (L1)
					(R1) edge[gray] (L1)
					(L21) edge[gray, bend right=0] (L2)
					(R2) edge[gray] (L2)
					
					(L21-1-t) edge[gray] (L21)
					(L21-0-t) edge[gray] (L21)
					(R1) edge[gray] (L21) 
					
					(L11) edge[blue, bend left] (L21-0)
					(L11) edge[blue] (L21-1) 
					
					(L21-1) edge[red, <->] (L21-0)
					
					node[below right of=R1, yshift=0.35cm, xshift=0.9cm] (la) {$(b)$}
					;
				\end{scope}
			
				\begin{scope} [xshift=16cm, yshift=0cm] 
					\path[->, thick]
					
					node[format, yshift=-0.cm] (L11) {\small{$L^{(1)}_1$}}
					node[format, right of=L11, xshift=0.15cm] (L21) {\small{$L^{(1)}_2$}}
					
					node[format, below of=L11, yshift=-0.75cm] (R1) {\small{$R_1$}}
					node[format, below of=L21, xshift=0cm, yshift=-0.75cm] (R2) {\small{$R_2$}}
					
					node[format, left of=R1] (L1) {\small{$L_1$}}
					node[format, right of=R2] (L2) {\small{$L_2$}}
					
					(L11) edge[blue] (L21)
					(R1) edge[blue] (L21)
					
					(L11) edge[gray, bend right=0] (L1)
					(R1) edge[gray] (L1)
					(L21) edge[gray, bend right=0] (L2)
					(R2) edge[gray] (L2)

					node[below right of=R1, yshift=0.35cm, xshift=0.cm] (la) {$(c)$}
					;
				\end{scope}
				
			\end{tikzpicture}
		}
	\end{center}
	\caption{ Graphs  used to illustrate the role of missingness interference in i.i.d. settings. }
	\label{fig:disscuss_interference}
\end{figure}
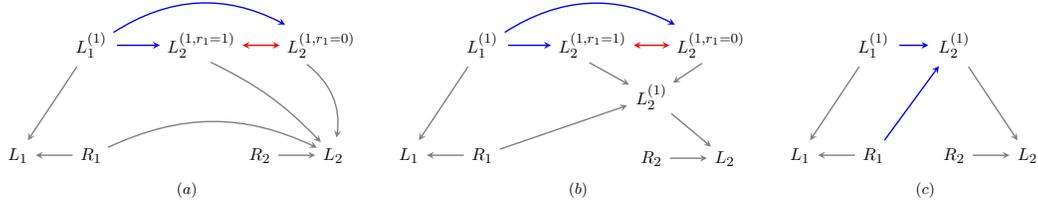

As an example, assume $L_1, L_2$ are  measurements during first and second followup visits, respectively ($i$ in $L_i$ is indexing variables of a single patient). In some situations, whether or not a patient shows up to their first followup can affect their future health; perhaps through mental self-empowerment or words of self-encouragement on treatment adherence. This can be captured via counterfactuals of the form $L^{( r_2=1, r_1)}_2,$ or simply $L^{(1, r_1)}_2$,  that reads as the measurement obtained at the second visit had the patient showed up to their first visit ($r_1=1$), or had they not $(r_1=0)$. This example is analogous to assuming $L_1$ is the measurement of one patient and $L_2$ is the measurement of another patient on the their first 
visit ($i$ in $L_i$ is indexing units for the same variable). Assuming an asymmetric relationship between the two patients, namely that only the measurement of the second patient depends on whether the first patient shows up to the visit or not and not vice versa, we can represent the relations using the ADMG shown in Fig.~\ref{fig:disscuss_interference}(a). This ADMG corresponds to the full data law with a single counterfactual version for $L_1$ (indexed by $R_1=1$) and two counterfactual versions for $L_2$ (indexed by $R_2=1, R_1=0$ and $R_2=1, R_1=1$). 

Similar to missingness interference in non-i.i.d. settings, the proxy variable $L_2$ can be defined as a deterministic function of $R_1, R_2$, $L^{(1, r_1=0)}_2$, and $L^{(1, r_1=1)}_2$ as follows: 
\begin{align*}
	L_2 \ \leftarrow  \
	\left\{ \begin{array}{cc}
		(1-r_1) \ L_2^{(1,r_1=0)} + r_1 \ L_2^{(1,r_1=1)} \ &\text{ if } \ r_2 = 1, \\
		? \ & \text{ if } \ r_2 = 0.
	\end{array} \right.
\end{align*}%
We can further introduce an intermediate step by first defining a single indexed counterfactual for $L_2$, i.e., $L^{(r_2=1)}_2$ or $L^{(1)}_2$ for short, as a deterministic function of $R_1, L^{(1, r_1=1)}_2, L^{(1, r_1=0)}_2$, and then defining the proxy $L_2$ as a deterministic function of $R_2, L^{(1)}_2$ only:
\begin{align*}
	L^{(1)}_2 &= (1-R_1) \times L^{(1, r_1=0)}_2 + R_1 \times L^{(1, r_1=1)}_2, \\
	L_2 \ &\leftarrow  \
	\left\{ \begin{array}{cc}
		 L_2^{(1)} \ &\text{ if } \ r_2 = 1, \\
		? \ & \text{ if } \ r_2 = 0.
	\end{array} \right.
\end{align*}%
The ADMG in Fig.~\ref{fig:disscuss_interference}(b) corresponds to the full data law with three counterfactual versions for $L_2$. The gray edges illustrate the consistency relations outlined above. 

The distinction of Fig.~\ref{fig:disscuss_interference}(a, b) from the classical i.i.d. missing data DAGs is the presence of multi-indexed counterfactuals, which indicates that, for instance in a given temporal ordering on variables, a variable later in the order may depend on whether an earlier variable is measured or not, regardless of whether sample realizations are dependent or not. Projecting out the multi-indexed counterfactuals from Fig.~\ref{fig:disscuss_interference}(b) yields the DAG in Fig.~\ref{fig:disscuss_interference}(c) where $R_1$ is pointing to $L^{(1)}_2$, a violation of edge restrictions in classical missing data graphs. 

In the above i.i.d. example, the target distribution $p(L^{(r_1=1)}_1, L^{(r_2=1,r_1=0)}_2)$ cannot be identified as a function of observed data law due to the inconsistency in assignments of $R_1$
, without imposing strong cross-world assumptions \footnote{Discussions on cross-world assumptions can be found in mediation literature. For example, see \citep{andrews2020insights}}. 
For instance, we could assume that the effect of $L^{(r_1=1)}_1$ on $L^{(r_2=1)}_2$ is fully mediated by a variable with no missingness, denoted by $X$. Then $p(L^{(r_1=1)}_1, L^{(r_1=0, r_2=1)}_2)$ can be identified as follows: 
\begin{align*}
	p(L^{(r_1=1)}_1, L^{(r_2=1,r_1=0)}_2) 
	&= \sum_x p(L^{(r_1=1)}_1 \mid x) \times p(L^{(r_2=1,r_1=0)}_2 \mid x, L^{(r_1=1)}_1 = l^{(r_1=1)}_1) \times p(x)\\
	&= \sum_x p(L_1 \mid x, R_1=1) \times p(L_2 \mid x, R_1=0, R_2=1) \times p(x).  
\end{align*}
In order to avoid such complications, edges from $\bf R$ to counterfactuals were excluded in prior literature on graphical models of missing data. Under this edge restriction, we are assuming $L_j^{(r_k, r_j=1)}=L_j^{(r_j=1)}, \forall R_j, R_k \in R.$ As shown above, this assumption can be relaxed by leveraging our developed framework on missingness interference entanglement.

\section{Simulations}
\label{sec:experiments}

We generate synthetic data and attempt to recover the ground truth in entangled missingness settings where the targets we are interested in are identified. In particular, we reserve our interest to the cases where missingness interference ($\mathcal{E}_3$) is present, as results for settings where missingness interference is absent have been discussed in literature \cite{nabi2022causal}. 

Data was simulated based on the graph in Fig.~\ref{fig:expmt-model}. There are three units in the network, and units 1 and 2 are neighbors. When the graph consists only the solid edges, we have an MCAR scenario. Inclusion of the dashed blue edges ($C_1 \rightarrow R_1$, $C_1 \rightarrow R_2$, $C_2 \rightarrow R_1$, $C_2 \rightarrow R_2$, $C_3 \rightarrow R_3$), yields a MAR scenario. Finally, inclusion of all the edges (adding $Z_3^{(1)} \rightarrow R_2, Z_3^{(1)} \rightarrow R_1$), results in an MNAR scenario. 

\begin{figure}[!t]
	\begin{center}
		\scalebox{0.87}{
			\begin{tikzpicture}[>=stealth, node distance=1.2cm]
				\tikzstyle{format} = [draw=none, thick, circle, minimum size=4mm,
				inner sep=0pt]

				\begin{scope} [xshift=-1cm]
					\path[->, thick]
					
					node[format] (y111) {\small{$Z_{1}^{(1,r_{2}=1)}$}}
					node[format, left of=y111, xshift=-1cm] (y110) {\small{$Z_{1}^{(1,r_{2}=0)}$}}
					node[format, below of=y111, xshift=-1.2cm, yshift=-0.3cm] (r1) {\small{$R_1$}}
					node[format, below of=r1, xshift=0cm] (y1) {\small{$Z_1$}}
					
					node[format, right of= y111, xshift=1.2cm] (y211) {\small{$Z_{2}^{(1,r_{1}=1)}$}}
					node[format, right of=y211, xshift=1cm] (y210) {\small{$Z_{2}^{(1,r_{1}=0)}$}}
					node[format, below of=y211, xshift=1cm, yshift=-0.3cm] (r2) {\small{$R_2$}}
					node[format, below of=r2, xshift=0cm] (y2) {\small{$Z_2$}}
					
					node[format, right of= y210, xshift=1cm] (y31) {\small{$Z_{3}^{(1)}$}}
					node[format, below of=y31, xshift=0cm, yshift=-0.3cm] (r3) {\small{$R_3$}}
					node[format, below of=r3, xshift=0cm] (y3) {\small{$Z_3$}}
					
					node[format, above of=y111, xshift=-1.1cm] (c1) {\small{$C_1$}}
					node[format, above of=y211, xshift=1cm] (c2) {\small{$C_2$}}
					node[format, above of=y31, xshift=0cm] (c3) {\small{$C_3$}}

					(y111) edge[black, opacity=0.3, bend left = 30] (y1)
					(y110) edge[black, opacity=0.3, bend right = 30] (y1)
					(y211) edge[black, opacity=0.3, bend right = 30] (y2)
					(y210) edge[black, opacity=0.3, bend left = 30] (y2)
					(y31) edge[black, opacity=0.3, bend right = 30] (y3)

					(c1) edge[blue] (y111)
					(c1) edge[blue] (y110)
					(c2) edge[blue] (y211)
					(c2) edge[blue] (y210)
					(c3) edge[blue] (y31)

					(r1) edge[black, opacity=0.3] (y1)
					(r2) edge[black, opacity=0.3] (y2)
					(r3) edge[black, opacity=0.3] (y3)

					(r2) edge[black, opacity=0.3] (y1)
					(r1) edge[black, opacity=0.3] (y2)
					
					(y111) edge[<->,red, bend right=0] (y110)
					(y211) edge[<->,red, bend left=0] (y210)
					(y111) edge[<->,red, bend left=0] (y211)
					(y210) edge[<->,red, bend left=0] (y31)
					
					(y110) edge[<->,red, bend right=15] (y211)
					(y110) edge[<->,red, bend left=15] (y210)
					(y110) edge[<->,red, bend left=18] (y31)
					(y111) edge[<->,red, bend right=15] (y210)
					(y111) edge[<->,red, bend left=15] (y31)
					(y211) edge[<->,red, bend right=15] (y31)
					
					(c1) edge[blue, dashed, bend left=0] (r1)
					(c1) edge[blue, dashed, bend left=10] (r2)
					
					(c2) edge[blue, dashed, bend right=10] (r1)
					(c2) edge[blue, dashed, bend left=0] (r2)
					
					(c3) edge[blue, dashed, bend left=25] (r3)

					(y31) edge[blue, dotted, bend left=10] (r1)
					(y31) edge[blue, dotted, bend left=5] (r2)
					
					;
				\end{scope}

			\end{tikzpicture}
		}
	\end{center}
	\caption{The model used to generate synthetic data for our experiments. }
	\label{fig:expmt-model}
\end{figure}
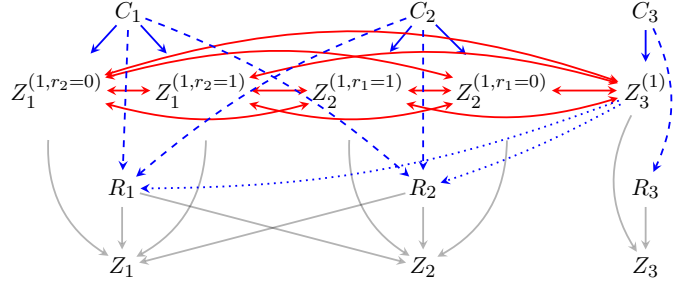

Parameters associated with the generative model are as follows: $p(C_1), p(C_2), p(C_3)$, all chosen to be univariate normal distributions, counterfactuals $Z_1^{(1,r_2=0)},$ $Z_1^{(1,r_2=1)}$, $Z_3^{(1,r_1=0)}$, $Z_3^{(1,r_1=1)}$ and $Z_3^{(1)}$ are drawn from a multivariate normal distribution. The means of the multivariate normal are given by linear functions of the corresponding parents $C$ in the graph, and covariance matrix has no zero entries; this ensures that all the counterfactuals are associated with each other. Missingness indicators $R_i$ are generated according to the scenario, using a binomial distribution where the binomial probability is chosen completely at random (MCAR) or is a logistic sigmoid of a linear function of the parents of $R_i$ (MAR, MNAR). Observed proxies $Z_i$ are a deterministic function of the counterfactuals and missingness indicators, as given by the consistency assumption.

We are interested in identifying the following functionals: $\E[Z_1^{(1,r_2=0)}]$, $\E[Z_1^{(1,r_2=1)}]$, $\E[Z_2^{(1,r_1=0)}]$, $\E[Z_2^{(1,r_1=1)}]$, under MCAR, MAR and MNAR conditions. In the given graph, since there are no e-colluders, we should be able to identify the full observability law (or its marginals) even in the MNAR case. We use an estimating equation of the form $\E[\frac{\I({\bf R} = {\bf 1})}{\pi({\bf C}, \tilde{{\bf Z}})} h(\tilde{{\bf Z}})]$ where $\pi({\bf C}, \tilde{{\bf Z}})$ is the inverse weight function composed of propensity scores and $h(.)$ is the functional we are interested in. The estimator uses regression models for the weights, and expectations are done empirically. 

We would like to compare the adjusted estimates we obtain with those obtained by an investigator who does not use entangled missingness ideas to really show why such a framework is necessary. 
First, when such an investigator, who is unaware of the various interactions between the units, notices missing data in the dataset, they are likely to adopt a naive, missing data approach to estimate counterfactuals, probably assuming a MAR mechanism on i.i.d data. This means that they are restricted to estimating $Z_1^{(1)}$ and $Z_2^{(1)}$, which do not correspond reliably to the underlying truth as units 1 and 2 have \textit{two} counterfactuals each. In order to replicate such an analysis, we estimate $Z_1^{(1)}$ and $Z_2^{(1)}$ using a MAR i.i.d assumption. Many estimators exist for such a scenario, including g-formula and inverse probability weight (IPW)-based estimators \cite{tsiatis06missing}. In this experiment, we use the doubly-robust \textit{augmented IPW} estimator given by $\frac{1}{N} \sum_{n=1}^{N} \bigg[\frac{R_{i,n} Z_{i,n}}{p(R_{i,n}|C_{i,n})} - \frac{\{R_{i,n} - p(R_{i,n}|C_{i,n})\}}{p(R_{i,n}|C_{i,n})} \E[Z_i | C_{i,n}] \bigg]$ , where $i$ represents the unit and $n$ is the sample. See \cite{tsiatis06missing} for more on these estimators. 

A total of $50k$ samples were generated from the network, and estimation was done over 50 bootstrap trials. The resulting estimates are shown in Table~\ref{tab:expmt}. The first four rows show our estimates for counterfactuals when properly adjusted for, followed by the last two rows that show the estimates when we completely ignore dependence between units, but account for missingness using MAR. 

To show how our adjusted estimates perform, we compare the bias of an estimate against the ground truth (which we have from model parameters), in two different scenarios - when we adjust for the network structure and the entanglement appropriately, and one where we do not adjust, for MAR and MNAR.\footnote{In MCAR, the bias would be the same for the two different approaches since we can treat the complete rows in the dataset as an unbiased, representative dataset.} The latter pertains to simply taking the conditional means for different missingness patterns.
Estimates based on complete rows without adjustment would be biased, as shown in the Fig.~\ref{fig:expmt-plots}. Estimates obtained after adjustment are labeled with an asterisk $(*)$ in the x-axis.

As we have pointed out here, there are any number of ways in which an investigator's approach might be biased, if they are not careful about considering \textit{both} dependence and missingness \textit{and} the precise way in which they interact and we have simply illustrated one way in which it can happen. 

\begin{table}[t]
	\caption{Estimates of counterfactuals over 50 bootstrap trials along with quantiles $q_{0.05}$ and $q_{0.95}$. When the investigator is unaware of entanglements, they are restricted to assuming an i.i.d missing data scenario and likely estimate $Z_1^{(1)}$ and $Z_2^{(1)}$ using a MAR assumption, as recorded in the last two rows, done here using an AIPW estimator. These values do not correspond in any reliable way to the four underlying counterfactuals which were estimated with the knowledge of entanglements and the corresponding adjustment required, and recorded in the first four rows.}
	\label{tab:expmt}
	\centering
	\begin{tabular}{cccc} 
		\toprule
		Estimate for & Mean &$q_{0.05}$& $q_{0.95}$ \\
		\midrule
	{$Z_1^{(1,r_2=1)}$} &  -2.942 & -3.009 & -2.889  \\[0.5em]
		{$Z_1^{(1,r_2=0)}$} & 3.138  & 3.108 &  3.164\\ [0.5em]
	{$Z_2^{(1,r_1=1)}$} & 10.816  & 10.773&  10.856 \\ [0.5em]
		$Z_2^{(1,r_1=0)}$ & -4.333 & -4.351 &-4.313  \\ [0.5em]
		$Z_1^{(1)}$ & 1.497 & 1.456 &  1.536\\ [0.5em]
		$Z_2^{(1)}$ & 2.252  & 2.138 & 2.373  \\ 
		\bottomrule
	\end{tabular} 
\end{table}

\begin{figure}
	
	\begin{subfigure}{0.8\linewidth}
		\centering
		\includegraphics[width=\linewidth]{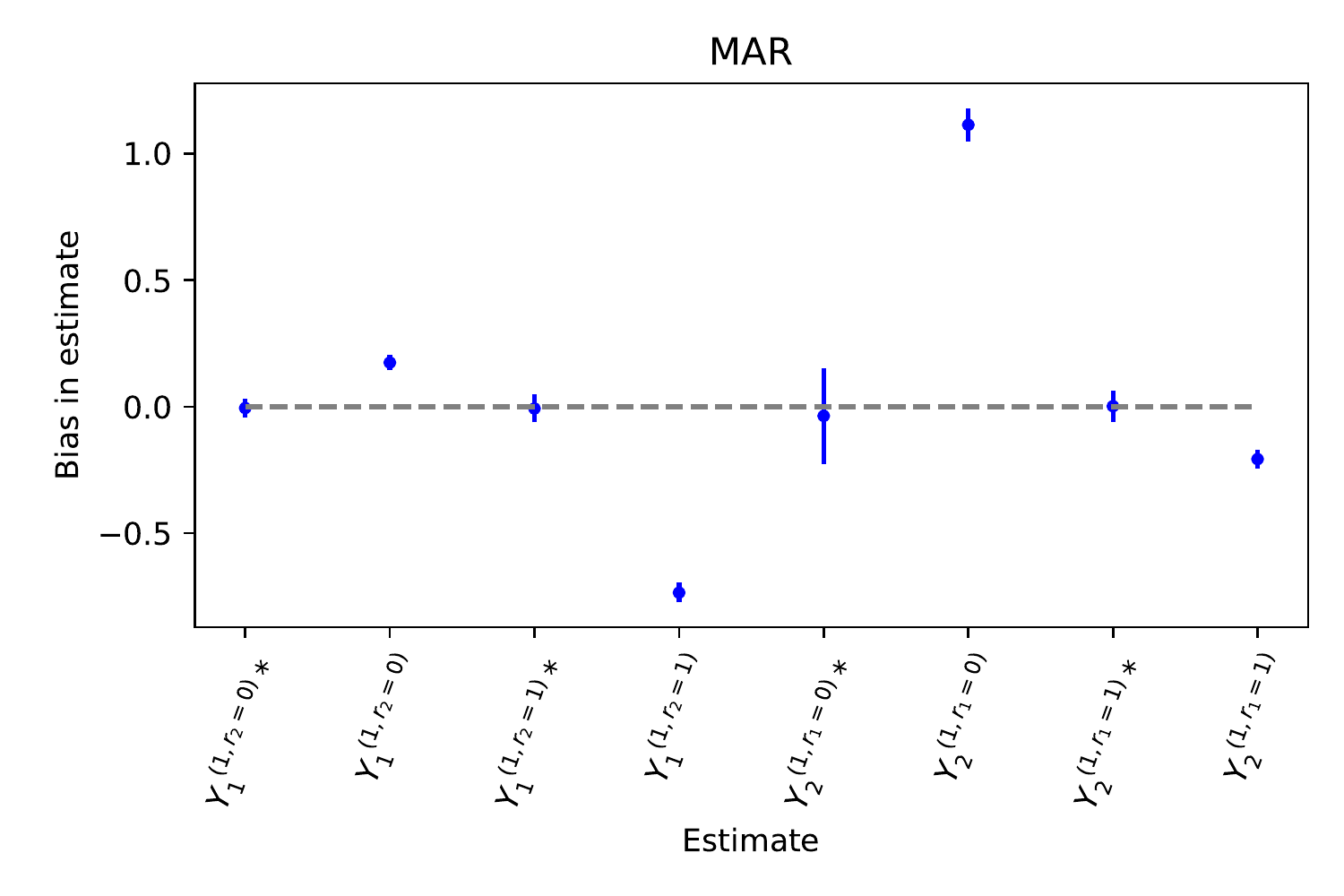}
	\end{subfigure}\hfill%
	\begin{subfigure}{0.8\linewidth}
		\centering
		\includegraphics[width=\linewidth]{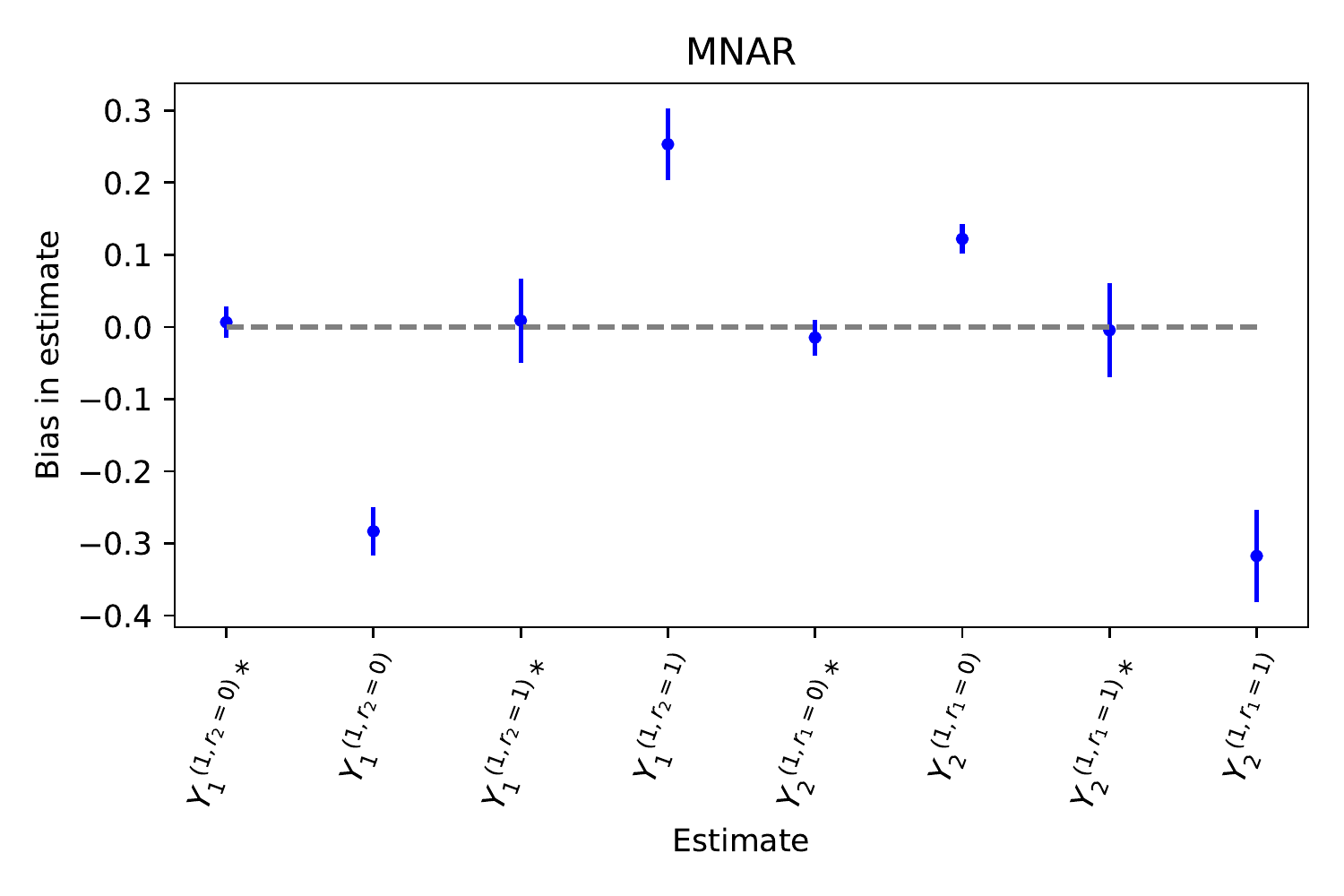}
		
	\end{subfigure}%
	\caption{Bias recorded in bootstrapped estimates of targets (shown in x-axis), MAR case (left) and MNAR (right). We compare our adjusted IPW estimates (denoted by an asterisk $(*)$ on the x-axis) to the unadjusted estimate, which is obtained by ignoring the network missingness structure underpinning the data. Error bars represent quantiles $q_{0.05}$ and $q_{0.95}$ across 50 boostrap samples.}
	\label{fig:expmt-plots}
\end{figure}

\clearpage
\section{Discussion}
\label{sec:discussion}

In this paper, we developed a graphical framework to model entanglements due to the simultaneous presence of data dependence and missing data. In non-i.i.d settings with partial interference, an entanglement could arise due to dependence exhibited among target law variables of different units \textit{(target law dependence)}, dependence of missingness indicators of one unit on target law variables of other units \textit{(missingness process dependence)}, or dependence of observed factuals of one unit on missingness indicators of other units \textit{(missingness interference)}. We have characterized these three entanglement scenarios mathematically and provided \textit{sound} and \textit{complete} full law identification theory under each scenario. We note that several open questions remain in identification of entangled missingness models that we have not addressed in this work. For instance, while we have focused on the full observability law, there are a larger number of models where specific targets or parameters such as the causal effect might still be identified when the full (observability) law is not. Even in i.i.d missing data models, there exist identified target laws where the corresponding full law is not identified; in fact, a complete characterization of target law identification remains an open problem. Finally, we have demonstrated that our model for missingness interference extends i.i.d. graphical missing data methods to allow for missingness indicators that may be causes of  counterfactual or fully observed random variables, a phenomenon that occurs frequently in real-world settings (especially with longitudinal data) but had not previously been considered in the missing data literature.



\clearpage

\begin{appendix}
	
\section{The Nested Markov Factorization}

The nested Markov factorization of $p({\bf V})$ relative to an ADMG $\G({\bf V})$ is defined with the use of conditional distributions known as kernels and their associated \textit{conditional} ADMGs (called \textit{CADMGs}) that are derived from $p({\bf V})$ and $G({\bf V})$ respectively, via repeated applications of the \textit{fixing} operator \cite{richardson17nested}. A CADMG $\G({\bf V}, {\bf W})$ is an ADMG whose
nodes can be partitioned into random variables ${\bf V}$ and fixed variables ${\bf W}$; with the restriction that only outgoing edges may be adjacent to variables in ${\bf W}$. The nested Markov factorization may also more generally be defined for a distribution $p({\bf V}|{\bf W})$ with respect to a CADMG ${\cal G}({\bf V},{\bf W})$, and links \textit{kernels}, mappings derived from $p({\bf V}|{\bf W})$ and CADMGs derived from ${\cal G}({\bf V},{\bf W})$ via a \textit{fixing} operation.

	{\bf Kernel:} A kernel $q_{\bf V}({\bf V} | {\bf W})$ is a mapping from values in ${\bf W}$ to normalized densities over ${\bf V}$ \citep{lauritzen96graphical}. A conditional distribution is a familiar example of a kernel, in that $\sum_{{\bf v} \in {\bf V}} q_{\bf V}({\bf v} | {\bf w}) = 1$. Conditioning and marginalization are defined in kernels in the usual way: For ${\bf A} \subseteq {\bf V}$, $q_{\bf V}({\bf A} | {\bf W}) \equiv \sum_{{\bf V} \setminus {\bf A}} q_{\bf V}({\bf V} | {\bf W})$ and $q_{\bf V}({\bf V} \setminus {\bf A} | {\bf A} \cup {\bf W}) \!\equiv\! \frac{q_{\bf V}({\bf V} | {\bf W})}{q_{\bf V}({\bf A} | {\bf W})}$. \\
	
	{\bf Fixability and the fixing operator:} A variable $V \in {\bf V}$ in a CADMG ${\G}$ is fixable if $\text{de}_{\G}(V) \cap \text{des}_{\G}(V) = \emptyset$. In other words, V is fixable if paths $V \leftrightarrow ... \leftrightarrow B$ and $V \to ... \to B$ do not both exist in ${\G}$ for any $B \in {\bf V} \textbackslash \{V \}$.
	
	We define a fixing operator $\phi_V({\cal G})$ for graphs, and a fixing operator $\phi_V(q; {\cal G})$ for kernels. Given a CADMG ${\cal G}({\bf V},{\bf W})$, with a fixable $V \in {\bf V}$, $\phi_V({\cal G}({\bf V},{\bf W}))$ yields a new CADMG ${\cal G}({\bf V} \setminus \{ V \}, {\bf W} \cup \{ V \})$ obtained from ${\cal G}({\bf V}, {\bf W})$ by moving $V$ from ${\bf V}$ to ${\bf W}$, and removing all edges with arrowheads into $V$.  Given a kernel $q_{{\bf V}}({\bf V} | {\bf W})$, and a CADMG ${\cal G}({\bf V}, {\bf W})$, the operator
	$\phi_V(q_{{\bf V}}({\bf V} | {\bf W}), {\cal G}({\bf V}, {\bf W}))$ yields a new kernel:
	$$q_{{\bf V} \setminus \{ V \}}({\bf V} \setminus \{ V \} | {\bf W} \cup \{ V \}) \equiv
	\frac{
		q_{{\bf V}}({\bf V} | {\bf W})
	}{
		q_{{\bf V}}(V | \mb_{\cal G}(V))
	}$$
	
	{\bf Fixing sequences:} A sequence $\langle V_1, \ldots, V_k \rangle$ is said to be \emph{valid} in ${\cal G}({\bf V},{\bf W})$ if $V_1$ fixable in ${\cal G}({\bf V}, {\bf W})$, $V_2$ is fixable in $\phi_{V_1}({\cal G}({\bf V}, {\bf W}))$, and so on.  If any two sequences $\sigma_1, \sigma_2$ for the same set ${\bf S} \subseteq {\bf V}$ are fixable in ${\cal G}$, they lead to the same CADMG. The graph fixing operator can be extended to a set ${\bf S}$: $\phi_{{\bf S}}({\cal G})$. This operator is defined as applying the vertex fixing operation in any valid sequence $\sigma$ for set ${\bf S}$.

	Given a sequence $\sigma_{{\bf S}}$, define $\eta(\sigma_{{\bf S}})$ to be the first element in $\sigma_{{\bf S}}$, and $\tau(\sigma_{{\bf S}})$ to be the subsequence of $\sigma_{{\bf S}}$ containing all elements but the first. Given a sequence $\sigma_{{\bf S}}$ on elements in ${\bf S}$ valid in ${\cal G}({\bf V}, {\bf W})$, the kernel fixing operator $\phi_{\sigma_{{\bf S}}}(q_{{\bf V}}({\bf V} | {\bf W}), {\cal G}({\bf V}, {\bf W}))$ is defined to be equal to $q_{{\bf V}}({\bf V} | {\bf W})$ if $\sigma_{{\bf S}}$ is the empty sequence, and
	$\phi_{\tau(\sigma_{{\bf S}})}( \phi_{\eta(\sigma_{{\bf S}})}(q_{{\bf V}}({\bf V} | {\bf W}); {\cal G}({\bf V}, {\bf W}))$, $\phi_{\eta(\sigma_{{\bf S}})}({\cal G}({\bf V}, {\bf W})))$ otherwise.\\
	
	{\bf Reachability:} Given a CADMG ${\cal G}({\bf V}, {\bf W})$, a set ${\bf R} \subseteq {\bf V}$ is called \emph{reachable} if there exists a sequence for ${\bf V} \setminus {\bf R}$ valid in ${\cal G}({\bf V},{\bf W})$. In other words, if ${\bf S}$ is fixable in $\G$, ${\bf V} \setminus {\bf S}$ is reachable. \\
	
	{\bf Intrinsic sets:} A set ${\bf R}$ reachable in ${\cal G}({\bf V},{\bf W})$ is \emph{intrinsic} in ${\cal G}({\bf V},{\bf W})$ if $\phi_{{\bf V} \setminus {\bf R}}({\cal G})$ contains a single district, ${\bf R}$ itself.  The set of intrinsic sets in a CADMG ${\cal G}$ is denoted by ${\cal I}({\cal G})$.\\
	
	{\bf Nested Markov factorization:} A distribution $p({\bf V}|{\bf W})$ is said to obey the \emph{nested Markov factorization} with respect to the CADMG ${\cal G}({\bf V},{\bf W})$ if there exists a set of kernels of the form $\{ q_{{\bf S}}({\bf S} | \pa_{\cal G}({\bf S}) ) : {\bf S} \in {\cal I}({\cal G}) \} \}$ such that for every valid sequence $\sigma_{{\bf R}}$ for a reachable set ${\bf R}$ in ${\cal G}$, we have:
	\begin{align*}
		\phi_{\sigma_{{\bf R}}}(p({\bf V}|{\bf W}); {\cal G}({\bf V},{\bf W})) 
		= \quad \prod_{\mathclap{{\bf D} \in {\cal D}(\phi_{{\bf R}}({\cal G}({\bf V},{\bf W})))}} \quad q_{{\bf D}}({\bf D} | \pas_{\cal G}({\bf D}) )
	\end{align*}
	
	If a distribution obeys this factorization, then for any reachable ${\bf R}$, any two valid sequences on ${\bf R}$ applied to $p({\bf V}|{\bf W})$ yield the same kernel $q_{{\bf R}}({\bf R} | {\bf V} \setminus {\bf R})$.  Hence, kernel fixing may be defined on sets, just as graph fixing.  In this case, for every
	${\bf D} \in {\cal I}({\cal G})$, $q_{{\bf D}}({\bf D} | \pas_{\cal G}({\bf D}) ) \equiv \phi_{{\bf V} \setminus {\bf D}}(p({\bf V}|{\bf W}); {\cal G}({\bf V},{\bf W}))$.
	
	The \emph{district factorization} or \emph{Tian factorization} of $p({\bf V}|{\bf W})$ results from the nested factorization:
	\begin{align*}
		p({\bf V}|{\bf W})
		&= \prod_{{\bf D} \in {\cal D}({\cal G}({\bf V},{\bf W}))} q_{{\bf D}}({\bf D} | \pas_{\cal G}({\bf D})) \\
		&
		= \prod_{{\bf D} \in {\cal D}({\cal G}({\bf V},{\bf W}))} \left( \prod_{D \in {\bf D}} p(D \mid \pre_{\prec}(D)) \right),
	\end{align*}
	where $\pre_{\prec}(D)$ is the set of predecessors of $D$ according to a topological total ordering $\prec$.  Each factor $\prod_{D \in {\bf D}} p(D \mid \pre_{\prec}(D))$ is only a function of ${\bf D} \cup \pa_{\cal G}({\bf D})$ under the nested factorization.
	
	An important result in \citep{richardson17nested} states that if $p({\bf V} \cup {\bf H}|{\bf W})$ obeys the factorization for a CDAG $\G({\bf V} \cup {\bf H},{\bf W})$, then $p({\bf V}|{\bf W})$ obeys the nested factorization for the latent projection CADMG ${\cal G}({\bf V},{\bf W})$.
	
	\textbf{{Identification}}: 
	Not every interventional distribution $p({\bf Y}({\bf a}))$ is identified in a hidden variable causal model.
	However, \emph{every} $p({\bf Y}({\bf a})|{\bf W})$ identified from $p({\bf V}|{\bf W})$ can be expressed as a modified nested factorization as follows:
	\begin{align*}
		\MoveEqLeft[1] p({\bf Y}({\bf a})|{\bf W}) \\
		&= \sum_{{\bf Y}^* \setminus {\bf Y}} \prod_{{\bf D} \in {\cal D}(
			{\cal G}_{{\bf Y}^*}
			)} p({\bf D} | \doo(\pas_{\cal G}({\bf D}) )) \vert_{{\bf A} = {\bf a}}\\
		&= \sum_{{\bf Y}^* \setminus {\bf Y}} \prod_{{\bf D} \in {\cal D}(
			{\cal G}_{{\bf Y}^*}
			)} \phi_{{\bf V} \setminus {\bf D}}(p({\bf V}|{\bf W}); {\cal G}({\bf V},{\bf W})) \vert_{{\bf A} = {\bf a}}
	\end{align*}
	where ${\bf Y}^* \equiv \an_{{\cal G}({\bf V}({\bf a}), {\bf W})}({\bf Y}) \setminus {\bf a}$. That is, $p({\bf Y}({\bf a})|{\bf W})$ is only identified if it can be expressed as a factorization, where every
	piece corresponds to a kernel associated with a set intrinsic in $\G({\bf V},{\bf W})$. Moreover, no piece in this factorization contains elements of $\bf A$ as random variables.
	
	\textbf{Binary Parameterization of Nested Markov Models:}
	From the nested factorization, intrinsic sets given their parents form the atomic units of the nested Markov model. Using this observation, a smooth parameterization of discrete nested Markov models was provided by \cite{evans14markovian}. We provide a brief description of how to derive the M\"{o}bius parameters of a binary nested Markov model.
	
For each district ${\bf D} \in \mathcal{D}(\G))$, consider all possible subsets ${\bf S} \subseteq {\bf S}$. If ${\bf S}$ is intrinsic 
in $\phi_{{\bf V}\setminus {\bf S}}(\G)$, define the head ${\bf H}$ of the intrinsic set to be all vertices in $\bf S$ that are childless in $\phi_{{\bf V}\setminus {\bf S}}(\G)$ and the tail $\bf T$ to be all parents of the head in the CADMG $\phi_{{\bf V}\setminus {\bf S}}(\G)$, excluding the head itself. Formally, ${\bf H} \equiv \{{V} \in {\bf S} | \ch_{\phi_{{\bf V}\setminus {\bf S}}(\G)} (V) = \emptyset\}$ and $T \equiv \pa_{\phi_{{\bf V}\setminus {\bf S}}(\G)}({\bf H})\setminus {\bf H}$. The corresponding set of M\"{o}bius parameters for this intrinsic head and tail pair parameterizes
the kernel $q_{{\bf S}}({\bf H} = 0 | {\bf T})$; i.e., the kernel where all variables outside the intrinsic set $\bf S$ are fixed, and all elements of the head are set to zero given the tail. 
Note that these parameters are, in general, variationally dependent (in contrast to variationally independent in the case of an ordinary DAG model) as the heads and tails in these parameter sets may overlap. The joint density for any query $p({\bf V} = {\bf v})$ can be obtained through the M\"{o}bius inversion formula; see \cite{lauritzen96graphical, evans14markovian} for details. We will denote $q_{{\bf S}}({\bf H} = {\bf 0} | {\bf T})$ simply as $q({\bf H} = {\bf 0} | {\bf T})$ as it is generally clear what variables are still random in the kernel corresponding to a given intrinsic set.

\textbf{Binary Parameterization of Missing Data Models:}
We use the parameterization described earlier to count the number of parameters required to
parameterize the full observability law of a missing data ADMG and its corresponding observed law. We then use this to reason that if the number of parameters in the full observability law exceeds those in the observed law, it is impossible to establish a map from the observed law to the full law. This in turn implies that such a full observability law is not identified. In the full observability, the deterministic factors, i.e., proxies given parents can be ignored as the probability of those events is always 1. However, while counting the observed law, we are careful to treat counterfactuals as unobserved and obtain the corresponding ADMG. The M\"{o}bius
parameters are then derived in a similar manner as before, but with additional constraint that if $Z_i$ appears in the
head of a parameter, and missingness indicators $R_i$ or ${\bf R}_{\aff(i)}$ appear in the tail, then the kernel must be
restricted to cases where $R_i = 1$ \textit{and} ${\bf R}_{\aff(i)} = {\bf 1}$. This is because, (1) when $R_i=0$, the probability of the head taking any value aside from those where $Z_i = \:\: ?$ is deterministically 0, and (2) cases where ${\bf R}_{\aff(i)}$ are set to values different from 1 are irrelevant to the identification of the full observability law. {The consideration that $R_i=1$ always holds, but that ${\bf R}_{\aff(i)} = {\bf 1}$ is only for identifying the full observability law. For other cases, considerations vary.}

	\section{Proofs}
	We restate the theorems and outline their proofs here.

\begin{thma}{\ref{thm-vector-mcar-mar-margins} }
	In a missing data ADMG $\G$ with missingness interference, valid single-world objects $h(\tilde{\bf Z};{\bf r})$ consisting of a set of counterfactuals $\bf Z' \equiv \bigcup_i \{Z_i^{(1,{\bf r}_{\aff(i)})}\} $, $i \in \{1, \cdots , n\}$ are identified 
	when either of these two conditions is satisfied: (1) ${\bf R}' \ci {\bf O}, {\tilde{{\bf Z}}}$ (MCAR), or (2) ${\bf R}' \ci  {\tilde{{\bf Z}}} | {\bf O}$ (MAR), where ${\bf R}'$ refers to the set 
	of all missingness indicators $R$ that index counterfactuals in $h(\tilde{\bf Z};{\bf r})$. The object $h(\tilde{\bf Z};{\bf r})$ is a function of $p({\bf Z}', {\bf R}, {\bf O})$, and the identifying functional is given by:
	\begin{align}
		p({\bf Z}', {\bf R}, {\bf O}) =  p({\bf Z}',{\bf O}) \times { p({\bf R}|{\bf O},{\bf Z}') }
		= \frac{p( {\bf Z}', {\bf R}={\bf r}, {\bf O})}{p({\bf R}={\bf r} \mid {\bf O})} \times {p({\bf R} \mid {\bf O})}
	\end{align}
	where propensity scores are obtained by simple m-separation or (d-separation) rules on ADMG (or DAG) factorization. 
	
	
\end{thma}

\begin{prf}
	${\bf Z}'$ is the set of counterfactuals in $h(\tilde{{\bf Z}}; {\bf r})$ such that they construct a valid single-world counterfactual in the world ${\bf R} = {\bf r}$. Further, let the set of all missingness indicators that index ${\bf Z}'$ be ${\bf R}'$, and that ${\bf R}' = {\bf r}'$ when ${\bf R} = {\bf r}$. We are interested in identifying a distribution $P({\bf Z}')$ or a function $h(\tilde{\bf Z};{\bf r})$ thereof. The crucial factor here is that no variable, whose missingness indicator is 0, is present in $h(.)$.
	
	In case (1), we can write $P({\bf Z}' ) = P({\bf Z}' \mid {\bf R}' = {\bf r}')$ since ${\bf R}' \ci {\tilde{{\bf Z}}}$ and ${\bf Z}' \subset  \tilde{{\bf Z}}$. And by consistency, we can replace all counterfactuals by their corresponding proxies and the object is identified.
	
	In case (2), we can write $P({\bf Z}' ) = P({\bf Z}' \mid {\bf O})\times P({\bf O}) = P({\bf Z}' \mid {\bf O}, {\bf R}' = {\bf r}')\times P({\bf O})$ since ${\bf R}' \ci {\tilde{{\bf Z}}} \mid {\bf O}$ and ${\bf Z}' \subset  \tilde{{\bf Z}}$. And by consistency, we can identify the distribution as we can replace all counterfactuals by their corresponding proxies.

\end{prf}

%
	
	\begin{thma}{\ref{thm-vector-mnar-11s-sound-complete} }
	In a missing data ADMG $\G$ with missingness interference, under Assumption \ref{assump:10-not-parent]}, the full-observability law $P(\tilde{\bf Z}^{({\bf r}={\bf 1})}, {\bf R})$ is identified 
	{if and only if}
	there is no e-colluding path for any $R \in {\bf R}$ in the ADMG. Further, if $\G$ is a missing data DAG, the full-observability law is identified 
	{if and only if} there is no e-colluder and no e-self-censoring. The identifying functional is given by
	\begin{align*}
		p(\tilde{\bf Z}^{({\bf r}={\bf 1})}, {\bf R}) =  p(\tilde{\bf Z}^{({\bf r}={\bf 1})}) \times \underbrace{p({\bf R} \mid \tilde{\bf Z}^{({\bf r}={\bf 1})})}_{g(p({\bf R},{\bf Z}))}
		= \frac{p( \tilde{\bf Z}^{({\bf r}={\bf 1})}, {\bf R}=1)}{\underbrace{p({\bf R}=1 \mid \tilde{\bf Z}^{({\bf r}={\bf 1})})}_{g(p({\bf R},{\bf Z}))\vert_{{\bf R}=1}}} \times \underbrace{p({\bf R} \mid \tilde{\bf Z}^{({\bf r}={\bf 1})})}_{g(p({\bf R},{\bf Z}))}.
	\end{align*}
	and missingness mechanism $p({\bf R} \mid \tilde{\bf Z}^{({\bf r}={\bf 1})})$ is identified using the OR parameterization given below:
	\begin{align*}
		p({\bf R} | \tilde{{\bf Z}}^{({\bf r} = {\bf 1})} ) = \frac{1}{\sigma} \times \prod_{k=1}^{K} p(R_k | R_{-k} = 1, \tilde{{\bf Z}}^{({\bf r} = {\bf 1})} ) \times \prod_{k=2}^{K}  \text{OR}(R_k, R_{\prec k} | R_{\succ k} = 1, \tilde{{\bf Z}}^{({\bf r} = {\bf 1})} )  
	\end{align*}
\end{thma}
\begin{prf}

	\textbf{Soundness:} The absence of e-colluding paths results in identification.
	
	The absence of a e-colluding path between $Z_i^{(1, {\bf r}_\aff(i))}$ and $R_k \in {\bf R}_{\aff(i)}$ 
	implies that  $Z_i^{(1, {\bf r}_{\aff(i)})} \notin \mb_{\G} (R_k)$ where 
	\begin{align*}
		\mb_{\G}(V) = \{\pa_{\G}(V), \dis_{\G}(V), \pa_{\G}(\dis_{\G}(V)), \ch_{\G}(V), \pa_{\G}(\ch_{\G}(V)), \\
		\dis_{\G}(\ch_{\G}(V)), \pa_{\G}(\dis_{\G}(\ch_{\G}(V)))\}
	\end{align*} 
	
	Let $\tilde{\bf Z}^{({\bf r}_k = {\bf 1})} = \{Z_i^{(1,{\bf r}_{\aff(i)} = {\bf 1})} : R_k \in {\bf R}_{\aff(i)} \} \cup Z_k^{(1,{\bf r}_{\aff(k)})}$. In words, $\tilde{\bf Z}^{({\bf r}_k= {\bf 1})}$ is the set of all counterfactuals that correspond to full observability, and are indexed (and influenced) by $R_k$.
	
	By Markov property, we have that $V \ci {\bf V} \setminus \mb_{\G}(V) | \mb_{\G}(V)$. Therefore, the absence of e-colluding paths implies the following assumptions:
	\begin{align*}
		R_k \ci \{Z_i^{(1, {\bf r}_{\aff(i)})} \in \tilde{\bf Z}^{({\bf r}_k = {\bf 1})}\} \:\:\ | \:\: \{  R \setminus R_k, \tilde{{\bf Z}}  \setminus Z_i^{(1,{\bf r}_{\aff(i)} = {\bf 1})}\} \:\:\:\: \text{ for } R_k \in  {\bf R}
	\end{align*}
	Given these assumptions, we can identify $p({\bf R} | \tilde{{\bf Z}}^{({\bf r} = {\bf 1)}}  )$ using the OR parameterization. The proof is similar to the identification proof of the no self-censoring model given in \cite{malinsky2021semiparametric} and the  representation of it in \cite{nabi2020full}. Since we are assuming $\ch_{\G}(\tilde{\bf Z}^{({\bf r}\neq{\bf 1})} ) \cap {\bf R} = \emptyset$, it suffices to only ID $p({\bf R} | \tilde{{\bf Z}}^{({\bf r} = {\bf 1})}  )$. 
	\begin{align*}
		p({\bf R} | \tilde{{\bf Z}}^{({\bf r} = {\bf 1})} ) = \frac{1}{\sigma} \times \prod_{k=1}^{K} p(R_k | {\bf R}_{-k} = 1, \tilde{{\bf Z}}^{({\bf r} = {\bf 1})} ) \times \prod_{k=2}^{K}  \text{OR}(R_k, {\bf R}_{\prec k} | {\bf R}_{\succ k} = 1, \tilde{{\bf Z}}^{({\bf r} = {\bf 1})} )  
	\end{align*}
	where notation and OR is consistent with earlier definition. 
	
	We have $p(R_k | {\bf R}_{-k} = {\bf 1}, \tilde{{\bf Z}}^{({\bf r} = {\bf 1})} ) =  p(R_k | {\bf R}_{-k} = 1, \tilde{{\bf Z}}^{({\bf r} = {\bf 1})} \setminus \tilde{\bf Z}^{({\bf r}_k = {\bf 1})})$. By consistency, every counterfactual past the conditioning bar is equal to the observed proxy. Then, pairwise OR terms are identified because OR$(R_k, R_i \mid {\bf R}_{-(k,i)}=1,\tilde{{\bf Z}}^{({\bf r} = {\bf 1})} ) $ is not a function of $\tilde{\bf Z}^{({\bf r}_i = {\bf 1})}$ or $\tilde{\bf Z}^{({\bf r}_k = {\bf 1})}$. Finally higher order terms are ID in similar ways \cite{nabi2020full, malinsky2021semiparametric}.
	
	\textbf{Completeness:}
	One approach to demonstrate completeness is to count the number of parameters required to parameterize the full-observability law of a missing data graph and its corresponding observed law and reason that if the former requires more parameters than the latter, it is impossible to identify all the parameters of the full-observability law uniquely. In order to do so, we assume all variables are binary, and adopt the binary parameterization of Nested Markov Models.

	First, we present two simple e-colluding path examples to show how the parameter counting argument proceeds. Thereafter, we provide the general argument for completeness for all graphs. 
	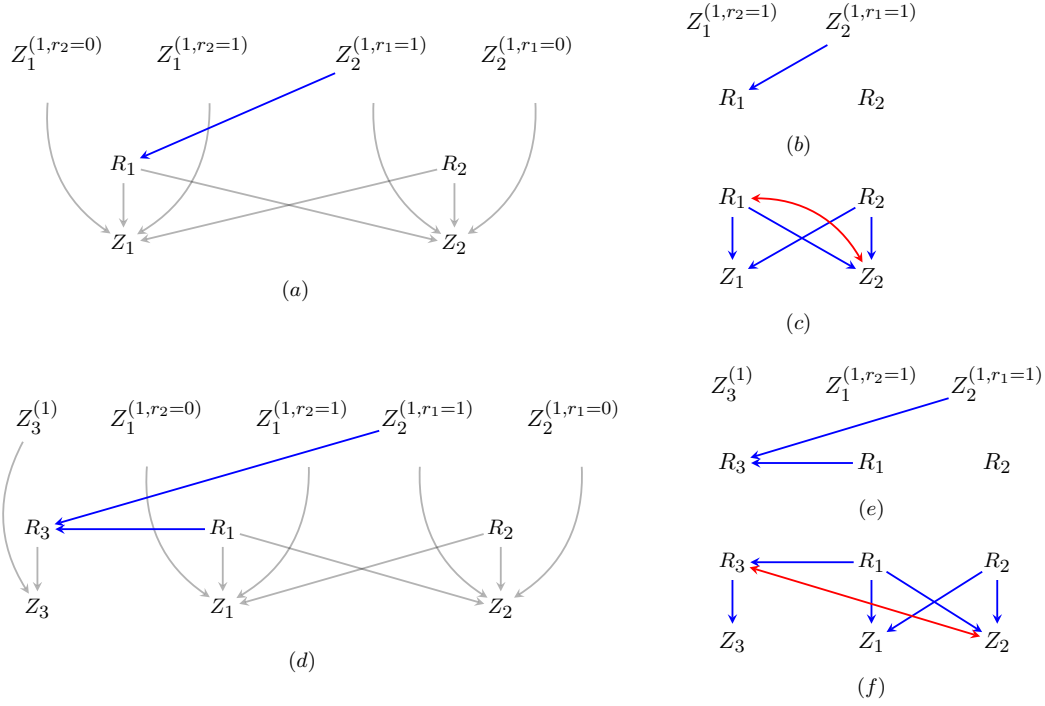
\begin{figure}[!t]
		\begin{center}
			\scalebox{0.88}{
				\begin{tikzpicture}[>=stealth, node distance=1.2cm]
					\tikzstyle{format} = [draw=none, thick, circle, minimum size=4mm,
					inner sep=0pt]
					
					\begin{scope} [xshift=-1cm]
						\path[->, thick]
						
						node[format] (y111) {\small{$Z_{1}^{(1,r_{2}=1)}$}}
						node[format, right of=y111, xshift=1.5cm] (y211) {\small{$Z_{2}^{(1,r_{1}=1)}$}}
						node[format, below of=y111, xshift=-1.2cm,yshift=-0.5cm] (r11) {$R_{1}$}
						node[format, below of=r11] (y11) {$Z_{1}$}
						
						node[format, left of=y111, xshift=-1cm] (y121) {\small{$Z_{1}^{(1,r_{2}=0)}$}}
						node[format, right of=y211, xshift=1cm] (y221) {\small{$Z_{2}^{(1,r_{1}=0)}$}}
						node[format, below of=y221,  xshift=-1.1cm, yshift=-0.5cm] (r21) {$R_{2}$}
						node[format, below of=r21] (y21) {$Z_{2}$}

						(r21) edge[black, opacity=0.3] (y11)
						(r11) edge[black, opacity=0.3] (y21)
						
						(r11) edge[black, opacity=0.3] (y11)
						(r21) edge[black, opacity=0.3] (y21)
						
						(y111) edge[black, opacity=0.3, bend left = 30] (y11)
						(y121) edge[black, opacity=0.3, bend right = 30] (y11)
						(y211) edge[black, opacity=0.3, bend right = 30] (y21)
						(y221) edge[black, opacity=0.3, bend left = 30] (y21)
						
						(y211) edge[blue] (r11)
						
						
						node[below of=y111, yshift=-2.4cm, xshift=1.4cm] (la) {$(a)$}
						;
					\end{scope}
					
					\begin{scope} [xshift=7cm,yshift=0.5cm]
						\path[->, thick]
						
						

						node[format] (y111) {\small{$Z_{1}^{(1,r_{2}=1)}$}}
						node[format, below of =y111] (r1) {\small{$R_1$}}
						node[format,right of=y111,xshift=0.9cm] (y211) {\small{$Z_{2}^{(1,r_{1}=1)}$}}
						node[format, below of =y211] (r2) {\small{$R_2$}}
						
						(y211) edge[blue] (r1)

						node[below of=y111, yshift=-0.7cm, xshift=1cm] (la) {$(b)$}
						;
					\end{scope}
					
					\begin{scope} [xshift=7cm,yshift=-2.2cm]
						\path[->, thick]
						
						

						node[format] (r1) {\small{$R_1$}}
						node[format, below of =r1] (y1) {\small{$Z_{1}$}}
						node[format,right of=r1,xshift=0.9cm] (r2) {\small{$R_2$}}
						node[format, below of =r2] (y2) {\small{$Z_{2}$}}
						
						(r1) edge[blue] (y1)
						(r2) edge[blue] (y2)
						
						(r2) edge[blue] (y1)
						(r1) edge[blue] (y2)
						(r1) edge[<->,red,bend left=30] (y2)
						
						node[below of=r1, yshift=-0.7cm, xshift=1cm] (la) {$(c)$}
						;
					\end{scope}	
					
					\begin{scope} [xshift=-2cm, yshift=-5.5cm]
						\path[->, thick]
						
						
						
						
						node[format,xshift=-1.5cm] (y31) {\small{$Z_{3}^{(1)}$}}
						node[format,below of = y31,yshift=-0.5cm] (r3) {$R_{3}$}
						node[format, below of=r3] (y3) {$Z_{3}$}
						
						node[format,xshift=2.5cm] (y111) {\small{$Z_{1}^{(1,r_{2}=1)}$}}
						node[format, right of=y111, xshift=0.7cm] (y211) {\small{$Z_{2}^{(1,r_{1}=1)}$}}
						node[format, below of=y111, xshift=-1.2cm,yshift=-0.5cm] (r11) {$R_{1}$}
						node[format, below of=r11] (y11) {$Z_{1}$}

						node[format, left of=y111, xshift=-1cm] (y121) {\small{$Z_{1}^{(1,r_{2}=0)}$}}
						node[format, right of=y211, xshift=1cm] (y221) {\small{$Z_{2}^{(1,r_{1}=0)}$}}
						node[format, below of=y221,  xshift=-1.1cm, yshift=-0.5cm] (r21) {$R_{2}$}
						node[format, below of=r21] (y21) {$Z_{2}$}

						(r21) edge[black, opacity=0.3] (y11)
						(r11) edge[black, opacity=0.3] (y21)
						
						(r11) edge[black, opacity=0.3] (y11)
						(r21) edge[black, opacity=0.3] (y21)
						
						(y111) edge[black, opacity=0.3, bend left = 30] (y11)
						(y121) edge[black, opacity=0.3, bend right = 30] (y11)
						(y211) edge[black, opacity=0.3, bend right = 30] (y21)
						(y221) edge[black, opacity=0.3, bend left = 30] (y21)
						
						
						(r3) edge[black, opacity=0.3] (y3)
						(y31) edge[black, bend right=30, opacity=0.3] (y3)
						
						(y211) edge[blue] (r3)
						(r11) edge[blue] (r3)
						
						node[below of=y111, yshift=-2.5cm, xshift=0cm] (la) {$(d)$}
						;
					\end{scope}
					
					\begin{scope} [xshift=7cm,yshift=-5cm]
						\path[->, thick]
						
						node[format] (y31) {\small{$Z_{3}^{(1)}$}}
						node[format, below of =y31] (r3) {\small{$R_3$}}
						node[format,right of=y31,xshift=0.9cm] (y11) {\small{$Z_{1}^{(1,r_{2}=1)}$}}
						node[format, below of =y11] (r1) {\small{$R_1$}}
						node[format,right of=y11,xshift=0.7cm] (y21) {\small{$Z_{2}^{(1,r_{1}=1)}$}}
						node[format, below of =y21] (r2) {\small{$R_2$}}
						
						(r1) edge[blue] (r3)
						(y21) edge[blue] (r3)

						node[below of=y11, yshift=-0.7cm, xshift=0cm] (la) {$(e)$}
						;
					\end{scope}
					
					\begin{scope} [xshift=7cm,yshift=-7.7cm]
						\path[->, thick]

						node[format] (r3) {$R_3$}
						node[format, below of =r3] (y3) {\small{$Z_3$}}
						node[format,right of=r3,xshift=0.9cm] (r1) {$R_1$}
						node[format, below of =r1] (y1) {\small{$Z_1$}}
						
						node[format,right of=r1,xshift=0.7cm] (r2) {$R_2$}
						node[format, below of =r2] (y2) {\small{$Z_2$}}
						
						(r1) edge[blue] (r3)
						(r1) edge[blue] (y2)
						(r2) edge[blue] (y1)
						
						(r1) edge[blue] (y1)
						(r2) edge[blue] (y2)
						(r3) edge[blue] (y3)
						
						(y2) edge[<->,red] (r3)
						

						node[below of=y1, yshift=0.5cm, xshift=0cm] (la) {$(f)$}
						;
					\end{scope}
					
				\end{tikzpicture}
			}
		\end{center}
		\caption{Examples where $P({\tilde{\bf{Z}}^{({\bf r} ={\bf 1})} }, {\bf R})$ is not identified. The proof is demonstrated using parameter counting. (a)-(c): Extended self-censoring (neighbor-censoring), (a) depicts the data generating process, (b) ADMG for $P({\tilde{\bf{Z}}^{({\bf r} ={\bf 1})} }, {\bf R})$, (c) ADMG for observed law. (d)-(f): Extended colluder, (d) depicts the data generating process, (e) ADMG for $P({\tilde{\bf{Z}}^{({\bf r} ={\bf 1})} }, {\bf R})$, (f) ADMG for observed law.}
		\label{fig:non-id-mobius}
	\end{figure}
	
	Consider the simple graph in Fig.~\ref{fig:non-id-mobius}(a) which has e-self-censoring $Z_{2}^{(1,r_{1}=1)} \rightarrow R_1$. To count the parameters required for the full-observability law, we follow the procedure in \cite{nabi2020full}. The binary parameterization of the full law of a missing data ADMG is the same as counting in an ordinary ADMG, with all the irrelevant counterfactuals projected out, except that deterministic factors $P(Z_i| R_j, Z_i^{(1,r_{\aff(i)})})$ can be ignored. This gives us the graph in Fig.~\ref{fig:non-id-mobius}(b). 
	The 5 parameters associated with this graph are  $q(Z_1^{(1,r_2=1)} = 0)$, $q(Z_2^{(1,r_1=1)} = 0)$, $q(R_2=0)$ and $q(R_1=0|Z_2^{(1,r_1=1)} =1 )$,  $q(R_1=0|Z_2^{(1,r_1=1)} =0 )$. 
	
	Next, we obtain the graph in (c) for the observed law, by projecting out all the counterfactuals in (a). Counting the observed law under full observability in entangled missingness settings has a special consideration: if $Z_i$ appears in the head of a parameter, and any of the corresponding missingness indicators $R_{\aff(i)}$ appear in the tail, the kernel must be restricted to cases where $R_i=1$ and ${\bf R}_{\aff(i)}= {\bf 1}$. This is because, (1) when $R_i=0$, the probability of the head taking any value aside from those where $Z_i = \:\: ?$ is deterministically 0, and (2) cases where ${\bf R}_{\aff(i)}$ are set to values different from 1 are irrelevant to the identification of the full observability law\footnote{The consideration that $R_i=1$ always holds, but that ${\bf R}_{\aff(i)} = {\bf 1}$ is only for identifying the full observability law. For other cases, considerations vary.}. The observed law in (c) can be parameterized using only 4 parameters, one each corresponding to $q(R_1=0)$, $q(R_2=0)$, $q(R_1=0, Z_2=0|R_2=1)$ and $q(Z_1=0|R_1= 1,R_2=1)$. That is one less than the full observability law and hence the latter may not be uniquely determined from data.
	
	Second, consider the graph in Fig.~\ref{fig:non-id-mobius}(d), which has an e-colluder $Z_2^{(1,r_1=1)} \rightarrow R_3 \leftarrow R_1$. The full observability law is shown in (e) and observed law in (f). Counting for (e) yields a total of 9 parameters: $q(Z_3^{(1)}=0)$, $q(Z_1^{(1,r_2=1)}=0)$, $q(Z_2^{(1,r_1=1)}=0)$, $q(R_1=0)$, $q(R_2=0)$ and finally, $q(R_3=0|R_1, Z_2^{(1,r_1=1)})$, which accounts for 4 parameters. The observed law, on the other hand, needs only 8 parameters: $q(R_1=0)$, $q(R_2=0)$, $q(R_3=0|R_1=1)$, $q(R_3=0|R_1=0)$, $q(Z_3=0|R_3=1)$, $q(Z_1=0|R_1=1,R_2=1)$, $q(Z_2=0|R_1=1,R_2=1)$ and $q(Z_2=0, R_3=0|R_1=1,R_2=1)$. Hence, it is not possible to uniquely map back to the full observability law from observed data.

	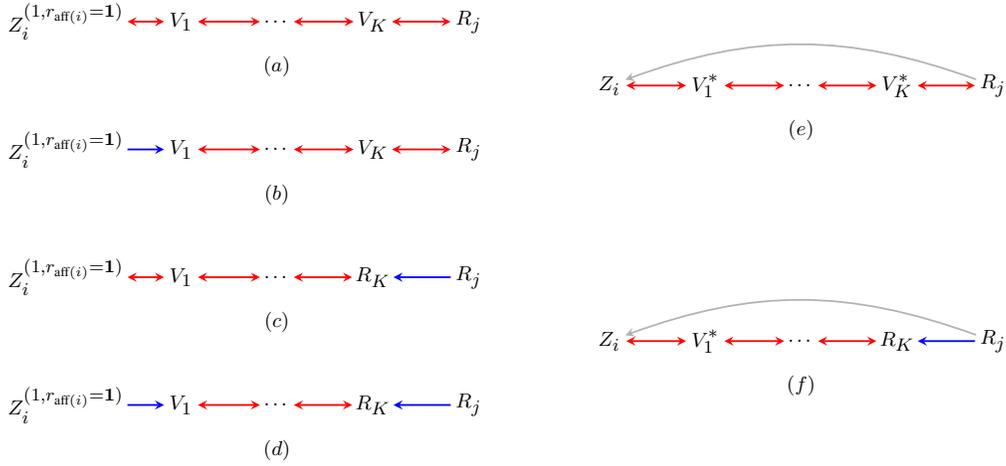
\begin{figure}[!t]
		\begin{center}
			\scalebox{0.85}{
				\begin{tikzpicture}[>=stealth, node distance=1.5cm]
					\tikzstyle{format} = [draw=none, thick, circle, minimum size=5.0mm,	inner sep=0pt]
					
					\begin{scope}[yshift=0cm]
						\path[->,  thick]
						node[format] (z) {$Z_i^{(1,r_{\aff(i)} = {\bf 1})}$}
						node[format, right of=z, xshift=0.3cm] (v1) {$V_1$}
						node[format, right of=v1] (v2) {$\cdots$}
						node[format, right of=v2] (v3) {$V_K$}
						node[format, right of=v3] (r) {$R_j$}

						(z) edge[<->,red] (v1)
						(v1) edge[<->,red] (v2)
						(v2) edge[<->,red] (v3)
						(v3) edge[<->,red] (r)
						
						node[below of=v2, yshift=0.8cm, xshift=0cm] (l) {$(a)$}				;
					\end{scope}
					
					\begin{scope}[yshift=-2cm]
						\path[->,  thick]
						node[format] (z) {$Z_i^{(1,r_{\aff(i)} = {\bf 1})}$}
						node[format, right of=z, xshift=0.3cm] (v1) {$V_1$}
						node[format, right of=v1] (v2) {$\cdots$}
						node[format, right of=v2] (v3) {$V_K$}
						node[format, right of=v3] (r) {$R_j$}

						(z) edge[blue] (v1)
						(v1) edge[<->,red] (v2)
						(v2) edge[<->,red] (v3)
						(v3) edge[<->,red] (r)
						
						node[below of=v2, yshift=0.8cm, xshift=0cm] (l) {$(b)$}				;
					\end{scope}
					
					\begin{scope}[yshift=-4cm]
						\path[->,  thick]
						node[format] (z) {$Z_i^{(1,r_{\aff(i)} = {\bf 1})}$}
						node[format, right of=z, xshift=0.3cm] (v1) {$V_1$}
						node[format, right of=v1] (v2) {$\cdots$}
						node[format, right of=v2] (v3) {$R_K$}
						node[format, right of=v3] (r) {$R_j$}

						(z) edge[<->, red] (v1)
						(v1) edge[<->,red] (v2)
						(v2) edge[<->,red] (v3)
						(v3) edge[<-,blue] (r)
						
						node[below of=v2, yshift=0.8cm, xshift=0cm] (l) {$(c)$}				;
					\end{scope}
					
					\begin{scope}[yshift=-6cm]
						\path[->,  thick]
						node[format] (z) {$Z_i^{(1,r_{\aff(i)} = {\bf 1})}$}
						node[format, right of=z, xshift=0.3cm] (v1) {$V_1$}
						node[format, right of=v1] (v2) {$\cdots$}
						node[format, right of=v2] (v3) {$R_K$}
						node[format, right of=v3] (r) {$R_j$}

						(z) edge[blue] (v1)
						(v1) edge[<->,red] (v2)
						(v2) edge[<->,red] (v3)
						(v3) edge[<-,blue] (r)
						
						node[below of=v2, yshift=0.8cm, xshift=0cm] (l) {$(d)$}				;
					\end{scope}
					
					\begin{scope}[yshift=-1cm,xshift=8.5cm]
						\path[->,  thick]
						node[format] (z) {$Z_i$}
						node[format, right of=z] (v1) {$V_1^{*}$}
						node[format, right of=v1] (v2) {$\cdots$}
						node[format, right of=v2] (v3) {$V_K^{*}$}
						node[format, right of=v3] (r) {$R_j$}

						(z) edge[<->,red] (v1)
						(v1) edge[<->,red] (v2)
						(v2) edge[<->,red] (v3)
						(v3) edge[<->,red] (r)
						(r) edge[black,opacity=0.3, bend right=20] (z)

						node[below of=v2, yshift=0.8cm, xshift=0cm] (l) {$(e)$}				;
					\end{scope}
					
					\begin{scope}[yshift=-5cm,xshift=8.5cm]
						\path[->,  thick]
						node[format] (z) {$Z_i$}
						node[format, right of=z] (v1) {$V_1^{*}$}
						node[format, right of=v1] (v2) {$\cdots$}
						node[format, right of=v2] (v3) {$R_K$}
						node[format, right of=v3] (r) {$R_j$}

						(z) edge[<->,red] (v1)
						(v1) edge[<->,red] (v2)
						(v2) edge[<->,red] (v3)
						(v3) edge[<-,blue] (r)
						(r) edge[black,opacity=0.3, bend right=20] (z)
						
						node[below of=v2, yshift=0.8cm, xshift=0cm] (l) {$(f)$}				;
					\end{scope}
					
				\end{tikzpicture}
				
			}
		\end{center}
		\caption{(a) e-colluding paths between $Z_i^{(1,r_{\aff(i)} = {\bf 1})}$ and $R_j$ where $R_j \in {\bf R}_{\aff(i)}$ ($i$ and $j$ are neighbors)
			(b) Projecting out $Z_i^{(1,r_{\aff(i)} = {\bf 1})}$
		}
		\label{fig:colluding-paths}
	\end{figure}
	Finally, we present the general argument for arbitrary graphs. 
	
	Assume that there are $n$ variables in $\tilde{{\bf Z}}$. For simplicity, assume that all counterfactuals are independent of each other, i.e, $Z_i^{(1,{\bf r}_{\aff(i)}={\bf r})} \ci Z_i^{(1,{\bf r}_{\aff(i)}= {\bf r'}) }$, when ${\bf r} \neq {\bf r'}$ and $Z_i^{(1,{\bf r}_{\aff(i)}={\bf r})} \ci Z_j^{(1,{\bf r}_{\aff(i)} = {\bf r'})}$, when ${\bf r}$ may or may not be equal to ${\bf r'}$. Graphically, there are no edges between counterfactuals \footnote{This assumption does not affect the generality of completeness for other types of models because this model is a submodel of others that involve dependence between counterfactuals, and completeness in this model guarantees completeness in others.}. And by Assumption \ref{assump:10-not-parent]}, $\tilde{{\bf Z}}^{(\bf r \neq 1)}$ is not a parent of any $R \in {\bf R}$, and can be projected out (adding no new edges) without affecting the Markov blanket of any variable in $\tilde{{\bf Z}}^{(\bf r = 1)}$, $\bf R$ or proxies $\bf Z$.
	
	Shown in Fig.~\ref{fig:colluding-paths} (a)-(d) are all possible e-colluding paths between $Z_i^{(1,{\bf r}_{\aff(i)} = {\bf 1})}$ and $R_j$. Here, $R_j \in {\bf R}_{\aff(i)}$. Assume there are $K$ variables, $V_1 \cdots V_K$, that lie on the smallest collider path between $Z_i^{(1,{\bf r}_{\aff(i)} = {\bf 1})}$ and $R_j$. For this to be the smallest collider path, we  require that $Z_l^{(1,{\bf r}_{\aff(l)} = {\bf 1})}$  or $R_l$ are not in $V_1, \cdots V_K$ when $i$ and $l$ share the same neighborhood, i.e., when $R_l \in {\bf R}_{\aff(i)}$ and $R_i \in {\bf R}_{\aff(l)}$. If not, we could truncate the path to have the smallest such path that goes between $Z_i^{(1,{\bf r}_{\aff(i)} = {\bf 1})}$ and $R_j$ such that $i$ and $j$ are neighbors. Fig.~\ref{fig:colluding-paths} (e) shows the projection of (a) and (b), and (f) shows the projection of (c) and (d). $V^{*} \in {\bf Z} \setminus Z_l, \:\: {\bf R} \setminus R_l$, $i$ and $l$ are neighbors.
	
	We now consider each of these paths (a)-(d) and their corresponding latent projections (e) and (f), as if they appear in a larger graph that is otherwise completely disconnected. Akin to what we did with the examples earlier, we count the number of M\"{o}bius parameters (as a function of $K$), and show that the full observability law always has more parameters than the observed law. When we place these colluding paths in a larger graph with arbitrary connectivity, the full observability law is still not identified because of the discrepancy arising from the g-colluding path alone. That is, any edge super graph (super model) is also not identified.  
	
	The following fact will be used towards counting parameters in a binary model: Given a bidirected chain $V_1 \leftrightarrow \cdots \leftrightarrow V_{K'}$ of length $K'$, the number of parameters required to parameterize this chain is $\frac{K'(K'+1)}{2}$, corresponding to parameters given by the following: 
	\begin{align*}
		q(V_1=0) \: q(V_1=V_2=0) \: \cdots q(V_1=\cdots V_{K'}=0) \: :& \: {K'} \text{ params} \\
		q(V_2=0) \: q(V_2=V_3=0) \: \cdots q(V_2=\cdots V_{K'}=0) \: :& \: {K'}-1 \text{ params} \\
		\cdots \\
		q(V_{K'}=0) \: :& \: 1  \text{ param}
	\end{align*}
	
	\textbf{Parameter counting for Fig.~\ref{fig:colluding-paths}(a), (b), (e)}
	\begin{enumerate}
		\item Number of M\"{o}bius parameters in Fig.~\ref{fig:colluding-paths}(a) is $ \frac{(K+2)(K+3)}{2}$
		\begin{itemize}
			\item It is a bidirected chain $Z_i^{(1,{\bf r}_{\aff(i)} = {\bf 1})} \leftrightarrow \cdots \leftrightarrow R_j$ of length ${K'}=K+2$.
		\end{itemize}
		\item Number of M\"{o}bius parameters in Fig.~\ref{fig:colluding-paths}(b) is $ \frac{(K+2)(K+3)}{2}$
		\begin{itemize}
			\item $q(Z_i^{(1,{\bf r}_{\aff(i)} = {\bf 1})} = 0)$ i.e. 1 parameter.
			\item A bidirected chain $V_2 \leftrightarrow \cdots \leftrightarrow V_K \leftrightarrow R_j$ of length $K'=K$, i.e. $\frac{K(K+1)}{2}$ parameters.
			\item Intrinsic sets involving $V_1$, i.e., $q(V_1 =0|Z_i^{(1,{\bf r}_{\aff(i)} = {\bf 1})} )$, $q(V_1 =V_2=0|Z_i^{(1,{\bf r}_{\aff(i)} = {\bf 1})} )$,  $\cdots$, $q(V_1 =V_2= \cdots R_j=0|Z_i^{(1,{\bf r}_{\aff(i)} = {\bf 1})} )$ which have $2$ parameters each, leading to $2*(K+1)$ parameters totally.
		\end{itemize}
		\item Number of M\"{o}bius parameters in Fig.~\ref{fig:colluding-paths}(e) is $ \frac{(K+2)(K+3)}{2} - 1$
		\begin{itemize}
			\item The bidirected chain $V_1^* \leftrightarrow \cdots \leftrightarrow R_j$ of length $K' = K+1$ has $\frac{(K+1)(K+2)}{2}$ parameters. It should be noted that for any $Z_l \in \{V_1 \cdots V_K\}$, in the head of a M\"{o}bius parameter, if $R \in R_l \cup {\bf R}_{\aff(l)}$ is the parameter's tail, $R$ is always set to 1 deterministically. Hence, it reduces to counting the simple bidirected chain.
			\item The number of intrinsic sets involving $Z_i$ is $K+1$ (and not $K+2$) since $R_i$ is not fixable, and the set $\{ Z_i, V_1^*, \cdots, V_K^*\}$ is not intrinsic. Each of these intrinsic sets corresponds to two parameters, so $2*(K+1)$ parameters.
		\end{itemize}
		
	\end{enumerate}
	
	\textbf{Parameter counting for Fig.~\ref{fig:colluding-paths}(c), (d), (f)}
	\begin{enumerate}
		\item Number of M\"{o}bius parameters in Fig.~\ref{fig:colluding-paths}(c) is $ \frac{(K+2)(K+3)}{2}$
		\begin{itemize}
			\item $q(R_j=0)$ i.e. 1 parameter.
			\item A bidirected chain $Z_i^{(1,{\bf r}_{\aff(i)} = {\bf 1})}  \leftrightarrow \cdots V_{K-1}$ of length $K' = K$, so $\frac{K(K+1)}{2}$ parameters.
			\item Intrinsic sets involving $R_K$, i.e., $q(R_K=0| R_j)$, $q(R_K=V_{K-1}=0| R_j)$, $\cdots$, $q(R_K= V_{K-1}= \cdots , Z_i^{(1,{\bf r}_{\aff(i)} = {\bf 1})} = 0 |R_j)$ , i.e, $2*(K+1)$ parameters. 
		\end{itemize}
		\item Number of M\"{o}bius parameters in Fig. \ref{fig:colluding-paths}(d) is $ \frac{(K+2)(K+3)}{2}$
		\begin{itemize}
			\item $q(Z_i^{(1,r_{\aff(i)} = {\bf 1})} = 0)$ i.e. 1 parameter.
			\item $q(R_j = 0)$, i.e., 1 parameter.
			\item A bidirected chain $V_2 \leftrightarrow \cdots \leftrightarrow V_{K-1}$ of length $K'=K-2$, i.e. $\frac{(K-2)(K-1)}{2}$ parameters.
			\item Intrinsic sets involving $V_1$ and not $R_K$, i.e., $q(V_1 =0|Z_i^{(1,{\bf r}_{\aff(i)} = {\bf 1})} )$, $q(V_1 =V_2=0|Z_i^{(1,{\bf r}_{\aff(i)} = {\bf 1})} )$,  $\cdots$, $q(V_1 =V_2= \cdots V_{K-1}=0|Z_i^{(1,{\bf r}_{\aff(i)} = {\bf 1})} )$ which give $2$ parameters each, leading to $2*(K-1)$ parameters.
			\item Intrinsic sets involving $R_K$ and not $V_1$, i.e., $q(R_K =0|Z_i^{(1,{\bf r}_{\aff(i)} = {\bf 1})} )$, $q(R_K =V_2=0|Z_i^{(1,{\bf r}_{\aff(i)} = {\bf 1})} )$,  $\cdots$, $q(R_K =V_2= \cdots V_{K-1}=0|Z_i^{(1,{\bf r}_{\aff(i)} = {\bf 1})} )$ which give $2$ parameters each, leading to $2*(K-1)$ parameters.
			\item The one intrinsic set involving both $V_1$ and $R_K$, i.e, $q(V_1=V_2=, \cdots = R_K = 0| R_j,Z_i^{(1,{\bf r}_{\aff(i)} = {\bf 1})} )$	corresponding to 4 parameters.
		\end{itemize}
		\item Number of M\"{o}bius parameters in Fig.~\ref{fig:colluding-paths}(f) is $ \frac{(K+2)(K+3)}{2} - 1$
		\begin{itemize}
			\item $q(R_j = 0)$, i.e., 1 parameter.
			\item The bidirected chain $Z_i \leftrightarrow V_1^* \leftrightarrow \cdots \leftrightarrow V_{K-1}$ of length $K' = K$ has $\frac{(K)(K+1)}{2}$ parameters. As before, for any $Z_l \in \{V_1 \cdots V_{K-1}\}$, in the head of a M\"{o}bius parameter, if $R \in R_l \cup {\bf R}_{\aff(l)}$ is the parameter's tail, $R$ is always set to 1 deterministically.
			\item Intrinsic sets involving $R_K$, i.e., $q(R_K = 0 |R_j)$, \\
			$q(R_K ,V_{K-1}= 0 |R_j)$, $\cdots ,q(R_K ,V_{K-1}, \cdots , V_1 = 0 |R_j)$ corresponding to $2*K$ parameters, and the intrinsic set $q(R_K ,V_{K-1}, \cdots , V_1 = 0, Z_i=0 |R_j=1)$ which only corresponds to 1 parameter (instead of 2) since $R_j \in {\bf R}_{\aff(i)}$ and has to be set to 1.
		\end{itemize}
		
	\end{enumerate}
\end{prf}
	\end{appendix}
	
	\clearpage

	\bibliography{references}

\begin{thebibliography}{89}
\providecommand{\natexlab}[1]{#1}
\providecommand{\url}[1]{\texttt{#1}}
\expandafter\ifx\csname urlstyle\endcsname\relax
  \providecommand{\doi}[1]{doi: #1}\else
  \providecommand{\doi}{doi: \begingroup \urlstyle{rm}\Url}\fi

\bibitem[Rubin(1976)]{rubin1976}
Donald~B. Rubin.
\newblock Inference and missing data.
\newblock \emph{Biometrika}, 63\penalty0 (3):\penalty0 581--592, 1976.

\bibitem[Little(2021)]{little2021missing}
Roderick~J Little.
\newblock Missing data assumptions.
\newblock \emph{Annual Review of Statistics and Its Application}, 8:\penalty0
  89--107, 2021.

\bibitem[Little and Rubin(2002)]{little2002statistical}
R.J.A. Little and D.B. Rubin.
\newblock \emph{Statistical Analysis with Missing Data}.
\newblock Wiley Series in Probability and Statistics. Wiley, 2002.

\bibitem[Glymour(2006)]{glymour2006using}
M~Maria Glymour.
\newblock Using causal diagrams to understand common problems in social
  epidemiology.
\newblock \emph{Methods in social epidemiology}, pages 393--428, 2006.

\bibitem[Daniel et~al.(2012)Daniel, Kenward, Cousens, and
  De~Stavola]{daniel2012using}
Rhian~M. Daniel, Michael~G. Kenward, Simon~N. Cousens, and Bianca~L.
  De~Stavola.
\newblock Using causal diagrams to guide analysis in missing data problems.
\newblock \emph{Statistical Methods in Medical Research}, 21\penalty0
  (3):\penalty0 243--256, 2012.

\bibitem[Martel~Garc{\'\i}a(2013)]{martel2013definition}
Fernando Martel~Garc{\'\i}a.
\newblock Definition and diagnosis of problematic attrition in randomized
  controlled experiments.
\newblock \emph{Available at SSRN 2302735}, 2013.

\bibitem[Mohan et~al.(2013)Mohan, Pearl, and Tian]{mohan13missing}
Karthika Mohan, Judea Pearl, and Jin Tian.
\newblock Graphical models for inference with missing data.
\newblock \emph{Advances in neural information processing systems}, 26, 2013.

\bibitem[Thoemmes and Rose(2014)]{thoemmes2014cautious}
Felix Thoemmes and Norman Rose.
\newblock A cautious note on auxiliary variables that can increase bias in
  missing data problems.
\newblock \emph{Multivariate Behavioral Research}, 49\penalty0 (5):\penalty0
  443--459, 2014.

\bibitem[Tian(2015)]{tian2015missing}
Jin Tian.
\newblock Missing at random in graphical models.
\newblock In \emph{Artificial Intelligence and Statistics}, pages 977--985.
  PMLR, 2015.

\bibitem[Shpitser(2016)]{shpitser2016consistent}
Ilya Shpitser.
\newblock Consistent estimation of functions of data missing non-monotonically
  and not at random.
\newblock \emph{Advances in Neural Information Processing Systems},
  29:\penalty0 3144--3152, 2016.

\bibitem[Bhattacharya et~al.(2019)Bhattacharya, Nabi, Shpitser, and
  Robins]{bhattacharya19mid}
Rohit Bhattacharya, Razieh Nabi, Ilya Shpitser, and James Robins.
\newblock Identification in missing data models represented by directed acyclic
  graphs.
\newblock In \emph{Proceedings of the Thirty Fifth Conference on Uncertainty in
  Artificial Intelligence (UAI-35th)}. AUAI Press, 2019.

\bibitem[Nabi et~al.(2020)Nabi, Bhattacharya, and Shpitser]{nabi2020full}
Razieh Nabi, Rohit Bhattacharya, and Ilya Shpitser.
\newblock Full law identification in graphical models of missing data:
  Completeness results.
\newblock In \emph{International Conference on Machine Learning}, pages
  7153--7163. PMLR, 2020.

\bibitem[Mohan and Pearl(2021)]{mohan2021graphical}
Karthika Mohan and Judea Pearl.
\newblock Graphical models for processing missing data.
\newblock \emph{Journal of the American Statistical Association}, pages 1--16,
  2021.

\bibitem[Scharfstein et~al.(2021)Scharfstein, Lee, McDermott, Campbell, Nunes,
  Matthews, and Shpitser]{scharfstein2021markov}
Daniel~O Scharfstein, Jaron~JR Lee, Aidan McDermott, Aimee Campbell, Edward
  Nunes, Abigail~G Matthews, and Ilya Shpitser.
\newblock Markov-restricted analysis of randomized trials with non-monotone
  missing binary outcomes: Sensitivity analysis and identification results.
\newblock \emph{arXiv preprint arXiv:2105.08868}, 2021.

\bibitem[Nabi and Bhattacharya(2022)]{nabi2022testability}
Razieh Nabi and Rohit Bhattacharya.
\newblock On testability and goodness of fit tests in missing data models.
\newblock \emph{arXiv preprint arXiv:2203.00132}, 2022.

\bibitem[Nabi et~al.(2022{\natexlab{a}})Nabi, Bhattacharya, Shpitser, and
  Robins]{nabi2022causal}
Razieh Nabi, Rohit Bhattacharya, Ilya Shpitser, and James Robins.
\newblock Causal and counterfactual views of missing data models.
\newblock \emph{arXiv preprint arXiv:2210.05558}, 2022{\natexlab{a}}.

\bibitem[Aronow and Samii(2013)]{aronow2013ace}
P.~M. Aronow and C.~Samii.
\newblock Estimating average causal effects under general interference.
\newblock \emph{Technical Report}, 2013.

\bibitem[Athey et~al.(2018)Athey, Eckles, and Imbens]{athey2018network}
S.~Athey, D.~Eckles, and G.W. Imbens.
\newblock Exact p-values for network interference.
\newblock \emph{Journal of the American Statistical Association}, 113
  (521):\penalty0 230--240, 2018.

\bibitem[Basse et~al.(2019)Basse, Feller, and Toulis]{basse2019interference}
G.~Basse, A.~Feller, and P.~Toulis.
\newblock Randomization tests of causal effects under interference.
\newblock \emph{Biometrika}, 106 (2):\penalty0 487--494, 2019.

\bibitem[Basse and Airoldi(2018)]{basse2018network}
G.~W. Basse and E.~M. Airoldi.
\newblock Model-assisted design of experiments in the presence of
  network-correlated outcomes.
\newblock \emph{Biometrika}, 105 (4):\penalty0 849--858, 2018.

\bibitem[Bowers et~al.(2013)Bowers, M, and C]{bowers2013interference}
J.~Bowers, F.~M. M, and P.~C.
\newblock Reasoning about interference between units: A general framework.
\newblock \emph{Political Analysis}, 21:\penalty0 97--124, 2013.

\bibitem[Cai et~al.(2019)Cai, Loh, and Crawford]{cai2019idcontagion}
X.~Cai, W.~W. Loh, and F.~W. Crawford.
\newblock Identification of causal intervention effects under contagion.
\newblock \emph{arXiv preprint arXiv:1912.04151}, 2019.

\bibitem[Eck et~al.(2022)Eck, Morozova, and Crawford]{eck2022randomization}
Daniel~J Eck, Olga Morozova, and Forrest~W Crawford.
\newblock Randomization for the susceptibility effect of an infectious disease
  intervention.
\newblock \emph{Journal of Mathematical Biology}, 85\penalty0 (4):\penalty0
  1--22, 2022.

\bibitem[Eckles et~al.(2017)Eckles, Karrer, and Ugander]{eckles2017design}
Dean Eckles, Brian Karrer, and Johan Ugander.
\newblock Design and analysis of experiments in networks: Reducing bias from
  interference.
\newblock \emph{Journal of Causal Inference}, 5\penalty0 (1), 2017.

\bibitem[Forastiere et~al.(2021)Forastiere, Airoldi, and
  Mealli]{forastiere2021identification}
Laura Forastiere, Edoardo~M Airoldi, and Fabrizia Mealli.
\newblock Identification and estimation of treatment and interference effects
  in observational studies on networks.
\newblock \emph{Journal of the American Statistical Association}, 116\penalty0
  (534):\penalty0 901--918, 2021.

\bibitem[Nabi et~al.(2022{\natexlab{b}})Nabi, Pfeiffer, Charles, and
  K{\i}c{\i}man]{nabi2022ads}
Razieh Nabi, Joel Pfeiffer, Denis Charles, and Emre K{\i}c{\i}man.
\newblock Causal inference in the presence of interference in sponsored search
  advertising.
\newblock \emph{Frontiers in big Data}, 5, 2022{\natexlab{b}}.

\bibitem[Graham et~al.(2010)Graham, Imbens, and Ridder]{graham2010measuring}
Bryan~S. Graham, Guido~W. Imbens, and Geert Ridder.
\newblock Measuring the effects of segregation in the presence of social
  spillovers: a nonparametric approach.
\newblock Technical report, National Bureau of Economic Research, 2010.

\bibitem[Halloran and Struchiner(1995)]{halloran1995causal}
M.~Elizabeth Halloran and Claudio~J. Struchiner.
\newblock Causal inference in infectious diseases.
\newblock \emph{Epidemiology}, pages 142--151, 1995.

\bibitem[Halloran and Hudgens(2012)]{halloran2012causal}
M~Elizabeth Halloran and Michael~G Hudgens.
\newblock Causal inference for vaccine effects on infectiousness.
\newblock \emph{The international journal of biostatistics}, 8\penalty0
  (2):\penalty0 1--40, 2012.

\bibitem[Hong and Raudenbush(2006)]{hong2006evaluating}
Guanglei Hong and Stephen~W. Raudenbush.
\newblock Evaluating kindergarten retention policy: A case study of causal
  inference for multilevel observational data.
\newblock \emph{Journal of the American Statistical Association}, 101\penalty0
  (475):\penalty0 901--910, 2006.

\bibitem[Hudgens and Halloran(2008{\natexlab{a}})]{hudgens2008toward}
Michael~G Hudgens and M~Elizabeth Halloran.
\newblock Toward causal inference with interference.
\newblock \emph{Journal of the American Statistical Association}, 103\penalty0
  (482):\penalty0 832--842, 2008{\natexlab{a}}.

\bibitem[Jagadeesan et~al.(2020)Jagadeesan, Pillai, and
  Volfovsky]{jagadeesan2020designs}
Ravi Jagadeesan, Natesh~S Pillai, and Alexander Volfovsky.
\newblock Designs for estimating the treatment effect in networks with
  interference.
\newblock \emph{The Annals of Statistics}, 48\penalty0 (2):\penalty0 679--712,
  2020.

\bibitem[Toulis and Kao(2013)]{toulis2013estimation}
Panos Toulis and Edward Kao.
\newblock Estimation of causal peer influence effects.
\newblock In \emph{International conference on machine learning}, pages
  1489--1497. PMLR, 2013.

\bibitem[Leung(2020)]{leung2020treatment}
Michael~P Leung.
\newblock Treatment and spillover effects under network interference.
\newblock \emph{Review of Economics and Statistics}, 102\penalty0 (2):\penalty0
  368--380, 2020.

\bibitem[Liu and Hudgens(2014)]{liu2014large}
Lan Liu and Michael~G Hudgens.
\newblock Large sample randomization inference of causal effects in the
  presence of interference.
\newblock \emph{Journal of the american statistical association}, 109\penalty0
  (505):\penalty0 288--301, 2014.

\bibitem[Papadogeorgou et~al.(2019)Papadogeorgou, Mealli, and
  Zigler]{papadogeorgou2019causal}
Georgia Papadogeorgou, Fabrizia Mealli, and Corwin~M Zigler.
\newblock Causal inference with interfering units for cluster and population
  level treatment allocation programs.
\newblock \emph{Biometrics}, 75\penalty0 (3):\penalty0 778--787, 2019.

\bibitem[Puelz et~al.(2019)Puelz, Basse, Feller, and Toulis]{puelz2019graph}
David Puelz, Guillaume Basse, Avi Feller, and Panos Toulis.
\newblock A graph-theoretic approach to randomization tests of causal effects
  under general interference.
\newblock \emph{arXiv preprint arXiv:1910.10862}, 2019.

\bibitem[Rosenbaum(2007)]{rosenbaum2007interference}
Paul~R. Rosenbaum.
\newblock Interference between units in randomized experiments.
\newblock \emph{Journal of the American Statistical Association}, 102\penalty0
  (477):\penalty0 191--200, 2007.

\bibitem[Rubin(1990)]{rubin1990comment}
Donald~B Rubin.
\newblock Comment: Neyman (1923) and causal inference in experiments and
  observational studies.
\newblock \emph{Statistical Science}, 5\penalty0 (4):\penalty0 472--480, 1990.

\bibitem[S{\"a}vje(2021)]{savje2021causal}
Fredrik S{\"a}vje.
\newblock Causal inference with misspecified exposure mappings.
\newblock \emph{arXiv preprint arXiv:2103.06471}, 2021.

\bibitem[S{\"a}vje et~al.(2021)S{\"a}vje, Aronow, and
  Hudgens]{savje2021average}
Fredrik S{\"a}vje, Peter Aronow, and Michael Hudgens.
\newblock Average treatment effects in the presence of unknown interference.
\newblock \emph{Annals of statistics}, 49\penalty0 (2):\penalty0 673, 2021.

\bibitem[Sobel(2006)]{sobel2006randomized}
Michael~E Sobel.
\newblock What do randomized studies of housing mobility demonstrate? causal
  inference in the face of interference.
\newblock \emph{Journal of the American Statistical Association}, 101\penalty0
  (476):\penalty0 1398--1407, 2006.

\bibitem[Tchetgen and VanderWeele(2012)]{tchetgen2012causal}
Eric J~Tchetgen Tchetgen and Tyler~J VanderWeele.
\newblock On causal inference in the presence of interference.
\newblock \emph{Statistical methods in medical research}, 21\penalty0
  (1):\penalty0 55--75, 2012.

\bibitem[Toulis et~al.(2018)Toulis, Volfovsky, and
  Airoldi]{toulis2018propensity}
Panos Toulis, Alexander Volfovsky, and Edoardo~M Airoldi.
\newblock Propensity score methodology in the presence of network entanglement
  between treatments.
\newblock \emph{arXiv preprint arXiv:1801.07310}, 2018.

\bibitem[VanderWeele(2010)]{vanderweele2010direct}
Tyler~J VanderWeele.
\newblock Direct and indirect effects for neighborhood-based clustered and
  longitudinal data.
\newblock \emph{Sociological methods \& research}, 38\penalty0 (4):\penalty0
  515--544, 2010.

\bibitem[Chang et~al.(2020)Chang, Deng, Jiang, and Long]{chang2020multiple}
Changgee Chang, Yi~Deng, Xiaoqian Jiang, and Qi~Long.
\newblock Multiple imputation for analysis of incomplete data in distributed
  health data networks.
\newblock \emph{Nature communications}, 11\penalty0 (1):\penalty0 1--11, 2020.

\bibitem[Smith et~al.(2017)Smith, Moody, and Morgan]{smith2017network}
Jeffrey~A Smith, James Moody, and Jonathan~H Morgan.
\newblock Network sampling coverage ii: The effect of non-random missing data
  on network measurement.
\newblock \emph{Social networks}, 48:\penalty0 78--99, 2017.

\bibitem[Gile and Handcock(2017)]{gile2017analysis}
Krista~J Gile and Mark~S Handcock.
\newblock Analysis of networks with missing data with application to the
  national longitudinal study of adolescent health.
\newblock \emph{Journal of the Royal Statistical Society: Series C (Applied
  Statistics)}, 66\penalty0 (3):\penalty0 501--519, 2017.

\bibitem[Almquist and Butts(2018)]{almquist2018dynamic}
Zack~W Almquist and Carter~T Butts.
\newblock Dynamic network analysis with missing data: theory and methods.
\newblock \emph{Statistica Sinica}, 28\penalty0 (3):\penalty0 1245--1264, 2018.

\bibitem[Sacerdote(2011)]{sacerdote2011peer}
Bruce Sacerdote.
\newblock Peer effects in education: How might they work, how big are they and
  how much do we know thus far?
\newblock In \emph{Handbook of the Economics of Education}, volume~3, pages
  249--277. Elsevier, 2011.

\bibitem[Pearl(2009{\natexlab{a}})]{pearl09causality}
Judea Pearl.
\newblock \emph{Causality: Models, Reasoning, and Inference}.
\newblock Cambridge University Press, 2 edition, 2009{\natexlab{a}}.
\newblock ISBN 978-0521895606.

\bibitem[Maathuis et~al.(2018)Maathuis, Drton, Lauritzen, and
  Wainwright]{maathuis2018handbook}
Marloes Maathuis, Mathias Drton, Steffen Lauritzen, and Martin Wainwright.
\newblock \emph{Handbook of Graphical Models}.
\newblock CRC Press, 2018.

\bibitem[Richardson and Robins(2013)]{thomas13swig}
Thomas~S. Richardson and Jamie~M. Robins.
\newblock Single world intervention graphs ({SWIG}s): A unification of the
  counterfactual and graphical approaches to causality.
\newblock \emph{preprint}, 2013.

\bibitem[Pearl(1988)]{pearl88probabilistic}
Judea Pearl.
\newblock \emph{Probabilistic Reasoning in Intelligent Systems}.
\newblock Morgan and Kaufmann, San Mateo, 1988.

\bibitem[Pearl(2009{\natexlab{b}})]{pearl2009causality}
Judea Pearl.
\newblock \emph{Causality}.
\newblock Cambridge University Press, 2009{\natexlab{b}}.

\bibitem[Hernan and Robins(2020)]{hernan2020whatif}
Miguel~A Hernan and James~M Robins.
\newblock \emph{Causal Inference: What If}.
\newblock CRC Press: Taylor and Francis Group, 2020.

\bibitem[Verma and Pearl(1990)]{verma90equiv}
Thomas~S. Verma and Judea Pearl.
\newblock Equivalence and synthesis of causal models.
\newblock Technical Report R-150, Department of Computer Science, University of
  California, Los Angeles, 1990.

\bibitem[Richardson(2003)]{richardson03markov}
Thomas~S. Richardson.
\newblock Markov properties for acyclic directed mixed graphs.
\newblock \emph{Scandinavial Journal of Statistics}, 30\penalty0 (1):\penalty0
  145--157, 2003.

\bibitem[Tian and Pearl(2002)]{tian02on}
Jin Tian and Judea Pearl.
\newblock On the testable implications of causal models with hidden variables.
\newblock In \emph{Proceedings of the Eighteenth Conference on Uncertainty in
  Artificial Intelligence (UAI-02)}, volume~18, pages 519--527. AUAI Press,
  Corvallis, Oregon, 2002.

\bibitem[Shpitser and Pearl(2006)]{shpitser06id}
Ilya Shpitser and Judea Pearl.
\newblock Identification of joint interventional distributions in recursive
  semi-{M}arkovian causal models.
\newblock In \emph{Proceedings of the Twenty-First National Conference on
  Artificial Intelligence (AAAI-06)}. AAAI Press, Palo Alto, 2006.

\bibitem[Huang and Valtorta(2006)]{huang06do}
Yimin Huang and Marco Valtorta.
\newblock Pearl's calculus of interventions is complete.
\newblock In \emph{Twenty Second Conference On Uncertainty in Artificial
  Intelligence}, 2006.

\bibitem[Richardson et~al.(2017)Richardson, Evans, Robins, and
  Shpitser]{richardson17nested}
Thomas~S. Richardson, Robin~J. Evans, James~M. Robins, and Ilya Shpitser.
\newblock Nested {M}arkov properties for acyclic directed mixed graphs, 2017.
\newblock {W}orking paper.

\bibitem[Bhattacharya et~al.(2022)Bhattacharya, Nabi, and
  Shpitser]{bhattacharya2022semiparametric}
Rohit Bhattacharya, Razieh Nabi, and Ilya Shpitser.
\newblock Semiparametric inference for causal effects in graphical models with
  hidden variables.
\newblock \emph{Journal of Machine Learning Research}, 23:\penalty0 1--76,
  2022.

\bibitem[Ogburn and VanderWeele(2014)]{ogburn14interference}
Elizabeth~L. Ogburn and Tyler~J. VanderWeele.
\newblock Causal diagrams for interference.
\newblock \emph{Statistical Science}, 29\penalty0 (4):\penalty0 559--578, 2014.

\bibitem[Hudgens and Halloran(2008{\natexlab{b}})]{hudgens08toward}
M.G. Hudgens and M.E. Halloran.
\newblock Toward causal inference with interference.
\newblock \emph{Journal of the American Statistical Association}, 103\penalty0
  (482):\penalty0 832--842, 2008{\natexlab{b}}.

\bibitem[{Tchetgen Tchetgen} and VanderWeele(2012)]{tchetgen12on}
Eric~J. {Tchetgen Tchetgen} and Tyler~J. VanderWeele.
\newblock On causal inference in the presence of interference.
\newblock \emph{Statistical Methods in Medical Research}, 21\penalty0
  (1):\penalty0 55--75, 2012.

\bibitem[Tchetgen~Tchetgen et~al.(2021)Tchetgen~Tchetgen, Fulcher, and
  Shpitser]{tchetgen17auto}
Eric~J Tchetgen~Tchetgen, Isabel~R Fulcher, and Ilya Shpitser.
\newblock Auto-g-computation of causal effects on a network.
\newblock \emph{Journal of the American Statistical Association}, 116\penalty0
  (534):\penalty0 833--844, 2021.

\bibitem[Dempster et~al.(1977)Dempster, Laird, and Rubin]{dempster77maximum}
A.P. Dempster, N.M. Laird, and D.B. Rubin.
\newblock Maximum likelihood from incomplete data via the {EM} algorithm.
\newblock \emph{Journal of the Royal Statistical Society, Series B},
  39:\penalty0 1--38, 1977.

\bibitem[Horton and Laird(1999)]{horton1999maximum}
Nicholas~J Horton and Nan~M Laird.
\newblock Maximum likelihood analysis of generalized linear models with missing
  covariates.
\newblock \emph{Statistical Methods in Medical Research}, 8\penalty0
  (1):\penalty0 37--50, 1999.

\bibitem[Rubin(1988)]{rubin1988overview}
Donald~B Rubin.
\newblock An overview of multiple imputation.
\newblock In \emph{Proceedings of the survey research methods section of the
  American statistical association}, pages 79--84. Citeseer Princeton, NJ, USA,
  1988.

\bibitem[Schafer(1999)]{schafer1999multiple}
Joseph~L Schafer.
\newblock Multiple imputation: a primer.
\newblock \emph{Statistical methods in medical research}, 8\penalty0
  (1):\penalty0 3--15, 1999.

\bibitem[Robins et~al.(1994)Robins, Rotnitzky, and Zhao]{robins94estimation}
James~M. Robins, Andrea Rotnitzky, and Lue~P. Zhao.
\newblock Estimation of regression coefficients when some regressors are not
  always observed.
\newblock \emph{Journal of the American Statistical Association}, 89:\penalty0
  846--866, 1994.

\bibitem[Li et~al.(2013)Li, Shen, Li, and Robins]{li2013weighting}
Lingling Li, Changyu Shen, Xiaochun Li, and James~M Robins.
\newblock On weighting approaches for missing data.
\newblock \emph{Statistical methods in medical research}, 22\penalty0
  (1):\penalty0 14--30, 2013.

\bibitem[Tchetgen et~al.(2018)Tchetgen, Wang, and Sun]{tchetgen2018discrete}
Eric J~Tchetgen Tchetgen, Linbo Wang, and BaoLuo Sun.
\newblock Discrete choice models for nonmonotone nonignorable missing data:
  Identification and inference.
\newblock \emph{Statistica Sinica}, 28\penalty0 (4):\penalty0 2069, 2018.

\bibitem[Wu and Carroll(1988)]{wu1988est}
Margaret~C Wu and R.~J. Carroll.
\newblock Estimation and comparison of changes in the presence of informative
  right censoring by modeling the censoring process.
\newblock \emph{Biometrics}, 175-188, 1988.

\bibitem[Wang et~al.(2014)Wang, Shao, and Kim]{wang2014instrumental}
Sheng Wang, Jun Shao, and Jae~Kwang Kim.
\newblock An instrumental variable approach for identification and estimation
  with nonignorable nonresponse.
\newblock \emph{Statistica Sinica}, pages 1097--1116, 2014.

\bibitem[Miao et~al.(2016)Miao, Ding, and Geng]{miao2016identifiability}
Wang Miao, Peng Ding, and Zhi Geng.
\newblock Identifiability of normal and normal mixture models with nonignorable
  missing data.
\newblock \emph{Journal of the American Statistical Association}, 111\penalty0
  (516):\penalty0 1673--1683, 2016.

\bibitem[Miao and Tchetgen~Tchetgen(2016)]{miao2016varieties}
Wang Miao and Eric~J Tchetgen~Tchetgen.
\newblock On varieties of doubly robust estimators under missingness not at
  random with a shadow variable.
\newblock \emph{Biometrika}, 103\penalty0 (2):\penalty0 475--482, 2016.

\bibitem[Sun et~al.(2018)Sun, Liu, Miao, Wirth, Robins, and
  Tchetgen]{sun2018semiparametric}
BaoLuo Sun, Lan Liu, Wang Miao, Kathleen Wirth, James Robins, and Eric
  J~Tchetgen Tchetgen.
\newblock Semiparametric estimation with data missing not at random using an
  instrumental variable.
\newblock \emph{Statistica Sinica}, 28\penalty0 (4):\penalty0 1965, 2018.

\bibitem[Jackson et~al.(2020)Jackson, Lin, and Yu]{jackson2020adjusting}
Matthew~O Jackson, Zhongjian Lin, and Ning~Neil Yu.
\newblock Adjusting for peer-influence in propensity scoring when estimating
  treatment effects.
\newblock \emph{Available at SSRN 3522256}, 2020.

\bibitem[Brown(1990)]{brown1990protecting}
C~Hendricks Brown.
\newblock Protecting against nonrandomly missing data in longitudinal studies.
\newblock \emph{Biometrics}, pages 143--155, 1990.

\bibitem[Mohan et~al.(2018)Mohan, Thoemmes, and Pearl]{mohan2018estimation}
Karthika Mohan, Felix Thoemmes, and Judea Pearl.
\newblock Estimation with incomplete data: The linear case.
\newblock In \emph{Proceedings of the International Joint Conferences on
  Artificial Intelligence Organization}, 2018.

\bibitem[Ebeling and Feistel(2011)]{ebeling2011physics}
Werner Ebeling and Rainer Feistel.
\newblock \emph{Physics of Self-organization and Evolution}.
\newblock John Wiley \& Sons, 2011.

\bibitem[Chen(2007)]{chen07semiparametric}
Hua~Yun Chen.
\newblock A semiparametric odds ratio model for measuring association.
\newblock \emph{biometrics}, 63:\penalty0 413--421, 2007.

\bibitem[Andrews and Didelez(2020)]{andrews2020insights}
Ryan~M Andrews and Vanessa Didelez.
\newblock Insights into the cross-world independence assumption of causal
  mediation analysis.
\newblock \emph{Epidemiology}, 32\penalty0 (2):\penalty0 209--219, 2020.

\bibitem[Tsiatis(2006)]{tsiatis06missing}
Anastasios Tsiatis.
\newblock \emph{Semiparametric Theory and Missing Data}.
\newblock Springer-Verlag New York, 1st edition edition, 2006.

\bibitem[Lauritzen(1996)]{lauritzen96graphical}
Steffan~L. Lauritzen.
\newblock \emph{Graphical Models}.
\newblock Oxford, U.K.: Clarendon, 1996.

\bibitem[Evans and Richardson(2014)]{evans14markovian}
Robin~J. Evans and Thomas~S. Richardson.
\newblock Markovian acyclic directed mixed graphs for discrete data.
\newblock \emph{Annals of Statistics}, pages 1--30, 2014.

\bibitem[Malinsky et~al.(2021)Malinsky, Shpitser, and
  Tchetgen~Tchetgen]{malinsky2021semiparametric}
Daniel Malinsky, Ilya Shpitser, and Eric~J Tchetgen~Tchetgen.
\newblock Semiparametric inference for nonmonotone missing-not-at-random data:
  the no self-censoring model.
\newblock \emph{Journal of the American Statistical Association}, pages 1--9,
  2021.

\end{thebibliography}
	
\end{document}